\documentclass[fleqn,usenatbib]{mnras}

\usepackage[T1]{fontenc}

\DeclareRobustCommand{\VAN}[3]{#2}
\let\VANthebibliography\thebibliography
\def\thebibliography{\DeclareRobustCommand{\VAN}[3]{##3}\VANthebibliography}

\usepackage{newtxtext,newtxmath}
\usepackage{graphicx}
\usepackage{gensymb}
\usepackage{subcaption}
\usepackage{bm}
\usepackage{fontawesome}
\usepackage{array}
\usepackage{orcidlink}

\newcommand{\PreserveBackslash}[1]{\let\temp=\\#1\let\\=\temp}
\newcolumntype{C}[1]{>{\PreserveBackslash\centering}p{#1}}
\newcolumntype{R}[1]{>{\PreserveBackslash\raggedleft}p{#1}}
\newcolumntype{L}[1]{>{\PreserveBackslash\raggedright}p{#1}}

\newcommand{\lya}{Ly\ensuremath{\alpha}}%
\newcommand{\lyb}{Ly\ensuremath{\beta}}%

\newcommand{\alphatitlebold}{\texorpdfstring{$\bm{\alpha}$}{\textbf{α}}}
\newcommand{\kms}{\text{s$\cdot$km}^{-1}}
\newcommand{\invkms}{\text{km$\cdot$s}^{-1}}
\newcommand{\invAA}{\AA$^{-1}$}
\newcommand\hi{H{\textsc{i}}}
\newcommand\siii{Si{\textsc{ii}}}
\newcommand\siiii{Si{\textsc{iii}}}
\newcommand\siiv{Si{\textsc{iv}}}
\newcommand\mgii{Mg{\textsc{ii}}}
\newcommand\civ{C{\textsc{iv}}}
\newcommand\caii{Ca{\textsc{ii}}}
\newcommand\ovi{O{\textsc{vi}}}
\newcommand{\pk}{$P_{1\mathrm{D},\alpha}$}

\newcommand{\pp}{$P_{\mathrm{pipeline}}$}
\newcommand{\pn}{$P_{\mathrm{noise}}$}
\newcommand{\pd}{$P_{\mathrm{diff}}$}
\newcommand{\pr}{$P_{\mathrm{raw}}$}
\newcommand{\ps}{$P_{\mathrm{SB}}$}
\newcommand{\pme}{$P_{\mathrm{metals}}$}
\newcommand{\pl}{$P_{\mathrm{\lya}}$}
\newcommand{\snr}{$\overline{\mathrm{SNR}}$}
\newcommand{\git}[2]{\href{https://github.com/#1}{\faGithub}\footnote{\label{#1}#2\url{https://github.com/#1}}}

\newcommand{\svo}{SV1}
\newcommand{\svt}{SV3}
\newcommand{\da}{M2}
\newcommand{\dafull}{DESI-M2}
\newcommand{\svtda}{SV3+M2}

\newcommand\CR[1]{{#1}}

\title[DESI FFT one-dimensional Lyman-\texorpdfstring{$\alpha$}{α} power spectrum]{The Dark Energy Spectroscopic Instrument: One-dimensional power spectrum from first Lyman-\alphatitlebold~forest samples with Fast Fourier Transform}

\author[C. Ravoux et al.]{Corentin Ravoux$^{1,2}$\orcidlink{0000-0002-3500-6635},
Marie Lynn Abdul Karim$^{2}$,
Eric Armengaud$^{2}$\orcidlink{0000-0001-7600-5148},
Michael Walther$^{3,4}$\orcidlink{0000-0002-1748-3745},
\newauthor
Naim G{\" o}ksel Kara{\c c}ayl{\i}$^{5,6,7,8}$\orcidlink{0000-0001-7336-8912}, 
Paul Martini$^{5,7,8}$\orcidlink{0000-0002-4279-4182},
Julien Guy$^{9}$,
Jessica Nicole Aguilar$^{9}$,
Steven Ahlen$^{10}$\orcidlink{0000-0001-6098-7247},
\newauthor
Stephen Bailey$^{9}$\orcidlink{0000-0003-4162-6619},
Julian Bautista$^{1}$\orcidlink{0000-0002-9885-3989},
Sergio Felipe Beltran$^{11}$\orcidlink{0000-0001-6324-4019}, 
David Brooks$^{12}$,
Laura Cabayol-Garcia$^{13}$,
\newauthor
Sol{\` e}ne Chabanier$^{9}$\orcidlink{0000-0002-5692-5243},
Edmond Chaussidon$^{2}$\orcidlink{0000-0001-8996-4874},
Jonás Chaves-Montero$^{13}$\orcidlink{0000-0002-9553-4261},
Kyle Dawson$^{14}$,
Rodrigo de la Cruz$^{11}$\orcidlink{0000-0001-9908-9129},
\newauthor
Axel de la Macorra$^{15}$\orcidlink{0000-0002-1769-1640},
Peter Doel$^{12}$,
Kevin Fanning$^{8}$\orcidlink{0000-0003-2371-3356},
Andreu Font-Ribera$^{13}$\orcidlink{0000-0002-3033-7312},
Jaime Forero-Romero$^{16,17}$\orcidlink{0000-0002-2890-3725},
\newauthor
Satya Gontcho A Gontcho$^{9}$\orcidlink{0000-0003-3142-233X},
Alma X. Gonzalez-Morales$^{11,18}$\orcidlink{0000-0003-4089-6924},
Calum Gordon$^{13}$,
Hiram K. Herrera-Alcantar$^{11}$\orcidlink{0000-0002-9136-9609},
\newauthor
Klaus Honscheid$^{5,6,8}$,
Vid Ir{\v s}i{\v c}$^{19}$\orcidlink{0000-0002-5445-461X}, 
Mustapha Ishak$^{20}$\orcidlink{0000-0002-6024-466X},
Robert Kehoe$^{20}$,
Theodore Kisner$^{9}$\orcidlink{0000-0003-3510-7134},
Anthony Kremin$^{9}$\orcidlink{0000-0001-6356-7424},
\newauthor
Martin Landriau$^{9}$\orcidlink{0000-0003-1838-8528},
Laurent Le Guillou$^{21}$\orcidlink{0000-0001-7178-8868},
Michael Levi$^{9}$\orcidlink{0000-0003-1887-1018},
Zarija Luki\'c$^{9}$,
Christophe Magneville$^{2}$,
\newauthor
Aaron Meisner$^{22}$\orcidlink{0000-0002-1125-7384},
Ramon Miquel$^{13,23}$,
John Moustakas$^{24}$\orcidlink{0000-0002-2733-4559},
Eva-Maria Mueller$^{25}$,
Andrea Muñoz-Gutiérrez$^{15}$,
\newauthor
Lucas Napolitano$^{26}$\orcidlink{0000-0002-5166-8671},
Jundan Nie$^{27}$\orcidlink{0000-0001-6590-8122},
Gustavo Niz$^{11,28}$\orcidlink{0000-0002-1544-8946},
Nathalie Palanque-Delabrouille$^{2,9}$\orcidlink{0000-0003-3188-784X},
\newauthor
Will Percival$^{29,30,31}$\orcidlink{0000-0002-0644-5727},
Ignasi P{\'e}rez-R{\`a}fols$^{13}$\orcidlink{0000-0001-6979-0125},
Matthew Pieri$^{32}$,
Claire Poppett$^{9,33,34}$,
Francisco Prada$^{35}$\orcidlink{0000-0001-7145-8674},
\newauthor
César Ramírez Pérez$^{13}$,
Graziano Rossi$^{36}$,
Eusebio Sanchez$^{37}$\orcidlink{0000-0002-9646-8198},
David Schlegel$^{9}$,
Michael Schubnell$^{38,39}$,
\newauthor
Hee-Jong Seo$^{40}$\orcidlink{0000-0002-6588-3508},
Francesco Sinigaglia$^{41,42}$,
Ting Tan$^{21}$,
Gregory Tarl{\' e}$^{39}$\orcidlink{0000-0003-1704-0781},
Ben Wang$^{43}$\orcidlink{0000-0003-4877-1659},
Benjamin Weaver$^{22}$,
\newauthor
Christophe Y{\` e}che$^{2}$\orcidlink{0000-0001-5146-8533},
Zhimin Zhou$^{27}$\orcidlink{0000-0002-4135-0977} \\
$^{1}$Aix Marseille Univ, CNRS/IN2P3, CPPM, Marseille, France\\
$^{2}$IRFU, CEA, Université Paris-Saclay, F-91191 Gif-sur-Yvette, France\\
$^{3}$Excellence Cluster ORIGINS, Boltzmannstrasse 2, D-85748 Garching, Germany \\
$^{4}$University Observatory, Faculty of Physics, Ludwig-Maximilians-Universit\"{a}t, Scheinerstr. 1, 81677 M\"{u}nchen, Germany\\
$^{5}$Center for Cosmology and AstroParticle Physics, The Ohio State University, 191 West Woodruff Avenue, Columbus, OH 43210, USA\\
$^{6}$Department of Physics, The Ohio State University, 191 West Woodruff Avenue, Columbus, OH 43210, USA\\
$^{7}$Department of Astronomy, The Ohio State University, 4055 McPherson Laboratory, 140 W 18th Avenue, Columbus, OH 43210, USA\\
$^{8}$The Ohio State University, Columbus, 43210 OH, USA\\
$^{9}$Lawrence Berkeley National Laboratory, 1 Cyclotron Road, Berkeley, CA 94720, USA\\
$^{10}$Physics Dept., Boston University, 590 Commonwealth Avenue, Boston, MA 02215, USA\\
$^{11}$Departamento de F\'{i}sica, Universidad de Guanajuato - DCI, C.P. 37150, Leon, Guanajuato, M\'{e}xico\\
$^{12}$Department of Physics \& Astronomy, University College London, Gower Street, London, WC1E 6BT, UK\\
$^{13}$Institut de F{\'i}sica d'Altes Energies (IFAE), The Barcelona Institute of Science and Technology, 08193 Bellaterra (Barcelona), Spain\\
$^{14}$Department of Physics and Astronomy, The University of Utah, 115 South 1400 East, Salt Lake City, UT 84112, USA\\
$^{15}$Instituto de F\'{\i}sica, Universidad Nacional Aut\'{o}noma de M\'{e}xico,  Cd. de M\'{e}xico  C.P. 04510,  M\'{e}xico\\
$^{16}$Departamento de F\'isica, Universidad de los Andes, Cra. 1 No. 18A-10, Edificio Ip, CP 111711, Bogot\'a, Colombia\\
$^{17}$Observatorio Astron\'omico, Universidad de los Andes, Cra. 1 No. 18A-10, Edificio H, CP 111711 Bogot\'a, Colombia\\
$^{18}$Consejo Nacional de Ciencia y Tecnolog\'{\i}a, Av. Insurgentes Sur 1582. Colonia Cr\'{e}dito Constructor, Del. Benito Ju\'{a}rez C.P. 03940, M\'{e}xico D.F. M\'{e}xico\\
$^{19}$Kavli Institute for Cosmology, Department of Physics, University of Cambridge, Madingley Road, Cambridge CB3 0HA, UK\\
$^{20}$Department of Physics, Southern Methodist University, 3215 Daniel Ave., Dallas, TX 75205, USA\\
$^{21}$Sorbonne Universit\'{e}, CNRS/IN2P3, Laboratoire de Physique Nucl\'{e}aire et de Hautes Energies (LPNHE), FR-75005 Paris, France \\
$^{22}$NSF's NOIRLab, 950 N. Cherry Ave., Tucson, AZ 85719, USA\\
$^{23}$Instituci\'{o} Catalana de Recerca i Estudis Avan\c{c}ats, Passeig de Llu\'{\i}s Companys, 23, 08010 Barcelona, Spain\\
$^{24}$Department of Physics and Astronomy, Siena College, 515 Loudon Road, Loudonville, NY 12211, USA\\
$^{25}$Department of Physics \& Astronomy, University of Sussex, Brighton BN1 9QH, UK\\
$^{26}$Department of Physics \& Astronomy, University  of Wyoming, 1000 E. University, Dept.~3905, Laramie, WY 82071, USA\\
$^{27}$National Astronomical Observatories, Chinese Academy of Sciences, A20 Datun Rd., Chaoyang District, Beijing, 100012, P.R. China\\
$^{28}$Instituto Avanzado de Cosmolog\'{\i}a A.~C. San Marcos 11 - Atenas 202. Magdalena Contreras, 10720. Ciudad de M\'{e}xico, M\'{e}xico\\
$^{29}$Department of Physics and Astronomy, University of Waterloo, 200 University Ave W, Waterloo, ON N2L 3G1, Canada\\
$^{30}$Perimeter Institute for Theoretical Physics, 31 Caroline St. North, Waterloo, ON N2L 2Y5, Canada\\
$^{31}$Waterloo Centre for Astrophysics, University of Waterloo, 200 University Ave W, Waterloo, ON N2L 3G1, Canada\\
$^{32}$Aix Marseille Univ, CNRS, CNES, LAM, Marseille, France\\
$^{33}$Space Sciences Laboratory, University of California, Berkeley, 7 Gauss Way, Berkeley, CA  94720, USA\\
$^{34}$University of California, Berkeley, 110 Sproul Hall \#5800 Berkeley, CA 94720, USA\\
$^{35}$Instituto de Astrof\'{i}sica de Andaluc\'{i}a (CSIC), Glorieta de la Astronom\'{i}a, s/n, E-18008 Granada, Spain\\
$^{36}$Department of Physics and Astronomy, Sejong University, Seoul, 143-747, Korea\\
$^{37}$CIEMAT, Avenida Complutense 40, E-28040 Madrid, Spain\\
$^{38}$Department of Physics, University of Michigan, Ann Arbor, MI 48109, USA\\
$^{39}$University of Michigan, Ann Arbor, MI 48109, USA\\
$^{40}$Department of Physics \& Astronomy, Ohio University, Athens, OH 45701, USA\\
$^{41}$Instituto de Astrof\'{i}sica de Canarias, C/ Vía L\'{a}ctea, s/n, E-38205 La Laguna, Tenerife, Spain\\
$^{42}$Universidad de La Laguna, Dept. de Astrof\'{\i}sica, E-38206 La Laguna, Tenerife, Spain\\
$^{43}$Department of Astronomy, Tsinghua University, 30 Shuangqing Road, Haidian District, Beijing, China, 100190\\
}

\date{Accepted XXX. Received YYY; in original form ZZZ}

\pubyear{2023}

\begin{document}
\label{firstpage}
\pagerange{\pageref{firstpage}--\pageref{lastpage}}
\maketitle

\begin{abstract}

We present the one-dimensional Lyman-$\alpha$ forest power spectrum measurement using the first data provided by the Dark Energy Spectroscopic Instrument (DESI). The data sample comprises $26,330$ quasar spectra, at redshift $z > 2.1$, contained in the DESI Early Data Release and the first two months of the main survey. We employ a Fast Fourier Transform (FFT) estimator and compare the resulting power spectrum to an alternative likelihood-based method in a companion paper. We investigate methodological and instrumental contaminants associated to the new DESI instrument, applying techniques similar to previous Sloan Digital Sky Survey (SDSS) measurements. We use synthetic data based on log-normal approximation to validate and correct our measurement. We compare our resulting power spectrum with previous SDSS and high-resolution measurements. With relatively small number statistics, we successfully perform the FFT measurement, which is already competitive in terms of the scale range. At the end of the DESI survey, we expect a five times larger Lyman-$\alpha$ forest sample than SDSS, providing an unprecedented precise one-dimensional power spectrum measurement.

\end{abstract}


\section{Introduction}

The Lyman-$\alpha$ (\lya) forest can be observed from the ground in the optical spectra of distant quasars at redshift between the end phase of reionization ($z\sim 6$) and the peak of galaxy formation $z \sim 2$. The \lya~forest consists of a series of \lya~absorption lines caused by intervening neutral hydrogen located at various redshifts between the quasar and the observer. The \lya~forest is a powerful probe of the underlying matter density field at redshift $z > 2$, together with the astrophysical state of the intergalactic medium \citep{gunn_density_1965,lynds_absorption-line_1971,meiksin_physics_2009,mcquinn_evolution_2016}. 

In particular, the small-scale distribution of neutral hydrogen ($\sim$ Mpc) is imprinted in the fluctuations of the \lya~forest along the line-of-sight that can be accessed by measuring the one-dimensional \lya~forest power spectrum (denoted \pk). This measurement is sensitive to the amplitude and slope of the matter power spectrum at redshift $z>2$. The impact of cosmological parameters on \pk~can only be accurately predicted using hydrodynamical simulations \citep{borde_new_2014,bolton_sherwood_2017,walther_simulating_2021,puchwein_sherwood-relics_2023}. The realization of those simulations is made arduous by the large dynamic range needed to model the \lya~forest adequately \citep{lukic_lyman-alpha_2015,chabanier_modeling_2022}. Fitting data measurements with those simulation predictions provides constraints on the cosmological parameters $\sigma_8$, $n_{\mathrm{s}}$, and $\Omega_{\mathrm{m}}$, as well as on parameters describing the thermal properties of the intergalactic medium. In particular, the simulations described in \citet{walther_simulating_2021} are able to predict \pk~with sufficient accuracy (at the 1\% level) when compared to expected uncertainties from the Dark Energy Spectroscopic Instrument (DESI) measurement.

Due to its sensitivity to the matter fluctuations at small scales, measurements of \pk~can constrain physics beyond the Standard Model, such as the mass of neutrinos, the mass of warm dark matter candidates, or a possible running of the spectral index due to primordial inflation physics. First, \pk~is well suited to constrain the sum of neutrino masses which damps the matter power spectrum at small scales \citep{lesgourgues_massive_2006,lesgourgues_neutrino_2012}. Stringent constraints are obtained by coupling \pk~with hydrodynamical simulations, and by combining it with the Cosmic Microwave Background (CMB) \citep{seljak_cosmological_2006,palanque-delabrouille_neutrino_2015,yeche_constraints_2017,palanque-delabrouille_hints_2020}. Secondly, several studies combined high- and moderate-resolution \pk~measurements to obtain constraints on the warm dark matter mass \citep{viel_constraining_2005,viel_how_2008,viel_warm_2013,baur_lyman-alpha_2016,yeche_constraints_2017,baur_constraints_2017,palanque-delabrouille_hints_2020}. The hydrodynamical simulations used in those study either directly model neutrinos as particles or using a rescaling of the matter power spectrum to account for neutrinos \citep{pedersen_massive_2020,pedersen_emulator_2020,pedersen_compressing_2023}. Finally, other exotic dark matter models such as fuzzy dark matter \citep{irsic_first_2017,armengaud_constraining_2017} can also be constrained using \pk~measurement.

Between the first \pk~measurements \citep{croft_recovery_1998,mcdonald_observed_2000,croft_towards_2002,kim_power_2004} and today, the large increase in observation capabilities brought numerous \lya~forest samples that can be split between moderate-resolution ($\lambda / \Delta \lambda \lesssim 5,000$) and high-resolution ($\lambda / \Delta \lambda \gtrsim 20,000$) observations. 

The computation of \pk~with high-resolution data sets such as SQUAD \citep{murphy_uves_2018}, KODIAQ \citep{omeara_first_2015,omeara_second_2017}, or XQ-100 \citep{lopez_xq-100_2016} are performed in \citet{viel_warm_2013,irsic_lyman-alpha_2016,walther_new_2018,day_power_2019,khaire_power_2019,boera_revealing_2019,gaikwad_consistent_2021,karacayli_optimal_2022}. Those measurements use high signal-to-noise quasars to probe the intergalactic medium at very small scales ($\sim 100$ kpc) but does not provide sufficient statistics to accurately measure the large-scale clustering ($> 5$ Mpc), needed for cosmological interpretation. 

Moderate-resolution surveys provide large numbers of \lya~forests, which yield smaller statistical uncertainties on \pk~estimates. However, the resolution of such spectrographs limits the reach of very small scales. The \pk~measurement with moderate-resolution spectrograph was first performed on a small sample of the Sloan Digital Sky Survey (SDSS) data in \citet{mcdonald_lyman-alpha_2006}. Subsequently, the increase of the \lya~forest statistic has largely improved this measurement with the Baryon Oscillation Spectroscopic Survey (BOSS) in \citet{palanque-delabrouille_one-dimensional_2013} and the extended Baryon Oscillation Spectroscopic Survey (eBOSS) in \citet{chabanier_one-dimensional_2019} using $43,751$ quasar spectra.

Several methods can be used to measure \pk~from \lya~forest samples. The most straightforward relies on the fast Fourier transform (FFT) and was applied in BOSS and eBOSS \citep{palanque-delabrouille_one-dimensional_2013,chabanier_one-dimensional_2019} analyses. The one-dimensional power spectrum can also be measured with configuration space estimators such as the quadratic maximum likelihood estimator (QMLE). This method has already been applied to moderate resolution observations \citep{mcdonald_lyman-alpha_2006,palanque-delabrouille_one-dimensional_2013} and more recently on high-resolution data in \citet{karacayli_optimal_2022}. The QMLE method is applied to the same data than used in the present paper and presented in a companion paper \citep{karacayli_optimal_2023}. The FFT method yields a straightforward calculation of \pk~and offers more control over the different calculation steps. Conversely, the more-complex QMLE estimation is not sensitive to gaps in the quasar spectra. The results between the two methods are presented in the companion paper and are in good agreement. FFT and QMLE results agree at 1\% level precision up to half the Nyquist frequency. 

The purpose of this work is to compute \pk~from the first DESI data, following the same methodology as in the latest eBOSS measurement in \citet{chabanier_one-dimensional_2019}. Using the same method facilitate the comparison between eBOSS and DESI. The \pk~is sensitive to instrumental properties such as noise and spectral resolution. As the telescopes used and the data are very different, it is essential to characterize the DESI instrument. We improve the algorithms and methodology used in \citet{chabanier_one-dimensional_2019} to account for systematic and instrumental differences between eBOSS and DESI. In particular, due to the spectral resolution improvement of DESI, our measurement allows accessing smaller scales than eBOSS.

The outline of this paper is as follows: Sec. \ref{sec:instrument_data} describes the DESI instrument and data processing used to perform this \pk~measurement. The \pk~pipeline is presented in Sec. \ref{sec:analysis}, and the characterization of the DESI instrument in Sec. \ref{sec:data_corrections}. We generate synthetic data to validate and correct our measurement in Sec. \ref{sec:synthetic_data}. The treatment of statistical and systematic uncertainties for the \pk~measurement is given in Sec. \ref{sec:uncertainties}. Finally, we present our measurement on DESI data, as well as a comparison to previous measurements in Sec. \ref{sec:results}, and conclude in Sec. \ref{sec:conclusion}.

\section{Instrument and data description}
\label{sec:instrument_data}

DESI has as objective to measure the spectra of 40 million galaxies and quasars in a footprint of $14,000~\mathrm{deg}^2$ over 5 years \citep{levi_desi_2013,desi_collaboration_desi_2016,abareshi_overview_2022}. This project aims to continue the cosmic mapping efforts started by SDSS, while drastically increasing its constraining power on the $\Lambda$CDM model and its possible extensions.

We first focus on the description of the data used for our measurement. In the following, we describe the data starting from the instrument (DESI), its associated spectroscopic pipeline, the different data acquisition phases, and the input catalogs of our study.

\subsection{DESI instrument}
\label{subsec:desi_instrument}

\begin{table}
\centering
\caption{Spectral range and effective resolving power ($R = \Delta \lambda / \lambda$) for each channel of the DESI spectrographs \citep{abareshi_overview_2022}.}
\begin{tabular}{lcc}
\hline
 \textbf{Channel} & \textbf{Spectral range (\AA)}  & \textbf{Resolving power} \\
\hline \\[-2mm]
Blue (B) & 3600 - 5930 & 2,000 – 3,200 \\
Red (R) & 5600 - 7720 & 3,200 – 4,100 \\
Near Infrared (Z) & 7470 - 9800 & 4,100 – 5,100 \\
\hline
\end{tabular}
\label{tab:desi_spectro}
\end{table}

The DESI instrument is mounted on the Mayall telescope, located on the Kitt Peak National Observatory (KPNO) in the Tohono O’odham Nation. The Mayall telescope is a reflective prime-focus telescope with a 4-meter diameter primary mirror. The DESI instrument \citep{desi_collaboration_desi_2016-1,miller_optical_2023} receives photons through an optical corrector designed to increase the field of view to $7.5~\mathrm{deg}^2$ on the focal plane. The focal plane system, composed of $5,000$ robotically controlled fibers, can quickly modify its configuration to aim at the targeted objects on a specific footprint \citep{silber_robotic_2022}. An optical fiber system redirects the light of the observed objects to a separate climate-controlled enclosure containing 10 spectrographs. Each spectrograph comprises 3 CCD cameras whose properties are given in the Tab. \ref{tab:desi_spectro}. In comparison to SDSS spectrographs, the effective resolving power ($\Delta \lambda / \lambda$) improved by at least a factor of two.

\subsection{DESI spectroscopic pipeline}
\label{subsec:desi_pipeline}

The high complexity of the DESI survey induces the need for advanced software pipelines and products, including the imaging from the DESI Legacy Imaging Surveys \citep{zou_project_2017,dey_overview_2019,schlegel_2023}, a pipeline to select the targets to observe \citep{myers_target_2022}, a pipeline to assign fibers \citep{raichoor_2023}, a pipeline to parse the survey and to optimize the observation strategy \citep{schlafly_survey_2023}, and an extensive spectroscopic reduction pipeline \citep{guy_spectroscopic_2022}.

This spectroscopic pipeline, called \texttt{desispec
} \git{desihub/desispec}{}, transforms the raw CCD images into spectra, and is detailed in \citet{guy_spectroscopic_2022}. Before extracting the spectra, the images are subtracted by dark and bias calibration frames to remove expected background sources, and to estimate the associated readout noise (noise estimation details are given in appendix \ref{appendix:noise_estimators}). The non-uniform CCD pixel response is corrected using a dedicated flat-field slit on the spectrograph and the CCD over-scan is removed. A dedicated software detects and flags cosmic rays or defective CCD pixels.

The spectral extraction is performed using the "spectroperfectionism" method~\citep{bolton_spectro-perfectionism_2010}, an optimal spectroscopic extraction that correctly models complex 2D point spread functions. This method provides the encoding for each fiber and each wavelength of the non-Gaussian instrument resolution into a resolution matrix (noted $\mathbf{R}$ in next sections) used to compute \pk. 

All the spectra are defined on the same wavelength grid without additional resampling. Consequently, the extracted spectra are linearly binned in observed wavelength with a constant separation $\Delta \lambda_{\mathrm{pix}} = 0.8$ \AA. Conversely, the SDSS/BOSS spectrographs were logarithmically binned with a constant spectral pixel size of $\Delta v = 69~\invkms$, which corresponds to $\Delta \lambda / \lambda = \Delta v/c = 2.3 \times 10^{-4}$, equivalent to $\Delta[\log(\lambda)] = 10^{-4}$ \citep{smee_multi-object_2013}.

Once the spectrum of each fiber has been extracted, several post-processing steps allow removing a variety of further observational effects. The non-uniform response of individual fibers as a function of wavelength is corrected with flat-field frames by observing a white screen attached to the telescope dome and illuminated with a LED array. For all exposures, some fibers are dedicated to observing the sky. The so-called sky spectra associated with these fibers provide the sky level and the intensity of atmospheric emission lines and are subtracted to the spectra associated with targets (sky subtraction). The transmission defaults of the atmosphere and telescope as a whole are corrected by the observation of calibrated star spectra. This step converts CCD units (number of electrons) to observed flux units. Finally, the spectrum of an object is obtained by coadding its different exposures. The resulting spectrum is expressed separately into the three spectrographs bands described in Tab. \ref{tab:desi_spectro}.

All the software of the pipeline employed for the analysis of DESI data are listed in the repository \texttt{desihub} \git{desihub}{}. In particular, the spectra analyzed in this article have been processed with the \texttt{fuji} \git{desihub/desispec/releases/tag/0.51.13}{} version of the spectroscopic pipeline.

\subsection{DESI data}
\label{subsec:desi_data}

\begin{figure*}
	\includegraphics[width=\textwidth]{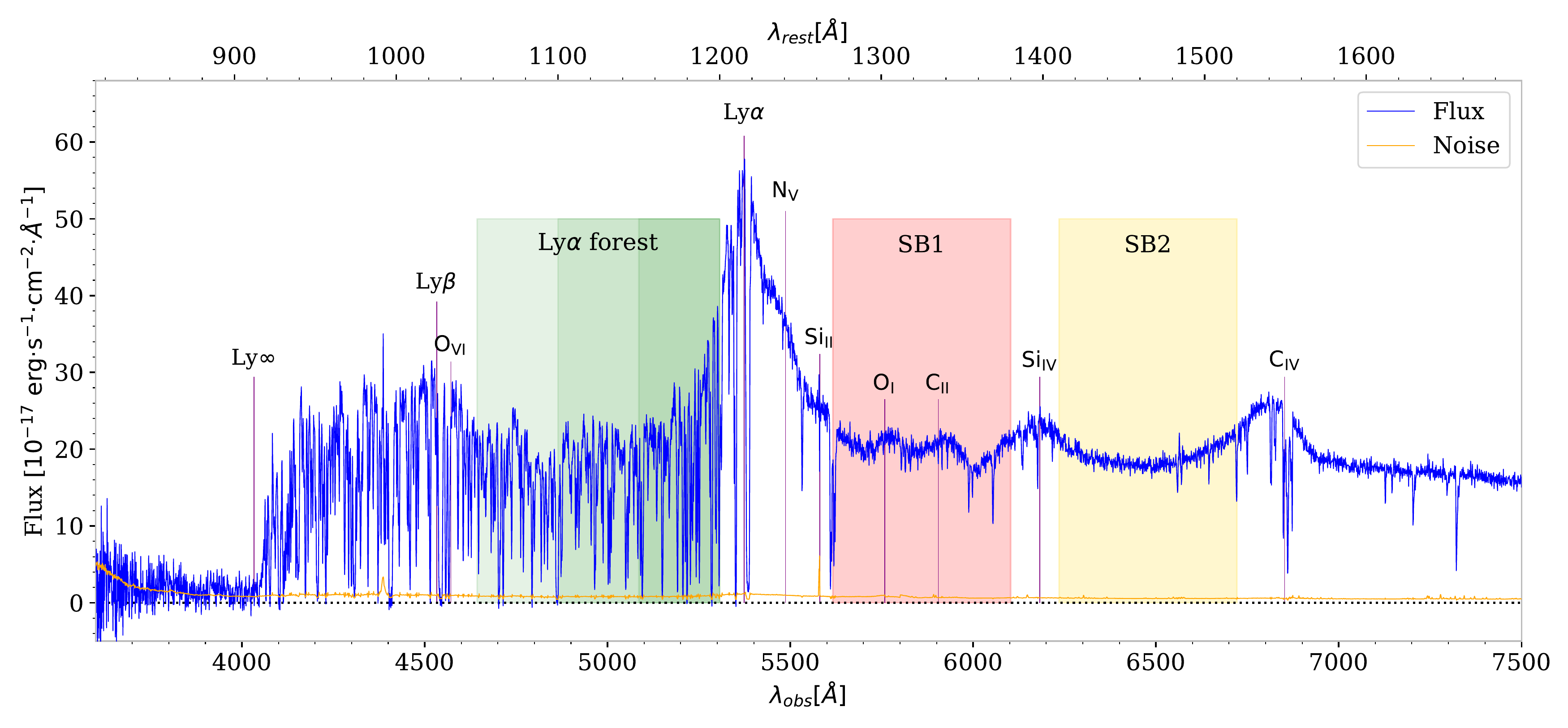}
    \caption{A particularly high-signal spectrum of a quasar located at a redshift $z=3.42$ measured by DESI with an exposure time of $2,300$ seconds. This quasar was observed on 12th April 2021, in the \svt~program, on DESI tile 221 (TARGETID = 39627746095137037, RA $= 217.263~\degree$, DEC $= -1.755~\degree$). The quasar flux is represented in blue and its noise in orange. The \lya~forest is shown in green. The side-band regions 1 and 2 pictured in red and yellow are used to estimate the forest contamination by metals.}
    \label{fig:spectra_quasar_desi}
\end{figure*}

\begin{table}
	\centering
    \caption{Summary of the DESI data sets used in this study, their associated acronyms, a subjective description, and the total number of quasar spectra whose redshifts are between $2.0$ and $5.0$.}
    \label{tab:desi_data_sets}
    \begin{tabular}{L{1.8cm}C{1.0cm}C{2.7cm}C{1.5cm}}
    \hline 
    \textbf{Data set} & \textbf{Acronym} & \textbf{Description} & \textbf{Quasar number} \\
    \hline \\[-2mm]
    Target Selection Validation & \svo & Small sample with deeper exposures (up to 16) than the end of DESI &  $12,355$ \\
    \hline \\[-2mm]
    One-Percent Survey & \svt &  Same density and depth as for the full 5yr DESI main survey & $12,686$\\
    \hline \\[-2mm]
    First two months of the Main Survey & \dafull~(\da) & Sparse data with small number of exposures & $87,373$ \\
    \hline
    \end{tabular}
\end{table}

The spectroscopic pipeline described previously was used to analyze data obtained over different periods. In this article, we use three data sets from the first observations of DESI described in Tab. \ref{tab:desi_data_sets}, which includes a total of $112,414$ quasar spectra whose redshifts are between $2.0$ and $5.0$.

The first two data sets 'Target Selection Validation' (\svo) \citep{desi_collaboration_validation_2023} and 'One-Percent Survey' (\svt) are part of the Early Data Release (EDR) whose complete description is given in \citet{desi_collaboration_early_2023}. 'Target Selection Validation' was conducted from December 2020 to March 2021 and includes a large number of exposures (up to 16) for the same targets. The objective of this survey was among others, to study extensively the survey performance as a function of instrumental depth and to build visual inspection truth tables. The \svo~data set also includes 'Secondary Tiles' as detailed in \citet{desi_collaboration_early_2023}. Following the completion of 'Target Selection Validation' observations, the 'One-Percent Survey' phase was dedicated more specifically to the evaluation of the survey design. The number of exposures is similar to the main survey at its end (between 4 and 5 for each \lya~quasars) and the goal was to determine the best strategy to cover the sky while limiting fiber loss.

The main DESI survey started in June 2021 and in this article we also use the first two months of data, named \dafull~(and noted \da~in this paper for conciseness), which is not present in the EDR but will be included in the Data Release 1. In the \da~data set, most quasars have only one exposure. While all three data sets are studied in this article, in the end we removed \svo~due to its different noise properties (see appendix \ref{appendix:data_set_comparison}). The final measurement is computed on \svtda~data set only.

\subsection{Input catalogs}
\label{subsec:catalogs}
\begin{figure}
	\includegraphics[width=\columnwidth]{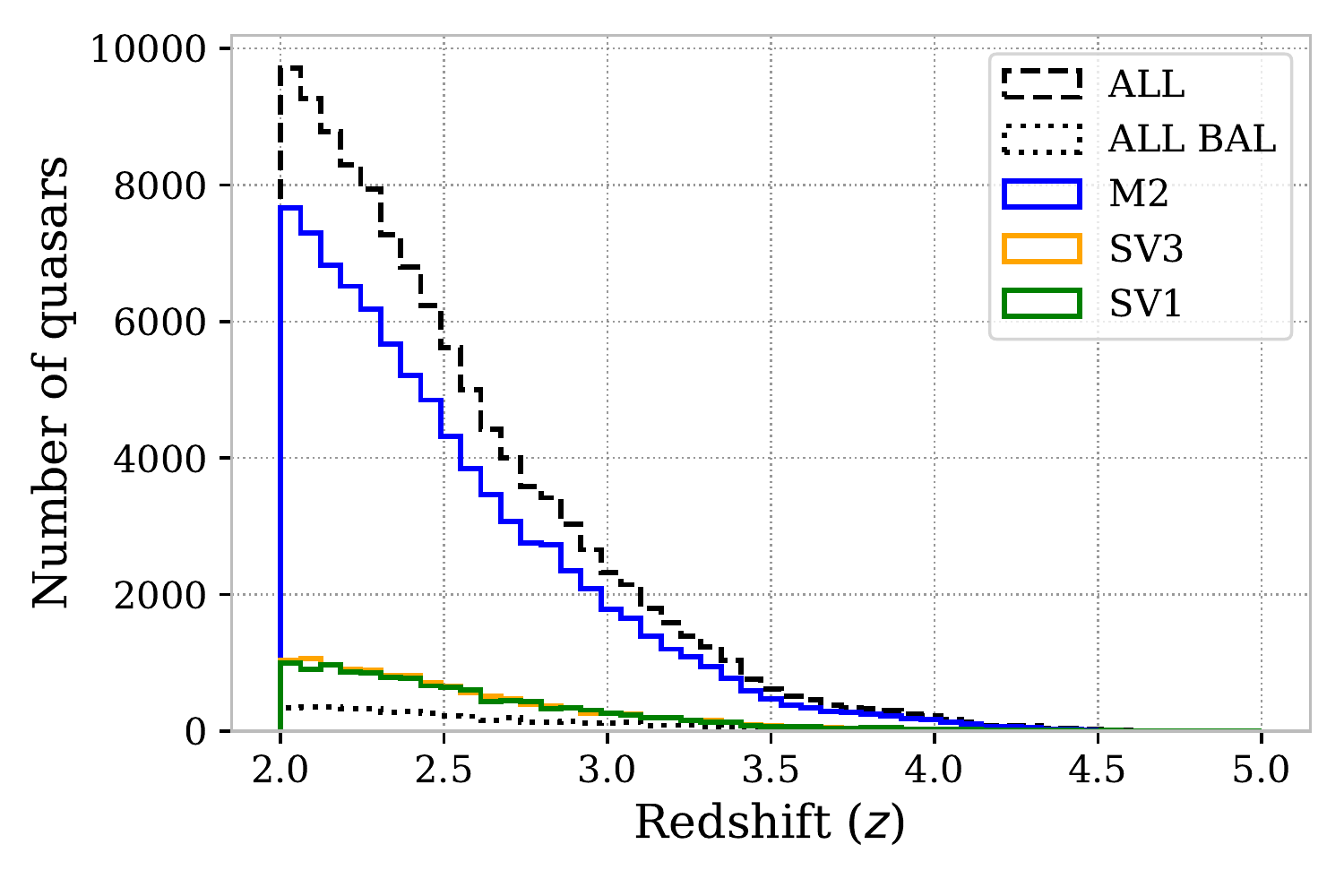}
    \caption{Histogram of the quasar redshift (with $z>2.0$) whose spectra are observed in the \svo, \svt, and \da~data sets. The histogram of the sum of data sets is shown in dashed line. Finally, the number of BAL quasars characterized by a balnicity index higher than zero, and not used in the \pk~computation pipeline, is shown with a dotted line.}
    \label{fig:quasar_redshift_hist}
\end{figure}

\begin{figure}
	\includegraphics[width=\columnwidth]{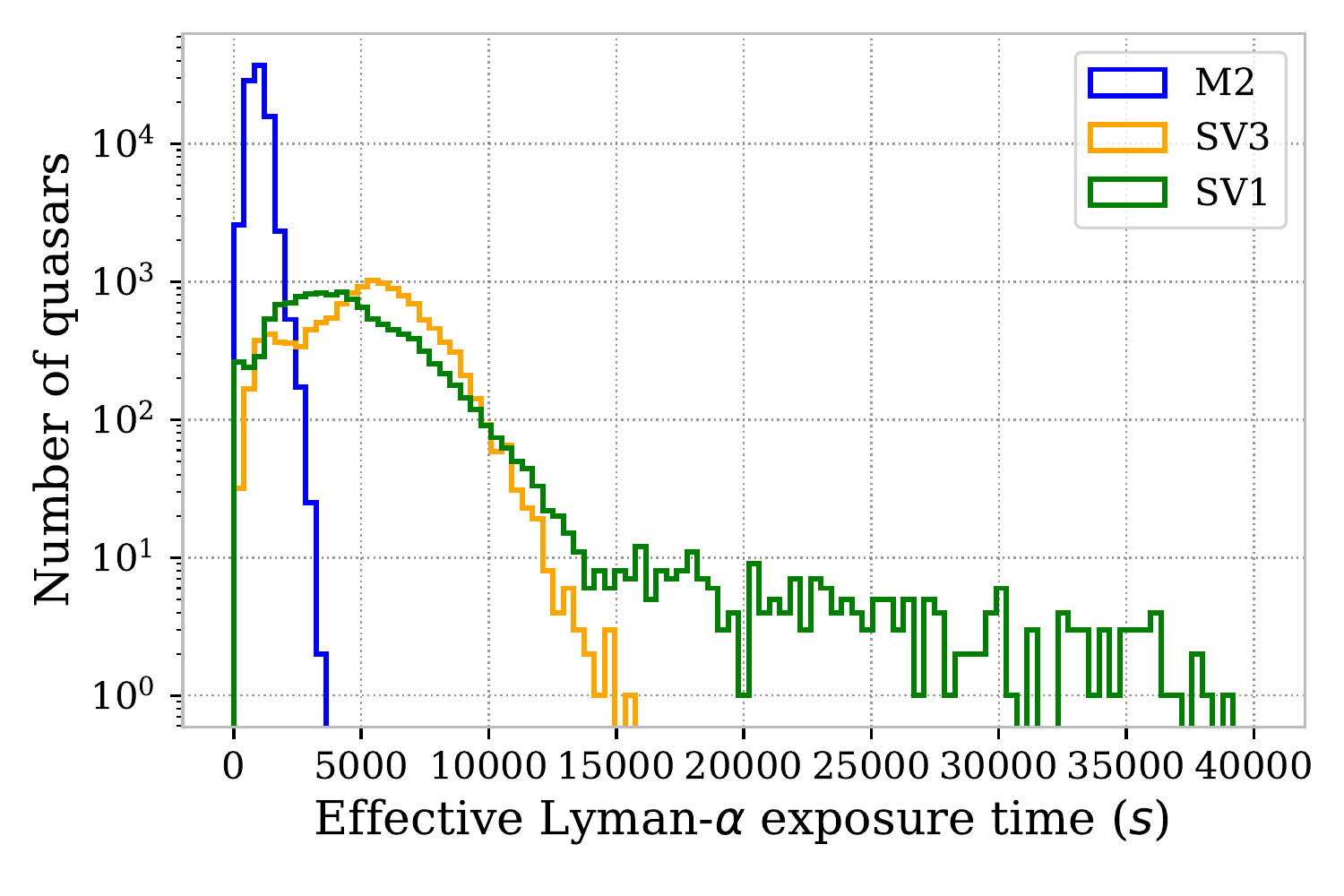}
    \caption{Normalized histogram of the effective exposure time in the \lya~forest region for the quasar spectra in the \svo, \svt, and \da~data sets. As mentioned in Sec. \ref{subsec:desi_data}, there is a wide disparity of exposure time for the three data sets. As a reference, the nominal time of one DESI exposure is set to $1200$ s.}
    \label{fig:quasar_time_hist}
\end{figure}

The input catalogs used to compute \pk~were obtained by applying specific procedures to the three data sets previously described.

The targeting of the quasars used in our study \citep{yeche_preliminary_2020,chaussidon_target_2022} was verified with visual inspection of subsets of early observations \citep{alexander_desi_2022}. For quasars, the DESI pipeline categorizes the observed spectra and estimates their redshifts using the \texttt{redrock} spectral template fitting software \git{desihub/redrock}{} \citep{bailey_2023,brodzeller_performance_2023}. In order to optimize the completeness of the quasar catalog while keeping a high purity, additionally a broad \mgii~line finder \texttt{mgii\_afterburner} \git{desihub/desispec/blob/master/py/desispec/mgii_afterburner.py}{} and a machine learning classifier applying deep convolutional neural networks \texttt{QuasarNP} \git{desihub/QuasarNP}{} \citep{busca_quasarnet_2018,farr_optimal_2020} are run after \texttt{redrock}. Both post-processing programs are run on all objects targeted by DESI as detailed in \citet{chaussidon_target_2022}.

An example of a quasar spectrum at redshift $z=3.42$ is given in Fig. \ref{fig:spectra_quasar_desi}. This spectrum, with a particularly high flux, is part of the \svt~data set which contains spectra with exposure time equivalent to the end of the DESI survey. The redshifts and effective exposure times for each data set considered in this paper are shown in Fig. \ref{fig:quasar_redshift_hist} and \ref{fig:quasar_time_hist} respectively. The effective exposure time accounts for nightly observing conditions by normalizing the real exposure time to a reference with airmass 1, zero galactic extinction, a $1.1''$ seeing
(FWHM), and zenith dark sky~\citep{guy_spectroscopic_2022}. The nominal exposure time of one exposure is defined to $1000~s$. The large differences in term of exposure time emphasize the need to treat the data sets differently, at least for noise properties. \svo~and \svt~contains a small number of forests but with heterogeneous exposure times, in opposition to \da, which contains many quasars only observed once.

Broad absorption line (BAL) quasars are specific quasars whose spectra exhibit consistent blueshifted absorptions associated with many spectral features. They are identified using the \texttt{baltools} \git{paulmartini/baltools}{} software. It consists of a $\chi^2$ minimizer algorithm that looks for blueshifted \civ~or \siiv~absorptions in an unabsorbed quasar model. The fit is performed for rest-frame wavelengths between $1,260$ and $2,400$ \AA. A quasar is considered a BAL type if its spectrum exhibits a region between \civ~and \siiv~emission lines, with at least 10\% flux decrement below the continuum and a width greater than $2,000~\invkms$. BAL quasars spectra ($4.18 \%$ in the total data set) are removed in the analysis performed in this work.

Damped Lyman-$\alpha$ absorbers (DLA) are regions within a quasar spectrum that show over-saturated absorption with prominent Lorentzian wings as the quasar flux intersects the dense, circumgalactic medium of an intervening (proto-)galaxy. They are a subclass of high-column density systems and are a significant contaminant of the \lya~forest signal, particularly because of their wings and the additional contamination by circumgalactic metal absorption lines \citep{mcdonald_physical_2005}. The correct modeling of such systems in simulations has been proven to be particularly complicated \citep{pontzen_damped_2008}.

We use a catalog resulting from the combination of a convolutional neural network (CNN) algorithm \texttt{desi-dlas} \git{cosmodesi/desi-dlas}{} \citep{parks_deep_2017,zou_statistical_2023} and a Gaussian process (GP) algorithm \git{jibanCat/gp_dla_detection_dr16q_public}{} \citep{ho_damped_2021}. \texttt{desi-dlas} is trained with SDSS spectra to identify candidate high-column density objects for rest-frame wavelengths between $900$ and $1,346$ \AA. It returns locations of high-column density systems in the spectra, as well as their \hi~column density and a confidence parameter. The GP finder provides similar output using the same training set and a Bayesian model selection. 
We only consider the high-column density objects with column density $N_{\mathrm{\hi}} > 10^{20.3}~\mathrm{cm}^{-2}$ (DLAs). In accordance with recommendations from \citet{parks_deep_2017,ho_damped_2021}, we consider CNN confidence level higher than 0.2 as valid DLA detections when the ratio between the quasar continuum and the noise is higher than 3. We take a confidence level limit of 0.3 when this ratio is lower than 3. For the GP model, a 0.9 minimal confidence level is applied. In the case when absorbers are detected by the two models, the combined DLA catalog uses $N_{\mathrm{\hi}}$ values and DLA redshifts from GP model.

Although DLAs by themselves constitute tracers of the matter distribution, they have an extended impact on the observed spectra. They increase the correlations of neighbor spectrum pixels, thus artificially increasing \pk~level. Therefore, we choose to mask the core of DLA regions of the spectra by fixing the transmitted flux fraction to its mean value for spectrum pixels where the DLA-induced absorption is larger than $20~\%$. In addition, the absorption in the Lorentzian damping wings that remain after the cut is corrected with a Voigt profile following \citet{bautista_measurement_2017,chabanier_completed_2021}.

Finally, we use a catalog of masks to account for atmospheric and Galactic emission lines, which has been adapted to DESI resolution \git{corentinravoux/p1desi/blob/main/etc/skylines/list_mask_p1d_DESI_EDR.txt}{}. The creation of this catalog is detailed in Sec. \ref{subsec:atmospheric_lines_p1d}.

\section{One-dimensional power spectrum estimation}
\label{sec:analysis}

The \pk~estimator is build using the data product described previously in two phases. First, the fitting of the continuum of quasars is used to convert the absolute received flux to a normalized quantity $\delta_{F}$. Secondly, \pk~is computed by employing a Fast Fourier Transform and by averaging the product of this transformation for all the selected \lya~forests.

\subsection{Continuum fitting}
\label{subsec:delta_extraction}

A standard normalized quantity used in the calculation of correlations and power spectra is the flux contrast $\delta_{F}$ of the \lya~forest, defined as

\begin{equation}
    \label{eq:delta_flux_continuum}
    \delta_{F}(\lambda) = \frac{F(\lambda)}{\overline{F}(\lambda)} - 1 =   \frac{f(\lambda)}{C_{\mathrm{q}}(\lambda,z_{\mathrm{q}})\overline{F}(\lambda)} - 1\,, 
\end{equation}

\noindent where $F$ is the transmitted flux fraction, and $\overline{F}(\lambda)$ is its average value, the mean transmission of the intervening IGM. 
Note that for the purpose of this work, we do not need to know the individual quasar continua $C_q$, but only the product $C_{\mathrm{q}}(\lambda,z_{\mathrm{q}}) \overline{F}(\lambda)$.
 Thus, similar to previous \lya~studies based on survey data \citep{bourboux_completed_2020,chabanier_one-dimensional_2019,ramirez-perez_lyman-alpha_2023}, we directly measure this product in our continuum fitting process using the \texttt{picca} \git{igmhub/picca}{} \citep{du_mas_des_bourboux_picca_2021} software package. This software also merges the quasar spectra over different bands to obtain $f$ over the all wavelength range. The continuum of each quasar is modeled as the product of a universal continuum $C$ common to all quasars, and a first-order polynomial term in wavelength:

\begin{equation}
    \label{eq:continuum_quasar}
    C_{\mathrm{q}}(\lambda,z_{\mathrm{q}}) = \left(a_{\mathrm{q}} + b_{\mathrm{q}}\lambda\right) C\left(\lambda_{\mathrm{rf}}=\frac{\lambda }{ (1+z_{\mathrm{q}})}\right)\,,
\end{equation}

\noindent where $a_{\mathrm{q}}$ and $b_{\mathrm{q}}$ are quasar-dependent constants. In previous studies \citep{palanque-delabrouille_one-dimensional_2013,chabanier_one-dimensional_2019}, $b_{\mathrm{q}}=0$ was assumed for all quasars, i.e. only a wavelength-independent normalization factor was taken into account. We add an additional linear wavelength-dependent term to account for the diversity of quasars after verifying that this change does not impact the mean level of our \pk~measurement. 

The $a_{\mathrm{q}}$ and $b_{\mathrm{q}}$ parameters for each quasar are determined along with $C$ by maximizing the following log-likelihood:

\begin{equation}
\ln \mathcal{L} = - \frac{1}{2} \sum_{i} \frac{\left[f_{i}-\overline{F}(\lambda_{i}) 
C_{\mathrm{q}}\left(\lambda_{i},z_{\mathrm{q}}, a_{\mathrm{q}}, b_{\mathrm{q}}\right)\right]^{2}}{\sigma_{\mathrm{q}}^2\left(\lambda_{i}\right)}-\ln \left[\sigma_{\mathrm{q}}^2\left(\lambda_{i}\right)\right]\,,
\end{equation}

\noindent where the sum is run over all the spectrum pixels of the quasar q, and $\sigma_{\mathrm{q}}$ is the standard deviation estimator of the flux $f$.

In contrast to analyses of the large-scale 3d correlation function such as \citet{bourboux_completed_2020}, we want all spectrum pixels to contribute equally to the continuum fitting and the \pk~computation, as the opposite could bias \pk. Therefore, we impose noise-independent weights in the continuum fitting procedure:

\begin{equation}
    \label{eq:noise_delta_extraction}
    \sigma_{\mathrm{q}}^{2}(\lambda) = \left(\overline{F}(\lambda) C_{\mathrm{q}}(\lambda)\right)^{2}
\end{equation}

This procedure is the same as in the previous \pk~analyses based on BOSS/eBOSS data \citep{palanque-delabrouille_one-dimensional_2013,chabanier_one-dimensional_2019}. The standard deviation associated with $\delta_{F}$ at the end of the continuum fitting procedure is defined by

\begin{equation}
\label{eq:error_delta}
\sigma_{\delta_{F}}(\lambda) =\frac{\sigma_{\mathrm{pip,q}}(\lambda)}{\overline{F}(\lambda) C_{\mathrm{q}}(\lambda)}\,,
\end{equation}

\noindent where $\sigma_{\mathrm{pip},\mathrm{q}}$ is the noise provided by the DESI spectroscopic pipeline detailed in appendix \ref{appendix:noise_estimators}.

The universal continuum $C$ and the $a_{\mathrm{q}}$ and $b_{\mathrm{q}}$ parameters are computed iteratively. In particular, $C$ is estimated from the average of all spectra, i.e. in a non-parametric way. During the entire fitting procedure, spectrum pixels that are masked due to the presence of a DLA or an atmospheric line are not considered in the fit. We use 7 iterations and have verified that the continuum fits are converged at this point.

The noise level of a \lya~forest is characterized by defining the average signal-to-noise ratio \snr~in the \lya~forest region:

\begin{equation}
\label{eq:mean_snr}
\overline{\mathrm{SNR}} = \left\langle \frac{f(\lambda)}{\sigma_{\mathrm{pip,q}}(\lambda)} \right\rangle_{\lambda}\,.
\end{equation}

Only \lya~forests with a \snr~larger than 1 are used in the continuum fitting procedure. This procedure is also restricted to the observed redshift range $3,600 < \lambda < 7,600$ \AA, to avoid the shorter wavelength range where a large fraction of the quasar spectra is absorbed by the atmosphere. We also select the rest-frame wavelength in the range $1,050 < \lambda_{\mathrm{rf}} < 1,180$ \AA, so that the measured contrasts are dominated by the \lya~forest. In particular, we try to avoid the \lyb~singlet and the \ovi~doublet emission regions respectively located at $\lambda_{\mathrm{\lyb}} = 1,025.72$ \AA~and $\lambda_{\mathrm{\ovi}} = (1031.912, 1037.613)$ \AA~in the rest-frame. 

The cut $\lambda_{RF} < 1,180$ \AA~facilitates the continuum fitting procedure and mitigates most of the proximity effect: close to a quasar, the neutral hydrogen fraction is indeed influenced by the quasar's UV radiation in addition to the extragalactic UV radiation background \citep{bajtlik_quasar_1988}.

As detailed in Sec. \ref{subsec:desi_pipeline}, the quasar spectra are linearly binned in observed wavelength with $\Delta \lambda_{\mathrm{pix}} = 0.8$ \AA. Note that when converting to rest-frame wavelength $\lambda_{\mathrm{rf}}$ the pixel size will be redshift dependent. For the continuum fitting process, we thus need to rebin our spectra to a uniform grid in $\lambda_{\mathrm{rf}}$. As the quasar continuum is relatively smooth and to avoid noisy continuum fits for analyses of relatively small data sets, we chose a grid for the common continuum $C$ that is 10 times coarser than the lowest redshift quasar pixels considered, i.e.:

\begin{equation}
    \label{eq:smoothing_continuum}
    \Delta \lambda_{\mathrm{pix,rf}} = 10 \frac{\Delta \lambda_{\mathrm{pix}}}{1 + z_{\mathrm{min}}}\,. 
\end{equation}

By taking $z_{\mathrm{min}} = 2.0$ as the lowest redshift, we obtain $\Delta \lambda_{\mathrm{pix,rf}} = 2.67$ \AA. With increasing size of the dataset, such a rebinning could be relaxed in future DESI measurements. At the end of the continuum fitting procedure, the stacking of all the \lya~contrasts is forced to be equal to zero to avoid introducing flux calibration errors.

\subsection{Fast Fourier transform power spectrum estimator}
\label{subsec:fft_estimator}

Conceptually, the $\delta_{F}$ quantity can be separated into different contributions in the \lya~forest region: 

\begin{equation}
\delta_{F} (\lambda)= \delta_{\mathrm{astro}} (\lambda) +\delta_{\mathrm{noise}} (\lambda)\,,
\end{equation}

\noindent where $\delta_{\mathrm{astro}}$ corresponds to the fluctuations caused by all the elements of the intergalactic medium (including \lya), and $\delta_{\mathrm{noise}}$ corresponds to noise fluctuations.

Considering instrumental effects, the true underlying flux contrast is modified on its way through the instrument in multiple ways. First, photons traverse the spectrograph leading to the output flux being convolved with the spectrograph line spread function $W(\lambda,\mathbf{R},\Delta \lambda_{\mathrm{pix}})$ which depends on the resolution matrix presented in Sec.~\ref{subsec:desi_pipeline}. The $W$ term also account for the signal pixelization as the photons are counted into CCD pixels of size $\Delta \lambda_{\mathrm{pix}}$. Finally, we need to account for noise sourcing from the processes of photon counting and readout. In total, we can write the measured flux contrast $\delta_{F}$ as

\begin{equation}
    \delta_{F} (\lambda)= \delta_{\mathrm{astro}} (\lambda)\circledast W(\lambda,\mathbf{R},\Delta \lambda_{\mathrm{pix}}) +\delta_{\mathrm{noise}}(\lambda)\,.
\end{equation}

We assume that the impact of the noise and the resolution term $W$ are decorrelated, i.e., that the noise contrast is not affected by the instrumental effects of pixelization and resolution. Verifying this assumption is beyond the scope of this paper.

The $\delta_{\mathrm{astro}}$ contrast can be decomposed between the \lya~signal and the one from all other elements of the intergalactic medium called metals. The contribution of those metal absorptions to $\delta_{F}$ can be decomposed into two parts. On the one hand, there are absorption lines with rest-frame wavelength $\lambda \gg \lambda_{\mathrm{\lya}} = 1,215.67$ \AA. Those lines can be independently observed redwards of the \lya~forest, using specific rest-frame spectral regions called side-bands SB1 ($1,270 < \lambda_{\mathrm{rf}} < 1,380$ \AA), and SB2 ($1,410 < \lambda_{\mathrm{rf}} < 1,520$ \AA), as shown in Fig. \ref{fig:spectra_quasar_desi}. We group the absorption of all those metals in a contrast noted $\delta_{\mathrm{metals}}$.

On the other hand, there are absorption lines with rest-frame wavelength $\lambda \lesssim \lambda_{\mathrm{\lya}}$, such as \siii~and \siiii~elements ($\lambda_{\mathrm{\siii}} = 1,190$ and $1,193$ \AA, and $\lambda_{\mathrm{\siiii}} = 1,206.50$ \AA). They cannot be observed independently of the \lya~forest, but will show absorption that is correlated with the \lya~absorption and will lead to an oscillatory feature in the estimated \pk. For those lines, we adopt the same approach developed in \citet{mcdonald_lyman-alpha_2006} and subsequently used in other analyses. We leave the features inside our power spectrum estimates to be corrected by fitting an additional oscillation during parameter inference. We note $\delta_{\mathrm{\lya}}$ the contrast containing the \lya~forest and the effect of those latter metals. Finally, the flux contrast can be expressed:

\begin{equation}
    \label{eq:contrast_p1d}
    \delta_{F} (\lambda)=  \left(\delta_{\mathrm{\lya}} (\lambda)+\delta_{\mathrm{metals}} (\lambda)\right)\circledast W(\lambda,\mathbf{R},\Delta \lambda_{\mathrm{pix}}) + \delta_{\mathrm{noise}}(\lambda)\,.
\end{equation}

The one-dimensional \lya~power spectrum (\pk) can be estimated from this decomposition by applying a fast Fourier transform (FFT) algorithm on $\delta_{F}$ of each \lya~forest. This method was applied in previous measurements \citep{croft_recovery_1998,palanque-delabrouille_one-dimensional_2013,chabanier_one-dimensional_2019}.

The FFT estimator is implemented in \texttt{picca}$^{\ref{igmhub/picca}}$ \citep{du_mas_des_bourboux_picca_2021}. For each \lya~forest, the raw power spectrum\footnote{This is not a proper power spectrum, strictly speaking, as it does not correspond to an average.} is defined from the Fourier transform of $\delta_{F}(r)$:

\begin{equation}
    \label{eq:praw}
    (2\pi) \delta_{\mathrm{D}}(k-k') P_{\mathrm{raw}}(k) = \delta_F(k) \delta_F(k') \,,
\end{equation}

\noindent where $\delta_{\mathrm{D}}$ is the one-dimensional Dirac distribution.

Applying a Fourier transform on equation (\ref{eq:contrast_p1d}), the raw power spectrum is expressed by

\begin{equation}
    \begin{aligned}
    \label{eq:praw_decomposition}
    P_{\mathrm{raw}}(k)=&\left(P_{\mathrm{\lya}}(k)+P_{\mathrm{\lya-\siii/\siiii}}(k)+P_{\mathrm {metals}}(k)\right) \\
    & \cdot W^{2}(k, \mathbf{R},\Delta \lambda_{\mathrm{pix}})+P_{\mathrm{noise}}(k)\,.
    \end{aligned}
\end{equation}

In this decomposition, \pl, \pme, and \pn~are the power spectra respectively associated to the contrasts $\delta_{\mathrm{\lya}}$, $\delta_{\mathrm {metals}}$, and $\delta_{\mathrm{noise}}$ with the same definition as in equation (\ref{eq:praw}). We assumed here that the \siii~and \siiii~power spectra are negligible and that all cross-correlation terms between the contrasts are null. The only non-neglected cross-term is $P_{\mathrm{\lya-\siii/\siiii}}(k) = 2 \left|\delta_{\mathrm{\lya}}(k) \delta_{\mathrm{\siii/\siiii}}(k)\right|$, which corresponds to the correlated absorptions of \lya~with either \siii~or \siiii. The oscillations induced by this term have a wavenumber $2\pi / (\lambda_{\mathrm{\lya}} - \lambda_{\mathrm{\siii/\siiii}})$ when "$k$" is expressed in \invAA.

The FFT estimator for the one-dimensional power spectrum is computed as an average over all available \lya~forests in the measurement sample. It is designed to match the sum of \lya~and \lya$-$\siii/\siiii~power spectra in equation (\ref{eq:praw_decomposition}), so that we define

\begin{equation}
P_{1\mathrm{D},\alpha}(k) = \left\langle P_{\mathrm{\lya}}(k) + P_{\mathrm{\lya-\siii/\siiii}}(k) \right\rangle\,,
\end{equation}

\noindent where $\langle . \rangle$ denotes the average over all the \lya~forests used for \pk~calculation. From equation (\ref{eq:praw_decomposition}), the estimator of \pk~is defined by

\begin{equation}
\label{eq:p1d_estimator}
P_{1\mathrm{D},\alpha}(k) = \left\langle\frac{P_{\mathrm{raw}}(k)-P_{\mathrm {noise}}(k)}{W^{2}(k, \mathbf{R},\Delta \lambda_{\mathrm{pix}})} \right\rangle-P_{\mathrm{metals}}(k)\,.
\end{equation}

The \pk~measurement is split in different redshift bins to take into account its evolution. The \lya~forest is split into sub-forests which correspond to consecutive and non-overlapping sub-regions of equal length. This procedure also reduces the correlations between the different redshift bins. We chose to cut \lya~forests into three sub-forests whose rest-frame wavelength boundaries are $\lambda_{\mathrm{rf}} = 1,093.3$ and $1,136.6$ \AA, so that the length of each sub-forest is $L_{\mathrm{sub}} = 43.3$ \AA. With this sub-forest separation, a single \lya~forest can contribute to up to 3 different redshift bins in the \pk~measurement. The sub-forest splitting constrains the minimal accessible observed wavenumber to $k_{\mathrm{min}} = 2\pi / (L_{\mathrm{sub}} (1 + z_{\mathrm{min}}))  = 0.0453$ \invAA, by taking the minimal redshift used. Each sub-forest spans at most $\Delta z = 0.2$ and we choose the same $\Delta z$ to define the redshift binning for \pk.

For observed wavelength $\lambda \lesssim 3,700$ \AA, the noise level is high in comparison to the spectra because of atmospheric absorptions. To minimize the impact of this noise, we remove the spectrum pixels for which the observed wavelength is lower than $3,750$ \AA, which corresponds to \lya~absorbers located at $z=2.085$. In the future, with a dedicated study to control the noise at shorter wavelength, the \pk~analysis can be extended to redshift $z \sim 2$.

In accordance with the eBOSS study \citep{chabanier_one-dimensional_2019}, we remove sub-forests shorter than 75 spectrum pixels due to a cut in the UV region or to the presence of a large DLA. We also do not consider the \lya~sub-forests with more than 120 masked spectrum pixels.

Unlike the analysis in \citet{chabanier_one-dimensional_2019}, we do not apply a second redshift dependent \snr~cut for the averaging of \pk. Instead, we develop and test a \snr~weighting scheme, as detailed in the appendix \ref{appendix:snr_weighting}. This procedure is used for all the article except in Sec. \ref{subsubsec:noise_asymptote} where the impact of the \snr~cut is investigated.

\section{DESI instrumental characterization}
\label{sec:data_corrections}

As the DESI instrument is new, we first focus the analysis on characterizing the instrumental effects on our \pk~measurement. In particular, we describe the impact of spectral resolution, instrumental noise, metal power spectrum, and atmospheric emission lines in the following.

\begin{figure}
	\includegraphics[width=\columnwidth]{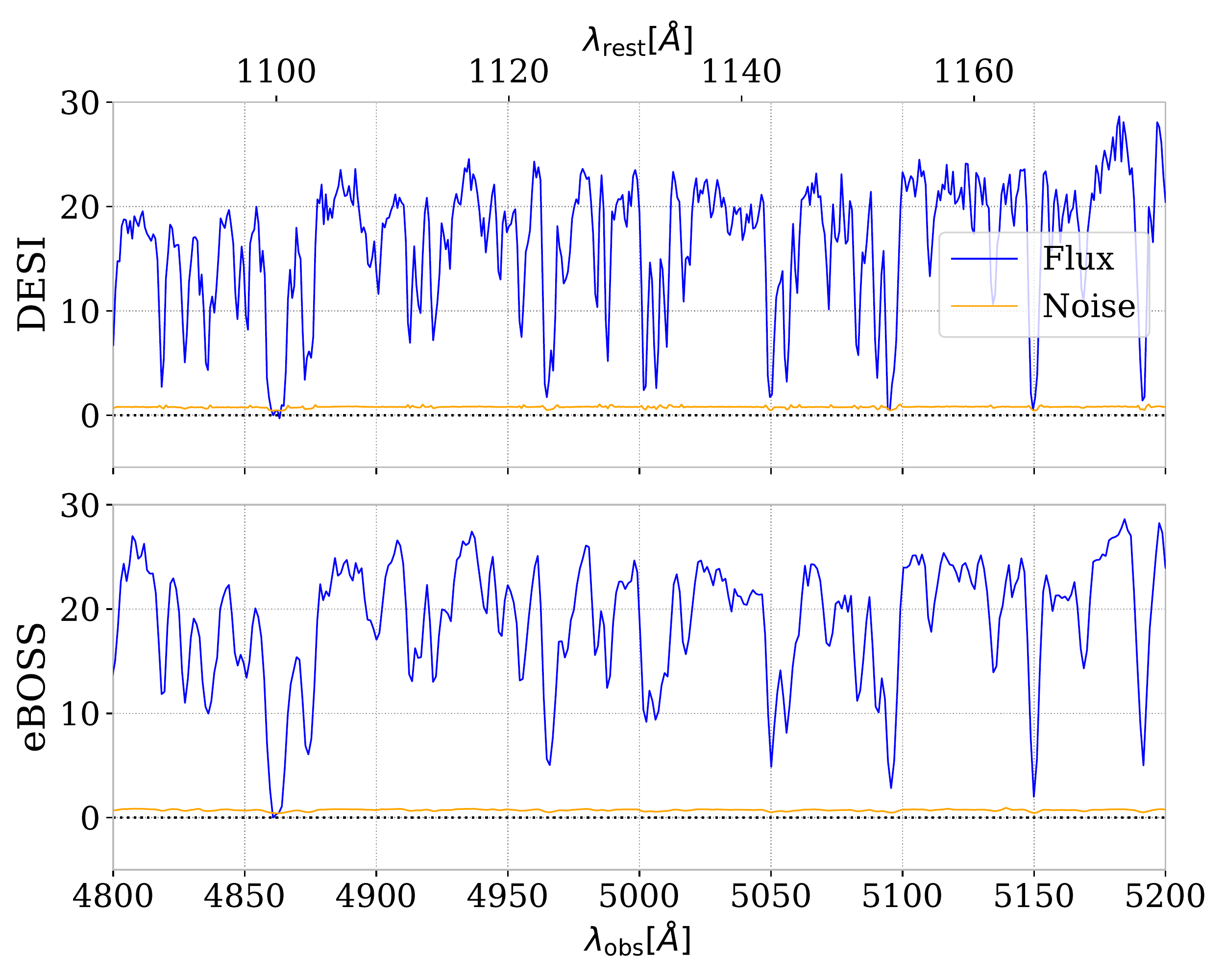}
    \caption{Illustration of the resolution improvement between eBOSS and DESI. The DESI spectrum of the quasar represented in Fig. \ref{fig:spectra_quasar_desi} is zoomed on a region of the \lya~forest on the top panel. The spectrum of the same quasar obtained by the Sloan Digital Sky Survey-IV (SDSS-IV) \citep{gunn_25_2006,smee_multi-object_2013,dawson_sdss-iv_2016,blanton_sloan_2017} in the 16$^{th}$ Data Release (DR16) \citep{ahumada_sixteenth_2020} of the extended Baryon Oscillation Spectroscopic Survey (eBOSS). The SDSS name associated to this quasar is 142903.03-014519.3. The DESI spectrum is obtained after 3 individual exposures for a total exposure time of $2,300$ seconds. The eBOSS spectrum have 11 exposures for a total of $6,300$ seconds. The large-scale absorption structures are similar but due to its improved spectroscopic resolution, the DESI spectrum clearly shows more details at small scales.}
    \label{fig:resolution_improvement}
\end{figure}

\begin{figure}
	\includegraphics[width=\columnwidth]{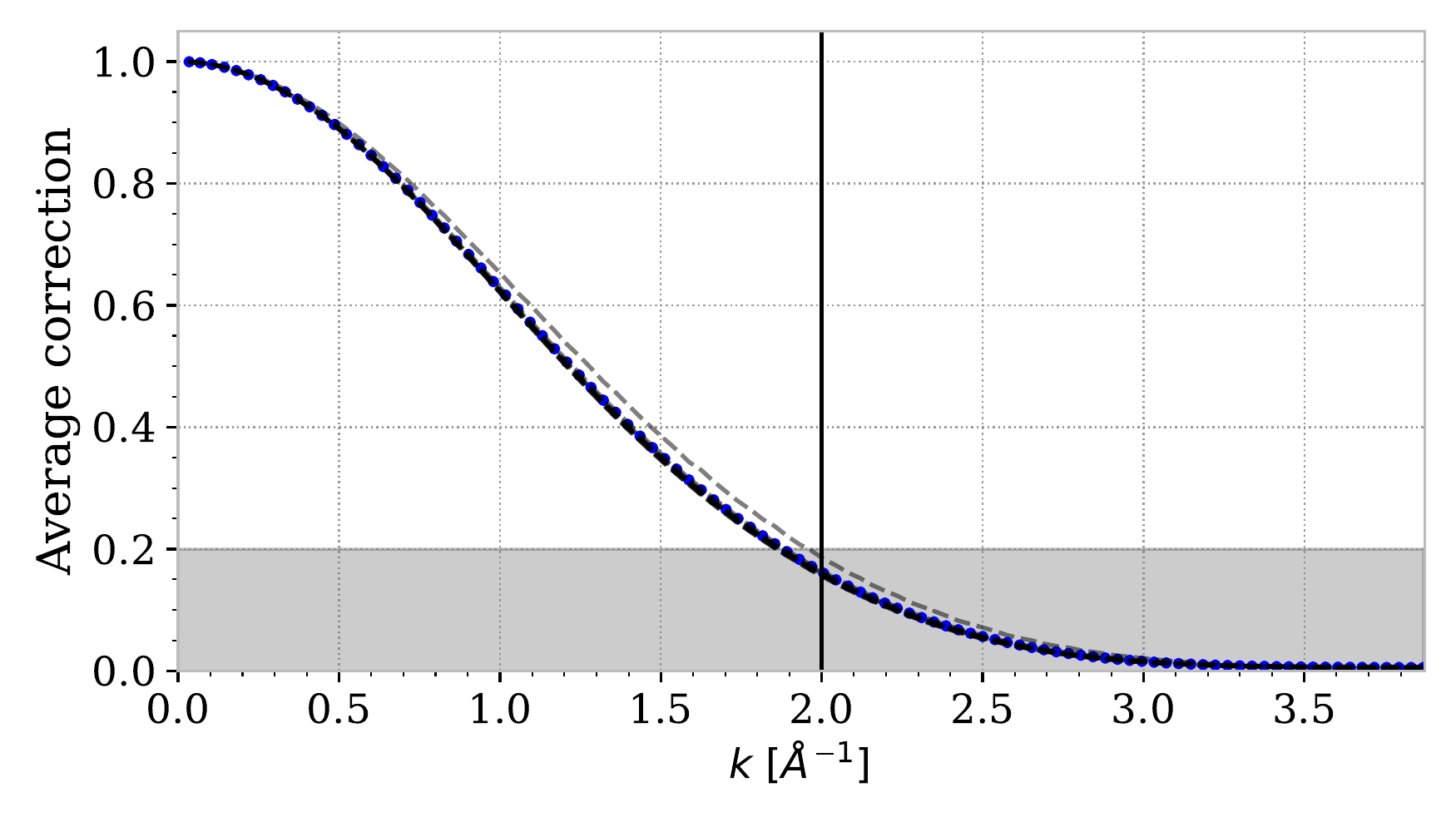}
    \caption{Average resolution correction $\left \langle W^{2}(k, \mathbf{R},\Delta \lambda_{\mathrm{pix}})\right \rangle = \left \langle \mathbf{R}^{2}(k) \cdot \right \rangle$ (blue points) using \lya~forest from \svtda~data set. This resolution correction is weakly dependent on the redshift range (shown by shaded dashed black lines along the blue points). Only one shaded black line is above the blue points ($z=3.8$). All the others are at the same level as the blue points. The mean value over all redshift bins is shown with the points. The shaded area represents the regime for which the associated impact of resolution and pixelization removes more than 80\% of the power spectrum. This criteria is chosen to define the maximal wavenumber of our \pk~measurement, shown with a vertical black line.}
    \label{fig:mean_resolution}
\end{figure}

\subsection{Spectrograph resolution}
\label{subsec:resolution}

\subsubsection{Resolution correction modeling}

As mentioned in Sec. \ref{subsec:desi_pipeline}, DESI spectra are linearly binned in observed wavelength, so that the natural unit for wavenumbers is \invAA. The maximal measurable wavenumber follows the Nyquist-Shannon limit: $k_{\mathrm{max}} = k_{\mathrm{Nyq}} = \pi / \Delta \lambda_{\mathrm{pix}} \simeq 3.92$ \invAA. 

For SDSS \pk~measurements \citep{croft_recovery_1998,mcdonald_lyman-alpha_2006,palanque-delabrouille_one-dimensional_2013,chabanier_one-dimensional_2019}, the pixelization is logarithmically binned in observed wavelength, making it suitable to express the power spectrum in Hubble velocity unit $v$, because $\Delta v \propto \Delta \lambda / \lambda = \Delta \log(\lambda)$. In eBOSS \citep{chabanier_one-dimensional_2019}, the resolution correction is modeled by a Gaussian function $W(k, \Delta \lambda) = \exp \left( -0.5 (k[\kms] \Delta v)^2\right)$, where $\Delta v$ is the spectral resolution in velocity units.

The spectroscopic resolution of DESI is improved with respect to the SDSS spectrographs. On the DESI blue band (see Tab. \ref{tab:desi_spectro}) where most of the \lya~forest are observed, the effective resolving power $R=\lambda/\Delta \lambda$ ranges from $2,000$ to $3,200$ \citep{abareshi_overview_2022}. In comparison, SDSS spectrographs had a $1,500 < R < 2,300$ in its blue band ($3,600 < \lambda < 6,350$ \AA). This improved resolution brings the opportunity to probe the clustering of matter at smaller scale by measuring the small fluctuations in the \lya~forest, as illustrated in Fig. \ref{fig:resolution_improvement}.

As described in Sec. \ref{subsec:desi_pipeline}, the DESI spectrograph resolution is entirely characterized by the resolution matrix $\mathbf{R}$ \citep{guy_spectroscopic_2022}. In opposition to SDSS, the resolution matrix also accounts for the pixelization of the signal. Consequently, we choose to express the resolution correction function $W$ in equation (\ref{eq:praw_decomposition}) directly as the Fourier transform of the resolution matrix that we note $\mathbf{R}(k)$.

% Old description of the resolution modeling, valid for quickquasars mocks:

% As this resolution matrix is doubly-pixelized over both columns and rows, we model $W$ in wavelength space such that:

% \begin{equation}
%     \mathbf{R}(\lambda) = W(\lambda, \mathbf{R}) \circledast \Lambda(\lambda,\Delta \lambda_{\mathrm{pix}})\,,
% \end{equation}

% \noindent where $\Lambda = \Pi \circledast \Pi$ is a triangle function, and $W$ is the binning-independent spectroscopic resolution kernel. By applying a Fourier transform, we can express the resolution correction in Fourier space:

% \begin{equation}
%   W(k,\mathbf{R}) = \frac{\mathbf{R}(k)}{\mathrm{sinc}^2\left(\frac{k \Delta \lambda_{\mathrm{pix}}}{2}\right)}\,.
% \end{equation}

Fig. \ref{fig:mean_resolution} shows the average correction due to resolution and pixelization for the \svtda~data set. This correction indicates that resolution and pixelization suppress more than $95~\%$ of the signal at $k > k_{\mathrm{res},95} = 2.73$ \invAA, and $98~\%$ at $k > k_{\mathrm{res},98} = 3.15$ \invAA. This observation drives the maximal wavenumber that of the \pk~measurement. We choose the conservative value $k_{\mathrm{max}} = 2$ \invAA, for which the average resolution corrections is equal to 80\%. We will extend this conservative limit in future studies after a full characterization of resolution on CCD pixel-level simulations.

\subsubsection{Validation with CCD image simulations}

\begin{figure}
    \centering
    \includegraphics[width =\columnwidth]{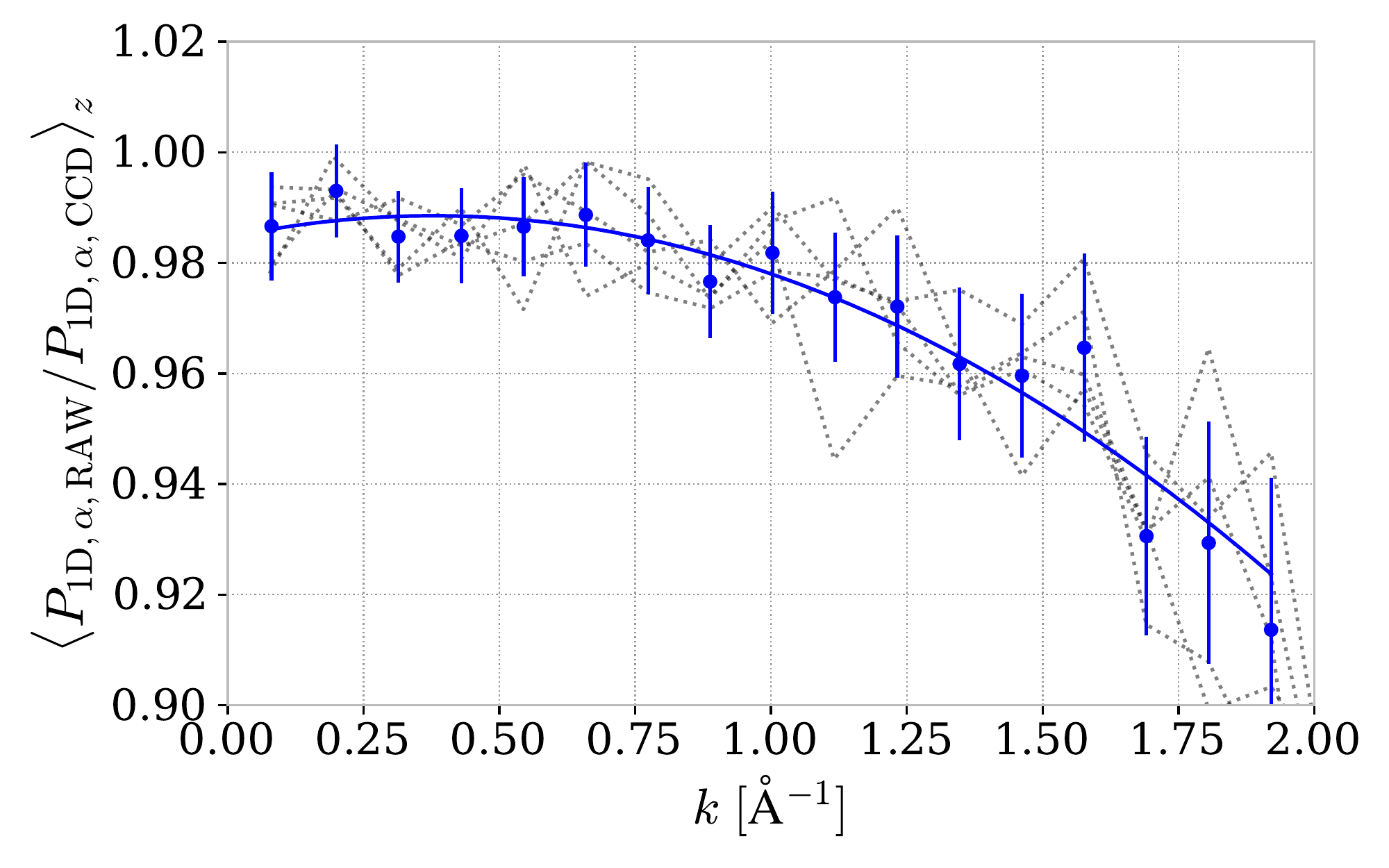}
    \caption{Ratios between the power spectrum obtained directly from \lya~transmissions (\textit{RAW}) and the one derived from the CCD image simulations (\textit{CCD}), for different redshift bins (light black) and averaged over all redshifts bins \CR{(blue points with error bars)}. Each power spectrum is re-binned by a factor 3 to reduce error bars. A second order polynomial is fitted to the ratio (blue continuous line). It is used to correct the miss-estimation of resolution in the \pk~measurement.}
    \label{fig:resolution_correction}
\end{figure}

We use CCD image simulations of the DESI instrument to verify our resolution modeling described previously. This method is also used for the QMLE estimation of \pk~in a companion paper~\citep{karacayli_optimal_2023}. Those simulations are built using the \texttt{desisim} \git{desihub/desisim}{} package developed to model and validate the spectroscopic extraction pipeline presented in Sec. \ref{subsec:desi_pipeline} and detailed in \citet{guy_spectroscopic_2022}. We use the \texttt{desisim} package to produce realistic realizations of two-dimensional spectroscopic images. Those images simulate various instrumental effects such as the different sources of noise detailed in appendix \ref{appendix:noise_estimators}, gain and bias of the CCD amplifiers, the throughput of each spectrograph, point spread function (PSF) of each fiber, and sky emission. 

Using the \texttt{desisim} package, we transform \lya~transmissions that follows a given input \pk~into realistic \lya~forests with apparent magnitudes representative of DESI quasar targets and noise that is representative of a single, 1000\,s exposure in nominal conditions. We simulate 45,000 \lya~forest spectra located over ten DESI tiles and process this simulated dataset with the spectroscopic pipeline. Since we only want to see the impact of spectral resolution, the true imposed noise level is used to reduce the data.

Two sets of \lya~contrasts $\delta_{F}$ are generated from this simulated quasar sample. The first set, called \textit{RAW}, is produced directly from \lya~transmissions. This type of realization
does not contain the effect of spectral resolution. In parallel, we create \lya~contrasts from the full CCD image simulations, noted \textit{CCD}, but using the true imposed continuum for each quasar to only see the impact of resolution. Finally, we run the \pk~FFT pipeline presented in Sec.~\ref{subsec:fft_estimator} with our resolution modeling on both \lya~contrast sets.

The ratio between \pk~obtained from \textit{RAW} and \textit{CCD} sets is shown in Fig. \ref{fig:resolution_correction}. As expected, the main difference between those measurements resides in the smallest scales ($k > 1.0$ \AA$ ^{-1}$). This ratio is not redshift dependent, as indicated by the light black curves in the background. We only consider the average overall redshifts. We checked that applying an additional pixelization correction $\mathrm{sinc}^{2} (0.5k \Delta \lambda_{\mathrm{pix}})$ similarly to eBOSS increases a lot the discrepancy between \textit{RAW} and \textit{CCD} power spectrum; thus confirming that the resolution matrix accounts the pixelization, at least partly. We derive a correction by fitting a second-order polynomial function to the following averaged ratio:

\begin{equation}
\label{eq:resolution_correction}
    A_{\mathrm{res}}(z,k) = \left\langle \frac{ P_{1\mathrm{D},\alpha,\mathrm{RAW}}(k,z)}{P_{1\mathrm{D},\alpha, \mathrm{CCD}}(k,z)}\right\rangle_{z}\,.
\end{equation}

This term is directly multiplied to our \pk~measurement to account for the miss-estimation of the resolution correction.

\subsection{Noise power spectrum measurement}
\label{subsec:noise}

\subsubsection{Comparison of noise estimators with high-wavenumber data}
\label{subsubsec:noise_asymptote}

The \pk~measurement is significantly impacted by the noise power spectrum at small scales. Thus, it is necessary to obtain an accurate estimate of this component to correct for it.

The noise power spectrum is estimated either directly from the pipeline noise, or by using an exposure difference method. A detailed description of the obtained noise power spectrum estimators, respectively noted \pp~and \pd, can be found in appendix \ref{appendix:noise_estimators}.

\begin{figure}
	\includegraphics[width=\columnwidth]{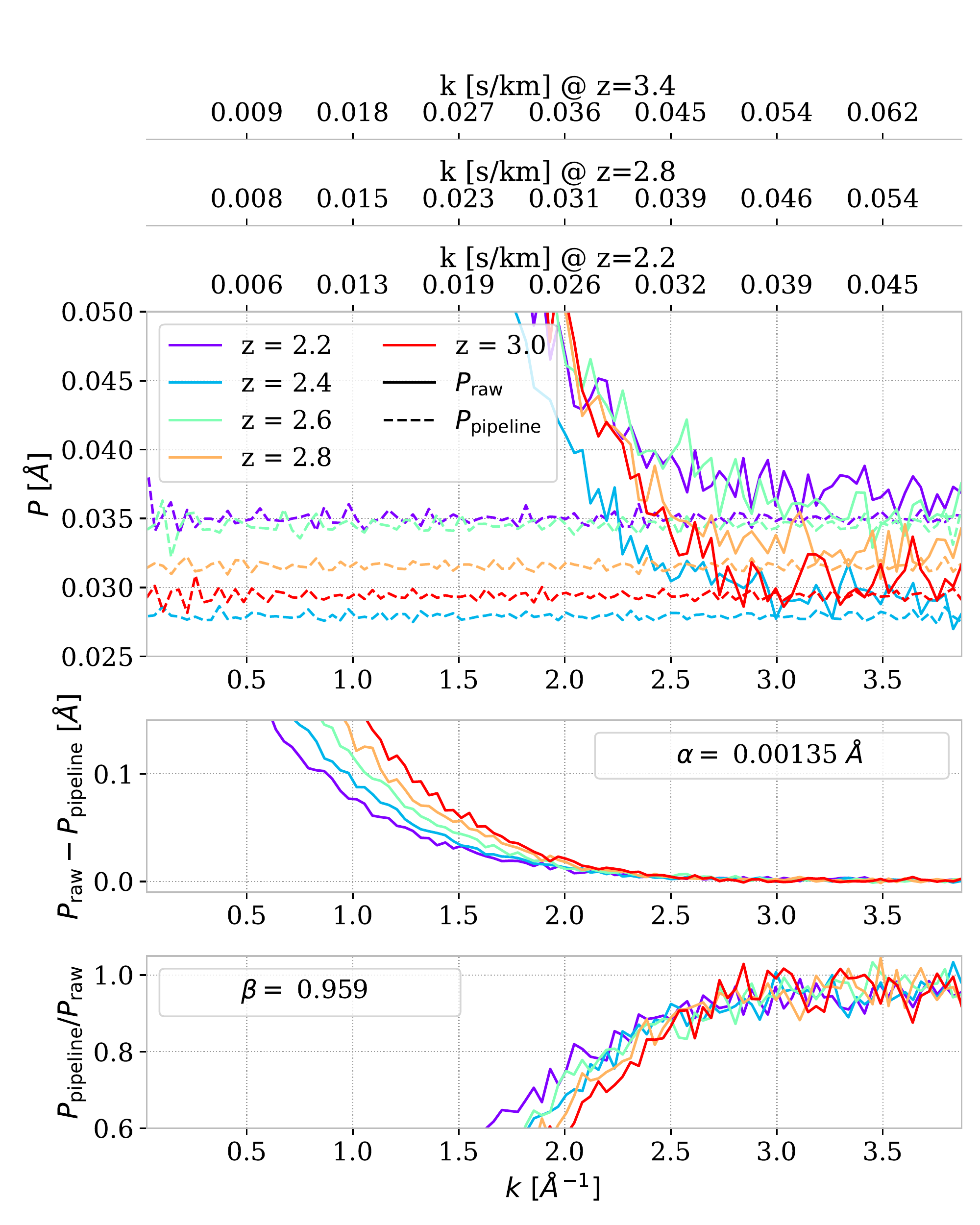}
    \caption{Measurement of the asymptotic difference ($\alpha$) and ratio ($\beta$) between noise (\pn) and raw (\pr) power spectra. Those terms are defined by equation \ref{eq:alpha_beta} and corresponds to the average of the difference and ratio of power spectra for large wavenumbers ($k > k_{\mathrm{res},98}$). Here, $\alpha$ and $\beta$ are measured for the pipeline noise (\pp), using the \svt~observations with a minimal \snr~cut of 3.}
    \label{fig:noise_asymptote}
\end{figure}

Additionally, the noise power spectrum level is determined by taking advantage of the combined effect of the resolution and the pixelization shown in Fig. \ref{fig:mean_resolution}, as well as the \lya~thermal broadening. Those effects erase essentially all "signal" power and thus at large wavenumbers, equation (\ref{eq:praw_decomposition}) simplifies into $P_{\mathrm{raw}}(k) \simeq P_{\mathrm{noise}}(k)$. 
The difference (or ratio) of \pr~and \pn~on the largest k-bins accessible can be used to validate the noise estimator and correct it empirically. We define the following asymptotic difference and ratio by averaging those quantities at large wavenumbers. We decide to use the criteria $k > k_{\mathrm{res},98}$ where $k_{\mathrm{res},98}$ is defined as the wavenumber for which the resolution and pixelization suppress more than $98 \%$ of the signal:

\begin{equation}
\label{eq:alpha_beta}
\begin{aligned}
\alpha &= \left \langle P_{\mathrm{raw}}(k) - P_{\mathrm{noise}}(k) \right \rangle_{k > k_{\mathrm{res},98}} \,, \\
\beta &= \left \langle \frac{P_{\mathrm{noise}}(k)}{P_{\mathrm{raw}}(k)} \right \rangle_{k > k_{\mathrm{res},98}}\,.
\end{aligned}
\end{equation}

Fig. \ref{fig:noise_asymptote} shows the measurement of $\alpha$ and $\beta$ on \svt~data set with \snr~$> 3$ (with \snr~defined by equation \ref{eq:mean_snr}), using the pipeline noise to compute \pn. Considering the observed statistical fluctuations, we notice that the asymptotic behavior of power spectra at high wavenumber enables a good measurement of $\alpha$ (respectively $\beta$), whose value is close to 0 (respectively 1). Additionally, we verified that the variation of $\alpha$ and $\beta$ as a function of redshift is small in comparison to the statistical fluctuations of the ratio and difference.

\begin{figure}
	\includegraphics[width=\columnwidth]{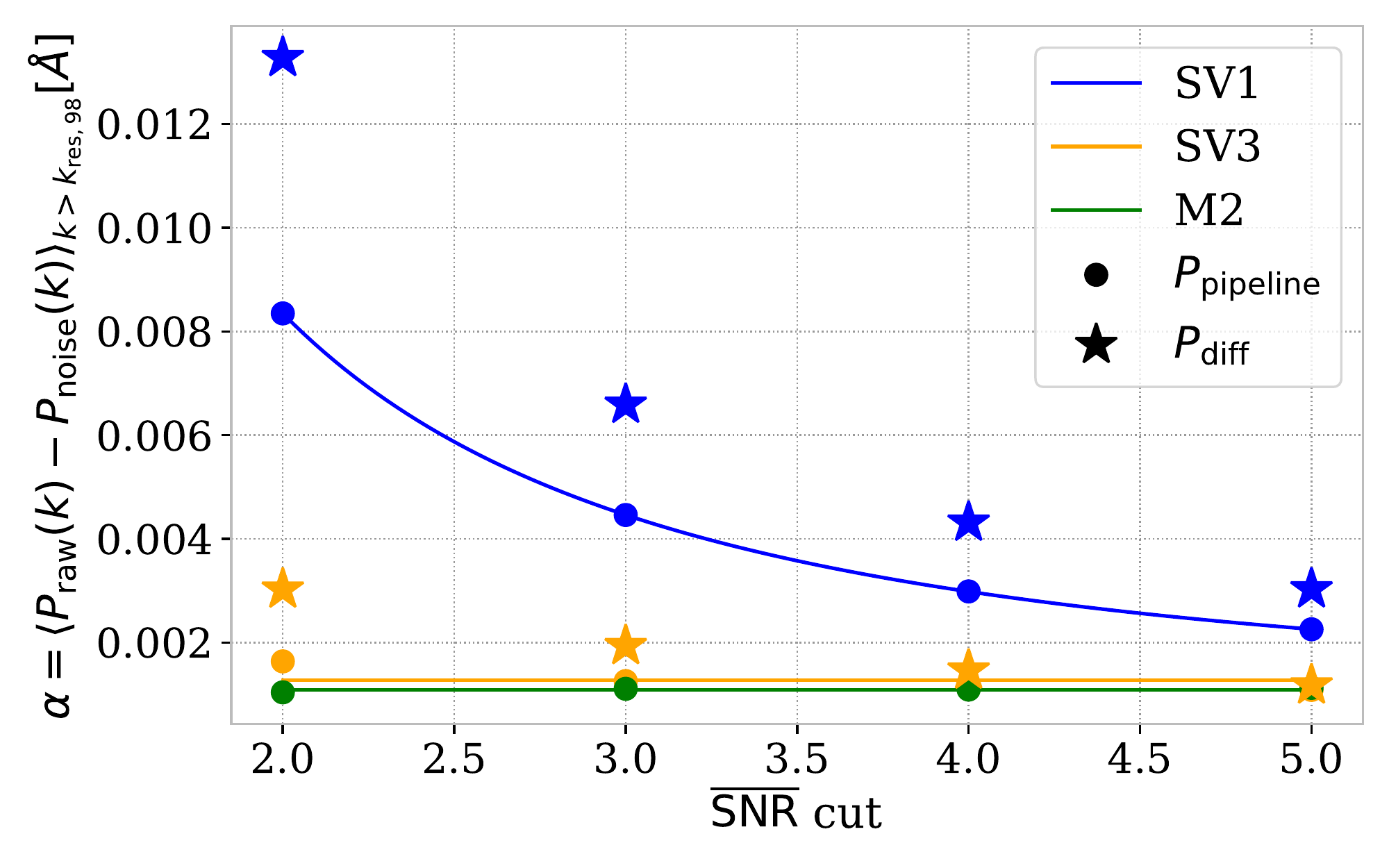}
    \caption{Asymptotic differences $\alpha$ between the noise and raw power spectra for \svo~(blue), \svt~(yellow), and \da~(green) data sets, as a function of the minimal \snr~cut. This difference is measured for both \pn~estimators from the pipeline (\pp, points) and from exposure differences (\pd, stars). The continuous lines are fits of the $\alpha$ values for pipeline noise, whose parameters are given in Tab. \ref{tab:noise_correction}.}
    \label{fig:diff_pipeline_comparison}
\end{figure}

\subsubsection{Characterization on DESI data sets}
\label{subsec:noise_data_set}

We compute the $\alpha$ and $\beta$ coefficients for the pipeline (\pp) and difference (\pd) noise estimators on \svo, \svt, and \da~data sets, while varying the applied \snr~cut.

Fig. \ref{fig:diff_pipeline_comparison} shows the measured $\alpha$ values on the \lya~forest regions for the different data sets and noise estimators. As previously stated, $\alpha$ should be equal to zero when the noise is perfectly estimated. The values of $\alpha$ are small compared to the absolute level of the noise power spectrum shown in Fig. \ref{fig:noise_asymptote}. The \svo~data set exhibits a \snr~cut dependence which is not present for \da~and \svt. We choose to make a data set-dependent correction to remove this residual noise. The miss-estimated noise is higher for data sets with a larger number of exposures (such as \svo). We think this originates from unaccounted common sources of noise coming from the statistical uncertainties in the CCD calibration data (dark current, pixel flat field), which explain why this effect increases with the number of exposures. The dependence in \snr~can be explained by the fact that this effect is amplified when we consider noisier spectra.

\begin{table}
	\centering
    \caption{Additive corrections applied to the pipeline noise for different spectral regions and data sets. An \snr~dependence is included in the case of \svo~only. The same parameters are used for both SB1 and SB2.}
    \label{tab:noise_correction}
    \begin{tabular}{l|c|c}
    \hline & & \\[-3mm]
    \textbf{Band} & \textbf{Data} & $\bm{P_{\mathrm{\textbf{noise,miss}}} = \alpha}$  \textbf{[\AA]} \\[1mm]
         \hline & & \\[-3mm]
    \lya & \svo~& $0.026 \times \left(\overline{\mathrm{SNR}}\right)^{-1.77} + 0.00076 $ \\
         & \svt~&  $0.00127$ \\
         & \da~& $0.00109$ \\
         \hline & & \\[-3mm] 
    Side-bands & \svo~& $0.018 \times \left(\overline{\mathrm{SNR}}\right)^{-1.52} + 0.000032 $ \\
      & \svt~& $0.00048$ \\
      & \da~& $0.00019$ \\
    \hline
    \end{tabular}
\end{table}

In Fig. \ref{fig:diff_pipeline_comparison}, the exposure difference noise estimator \pd~is shown for \svo~and \svt. In \da, quasars are observed with one, or a small number of exposures; thus, \pd~can not be reliably computed. For the \svo~and \svt~observations, the difference noise power spectrum exhibits the same trend as the pipeline noise estimate. However, \pd~consistently underestimates the noise level compared to \pp. We think that this under-estimation is due to the fact that \pd~is not accounting for all the common sources of noise between exposures. As a consequence, we only consider the \pp~estimation from now on.

In appendix \ref{subsec:noise_add}, we perform additional studies to characterize the additivity of the miss-estimated noise and its behavior for different spectra regions. Those regions, called side-bands, are used in the next section for metal power spectrum estimation. Tab. \ref{tab:noise_correction} summarizes the corrections to the pipeline noise we computed for the different data sets. For \svt~and \da, we choose to apply a constant additive correction ($\alpha$). For \svo, given the observed dependence of $\alpha$ as a function of the minimal \snr~cut, we fit it with a power-law.

\subsection{Side-band power spectrum}
\label{subsec:side_band}

Following previous \pk~studies \citep{mcdonald_lyman-alpha_2006,palanque-delabrouille_one-dimensional_2013,chabanier_one-dimensional_2019}, special spectrum regions, called side-bands and that are devoid of \lya~absorption, are used to statistically estimate the power spectrum components \pme~caused by metal absorptions in the \lya~forest. The resulting signal from the side-bands contains information about the abundance, temperature and clustering of metals in the intergalactic medium. In our study, we aim at creating a model to closely reproduce the side-band power spectrum (\ps) so that we can statistically subtract it in the measurement of \pk~in equation (\ref{eq:p1d_estimator}).

We define the side-bands SB1 ($1,270 < \lambda_{\mathrm{rf}} < 1,380$ \AA) and SB2 ($1,410 < \lambda_{\mathrm{rf}} < 1,520$ \AA). In both side-bands, the fraction of transmitted flux contrast can be expressed similarly to equation (\ref{eq:contrast_p1d}):

\begin{equation}
    \left.\delta_{F} (\lambda)\right|_{\mathrm{SB}}=\left.\delta_{\mathrm {metals}} (\lambda)\right|_{\mathrm{SB}} \circledast W(\lambda,\mathbf{R},\Delta \lambda_{\mathrm{pix}}) + \left.\delta_{\mathrm{noise}}(\lambda)\right|_{\mathrm{SB}}\,.
\end{equation}

The $\left.\delta_{\mathrm{metals}} (\lambda)\right|_{\mathrm{SB}}$ contrast contains all the fluctuations caused by metals with rest-frame absorption wavelength higher than $1,380$ \AA~for SB1, and $1,520$ \AA~for SB2. Similarly to the calculation of \pk, the side-band power spectrum writes:

\begin{equation}
P_{\mathrm{SB}}(k) = \left\langle\frac{\left.P_{\mathrm{raw}}(k)\right|_{\mathrm{SB}}-\left.P_{\mathrm {noise}}(k)\right|_{\mathrm{SB}}}{W^{2}(k, \mathbf{R},\Delta \lambda_{\mathrm{pix}})} \right\rangle\,.
\end{equation}

The main difference between both side-bands is that SB1 contains \siiv~absorption, which is not present in SB2. Consequently, we use the side-band power spectrum of SB1 to estimate \pme~in equation (\ref{eq:p1d_estimator}), and the SB2 power spectrum as a consistency check. The side-band power spectrum is computed at the same observed wavelength range as \pk. Because of the higher rest-frame absorption wavelength, the quasars employed to calculate the side-band power spectrum are at a lower redshift than the sample used for \lya. In particular, quasars at $z<2.0$ are employed to calculate \ps~in the lowest redshift bins of \pk.
However, the metals in the intergalactic medium which produces the absorptions responsible for the side-band power spectrum (\ps) are at the same redshift as those which produced the metal power spectrum (\pme) in the \pk~calculation for \lya. Consequently, the redshift dependence of metal absorptions is correctly taken into account.

\begin{figure}
	\includegraphics[width=\columnwidth]{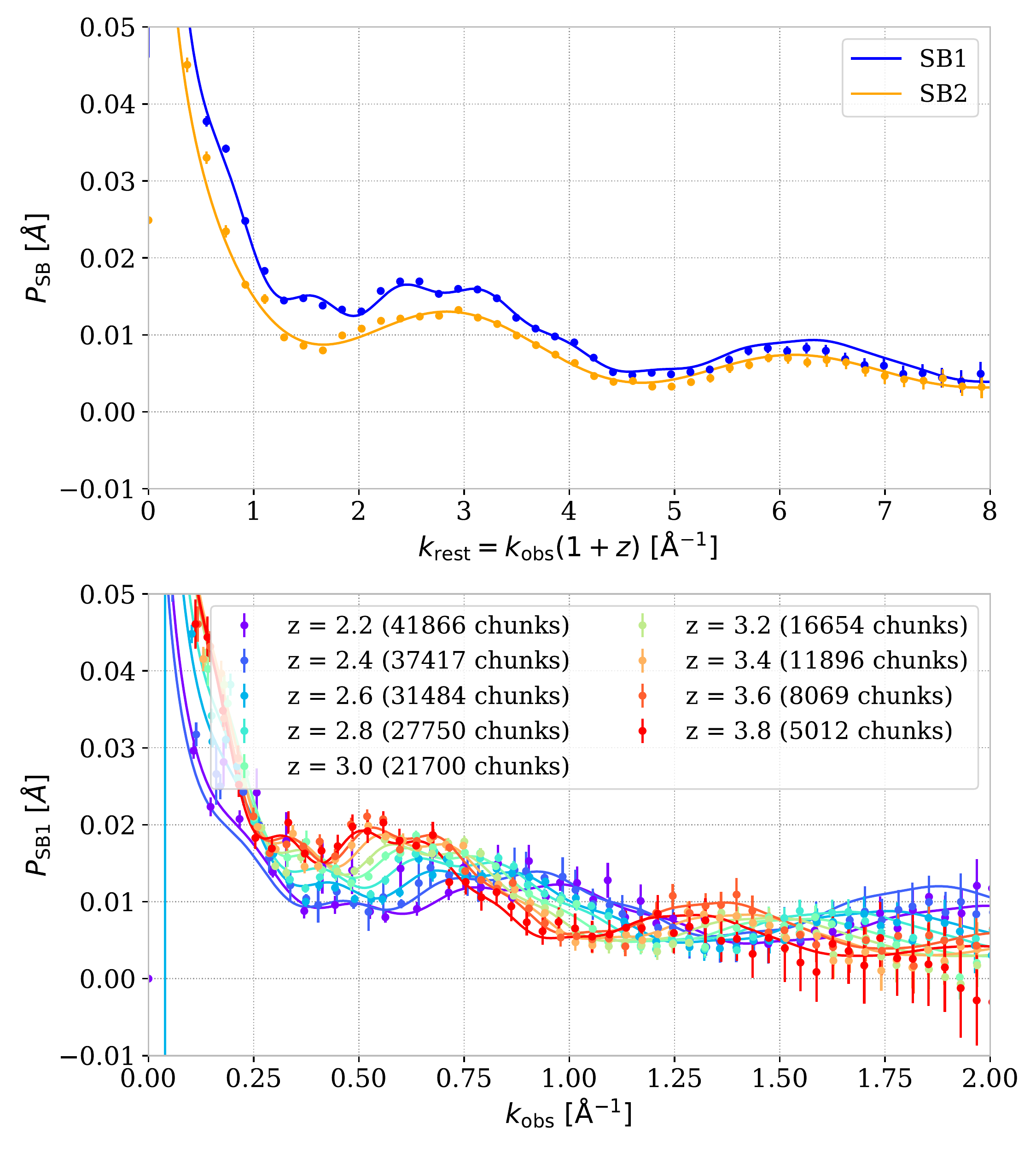}
    \caption{One-dimensional power spectra measured in the side-band regions SB1 and SB2, using the \svtda~data set after applying the dataset-dependent noise correction. (top) Average of \ps~over all redshift bins in the rest wavenumber frame $k_{\mathrm{rest},i} = k_{\mathrm{obs},i} \times (1+z)$. The fitted model represented in continuous line is given by equation (\ref{eq:side_band_model}). (bottom) \ps~on side-band SB1 as a function of redshift and observed wavenumber. Each redshift bin is fitted using the product between equation (\ref{eq:side_band_model}), with parameters fixed from previous fit, and a first-order polynomial function.}
    \label{fig:side_band}
\end{figure}

We note that the method we use to remove metal contribution is not perfect. In particular, the blending of metals with \lya~emission pointed out in \citet{day_power_2019} is not fully accounted here. This second-order effect should be included in future studies for which the precision level will significantly improve. 

The measurement of the side-band power spectrum using the \svtda~data set is shown in Fig. \ref{fig:side_band}. The top panel shows the stack of \ps~for all redshift bins as a function of wavenumber in rest-frame $k_{\mathrm{rest}} = (1+z) \times k_{\mathrm{obs}}$. The bottom panel shows separated redshift bins as a function of observed wavenumber. A total number of $201,849$ and $276,279$ sub-forests were used for SB1 and SB2 respective measurements.

In top panel of Fig. \ref{fig:side_band}, the average side-band power is lower for the SB2 than SB1, as expected by the addition of \siiv~absorptions. 

In the eBOSS measurement \citep{chabanier_one-dimensional_2019}, a sixth-degree polynomial is used to fit the \CR{shape of the side-band power}. For our measurement, we exploit the stacked \ps~profile to design a more physically motivated model.

A complete list of metals present in the \lya~forest and that impact \pk~is given in \citet{pieri_probing_2014,yang_metal_2022} and the strongest absorptions are from \siiii, \siii, \siiv, and \civ.

The emission peaks, and consequently absorption peaks, of \siiv~and \civ~are actually two doublets. Their rest-frame wavelength given by NIST \citep{kramida_nist_2021} are $\lambda_{\text{\siiv}^{\mathrm{a}}} = 1,393. 76$ \AA, $\lambda_{\text{\siiv}^{\mathrm{b}}} = 1,402.77$ \AA, $\lambda_{\text{\civ}^{\mathrm{a}}} = 1,548.202$ \AA, and $\lambda_{\text{\civ}^{\mathrm{b}}} = 1,550.774$ \AA.
The presence of an absorption doublet in the side-band creates a peak in the one-dimensional correlation function, which translates into an oscillatory pattern in the power spectrum, whose periodicity depends on the doublet separation. This effect is studied more in detail on the same dataset in \citet{karacayli_framework_2023} to determine cosmic ion abundance. In the top panel of Fig. \ref{fig:side_band}, both side-band power spectra display a large oscillation caused by the \civ~doublet. As expected, the SB1 power spectrum shows an additional oscillation induced by \siiv~doublet absorptions. These considerations lead us to model \ps~as the sum of a power-law including all-metal contributions and oscillations due to \siiv~and \civ~doublets:

\begin{equation}
    \label{eq:side_band_model}
    P_{\mathrm{SB,m}} = A \times k^{-\epsilon} + \sum_{i} P_{\mathrm{doublet},i}(k,A_{i},a_{i},k_{i},\psi_{i})\,.
\end{equation}

Oscillations induced by a doublet have a frequency characterized by the rest-frame wavenumber $k_{\mathrm{rest},i} = 2\pi/ \delta \lambda_{i}$ where $\delta \lambda_{i}$ is the doublet separation in  \AA. We choose to use damped sinusoidal functions to model the doublet oscillations as follows:

\begin{equation}
    P_{\mathrm{doublet},i}(k,A_{i},a_{i},k_{i},\psi_{i}) = A_{i} e^{-a_{i} k}  \sin\left(2\pi \left(\frac{k}{k_{i}}\right) + \psi_{i} \right) \,,
\end{equation}

\noindent where $k_{i}$ is a free parameter with a uniform prior centered around $k_{\mathrm{rest},i}$. 
 
The result of the fit on the redshift-averaged \ps, taking into account the oscillations of \civ~and \siiv~for SB1, and only \civ~for SB2, is shown in the top panel of Fig. \ref{fig:side_band}.

The SB1 fitted function is used to derive the redshift dependence of the side-band power spectrum, shown in the bottom panel of Fig. \ref{fig:side_band}. For each redshift bin, we fit a product between the global SB1 fitted function (expressed in the observational wavelength frame), and a first-order polynomial. As \ps~may also include other uncorrelated contaminations besides metals, we do not seek to interpret the fitted values for each power spectrum. In particular, we note that the $k_{i}$ are systematically shifted in comparison to their doublet oscillation frequency $k_{\mathrm{rest},i}$. For SB1, the fitted values are $k_{\mathrm{\civ}} = 3.32$~\invAA and $k_{\mathrm{\siiv}} = 0.812$~\invAA whereas rest-frame wavenumber are $k_{\mathrm{rest},\mathrm{\civ}} = 2.44$~\invAA and  $k_{\mathrm{rest},\mathrm{\siiv}} = 0.697$~\invAA. For SB2, we obtain $k_{\mathrm{\civ}} = 3.23$~\invAA. We think that this effect might be due to the blended impact of all metals present in the intergalactic medium. Thus, it is necessary to vary $k_{i}$ parameters to closely fit our data.

This \ps~measurement already represents a clear improvement with respect to that of BOSS \citep{palanque-delabrouille_one-dimensional_2013}, and eBOSS \citep{chabanier_one-dimensional_2019}: by eye, \siiv~and \civ~induced oscillatory patterns are seen even for individual redshift bins, and \ps~is essentially a decreasing function of wavenumber, even at high wavenumber. This indicates an improvement in the noise modeling.

\subsection{Atmospheric and Galactic emission lines}
\label{subsec:atmospheric_lines_p1d}

\begin{figure}
	\includegraphics[width=\columnwidth]{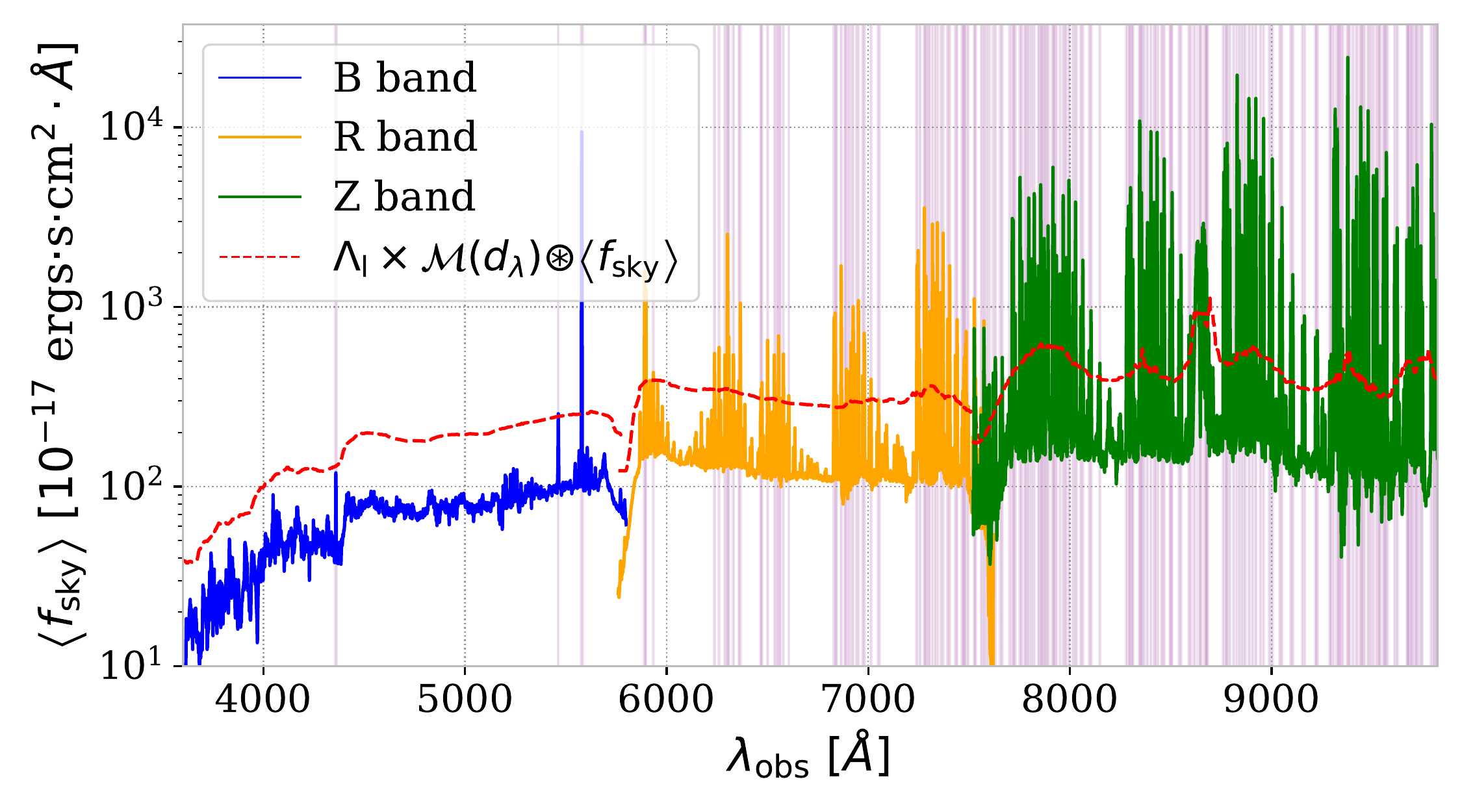}
    \caption{Average of the sky spectra of 15,000 sky fibers with optimal observation conditions (speed $> 105$, effective exposure time $> 1,100$ sec, seeing $< 1.05$ deg, and airmass $< 1.3$). These sky spectra originate from three exposures in the \svo~and \svt~data sets. The different spectral bands of DESI are represented (B in blue, R in orange, and Z in green). The median smoothing of this average sky spectrum multiplied by a threshold $\Lambda_{\mathrm{l}} = 2.5$, shown as a dashed red line, is used to select atmospheric emission lines we want to mask (light purple lines). The line located at $4,360$ \AA~was added manually, considering its large impact on noise seen in Fig. \ref{fig:vi_atmospheric_lines}.}
    \label{fig:sky_fiber}
\end{figure}

\begin{figure}
	\includegraphics[width=\columnwidth]{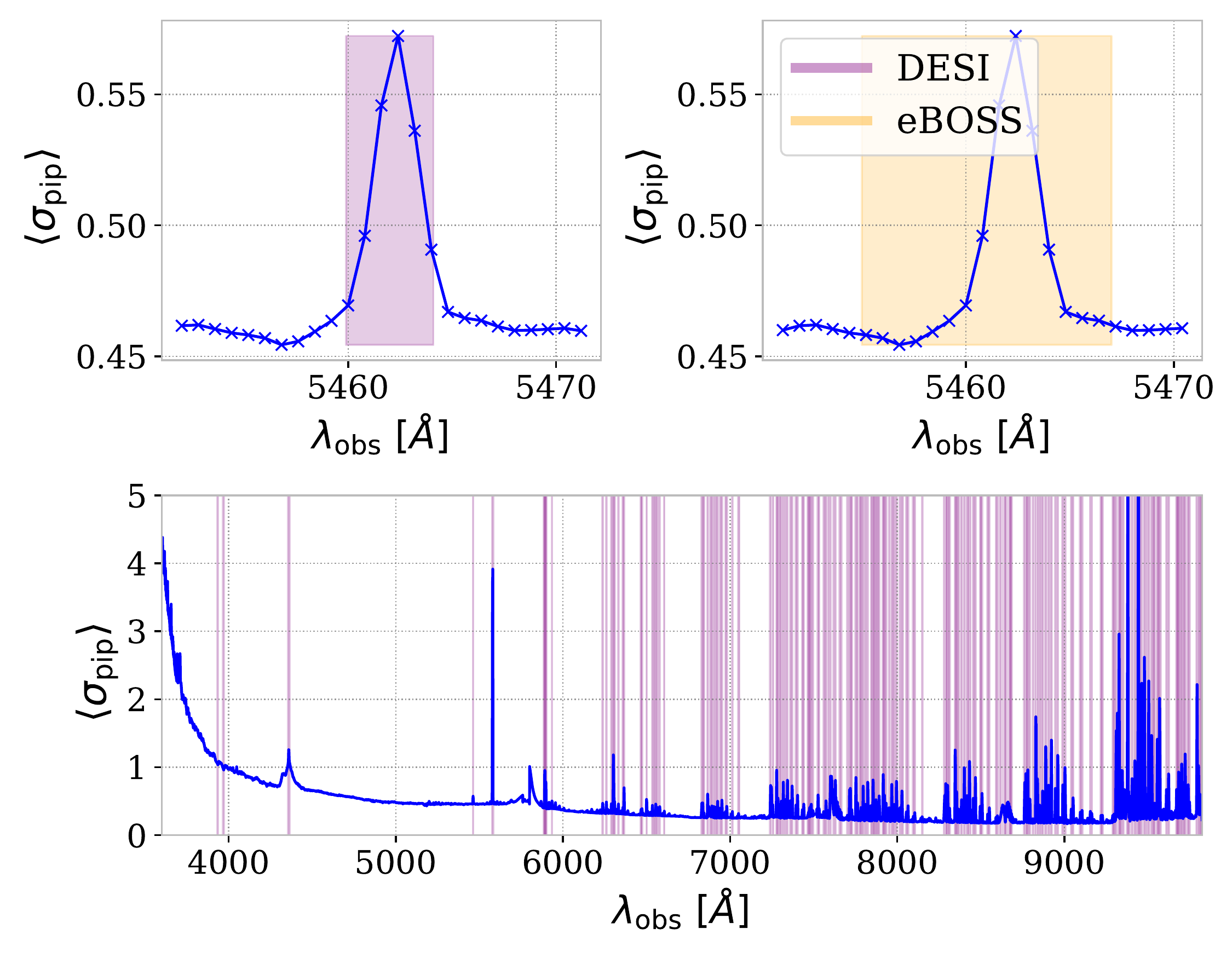}
    \caption{Comparison between the atmospheric emission line mask catalogs used for eBOSS$^{\ref{igmhub/picca/blob/master/etc/list_veto_line_Pk1D.txt}}$ and the new DESI catalog we designed in our study. \CR{(bottom) Average of the DESI pipeline noise for 125,477 objects categorized by \texttt{redrock} as quasars or luminous red galaxies on \svt~observations (blue curve). The DESI $(\Lambda_{\mathrm{l}} = 2.5,\Lambda_{\mathrm{w}} = 1.2)$ line catalog is shown on top with purple vertical lines. (top) Zoom on a specific atmospheric emission line on the DESI stacked noise (blue curve). The mask used on this specific line for DESI is shown in the left panel in purple and for eBOSS in orange on the right panel. The DESI mask decreases the masked length in accordance with stacked DESI noise.}}
    \label{fig:vi_atmospheric_lines}
\end{figure}

Atmospheric emission lines are corrected from DESI spectra by the spectral extraction pipeline as described in Sec. \ref{subsec:desi_pipeline}. The average of 15,000 sky spectra on exposures with optimal observing conditions, noted $\left\langle f_{\mathrm{sky}} \right\rangle$, is shown in Fig. \ref{fig:sky_fiber}. 

The noise of spectrum pixels associated with intense atmospheric lines is strongly increased. It induces additional oscillations in the \lya~contrasts and increases the level of \pk. We need to correct this effect as those atmospheric lines are not linked to IGM physics. We choose to mask the major atmospheric lines as in previous measurements \citep{palanque-delabrouille_one-dimensional_2013,chabanier_one-dimensional_2019}. The catalog of lines in these studies \git{igmhub/picca/blob/master/etc/list_veto_line_Pk1D.txt}{} was adapted to the spectral resolution of the SDSS instrument. The improved resolution of DESI makes it possible to reduce the masking size for narrow atmospheric lines, decreasing the impact of masking on the \pk~measurement

\begin{figure}
	\includegraphics[width=\columnwidth]{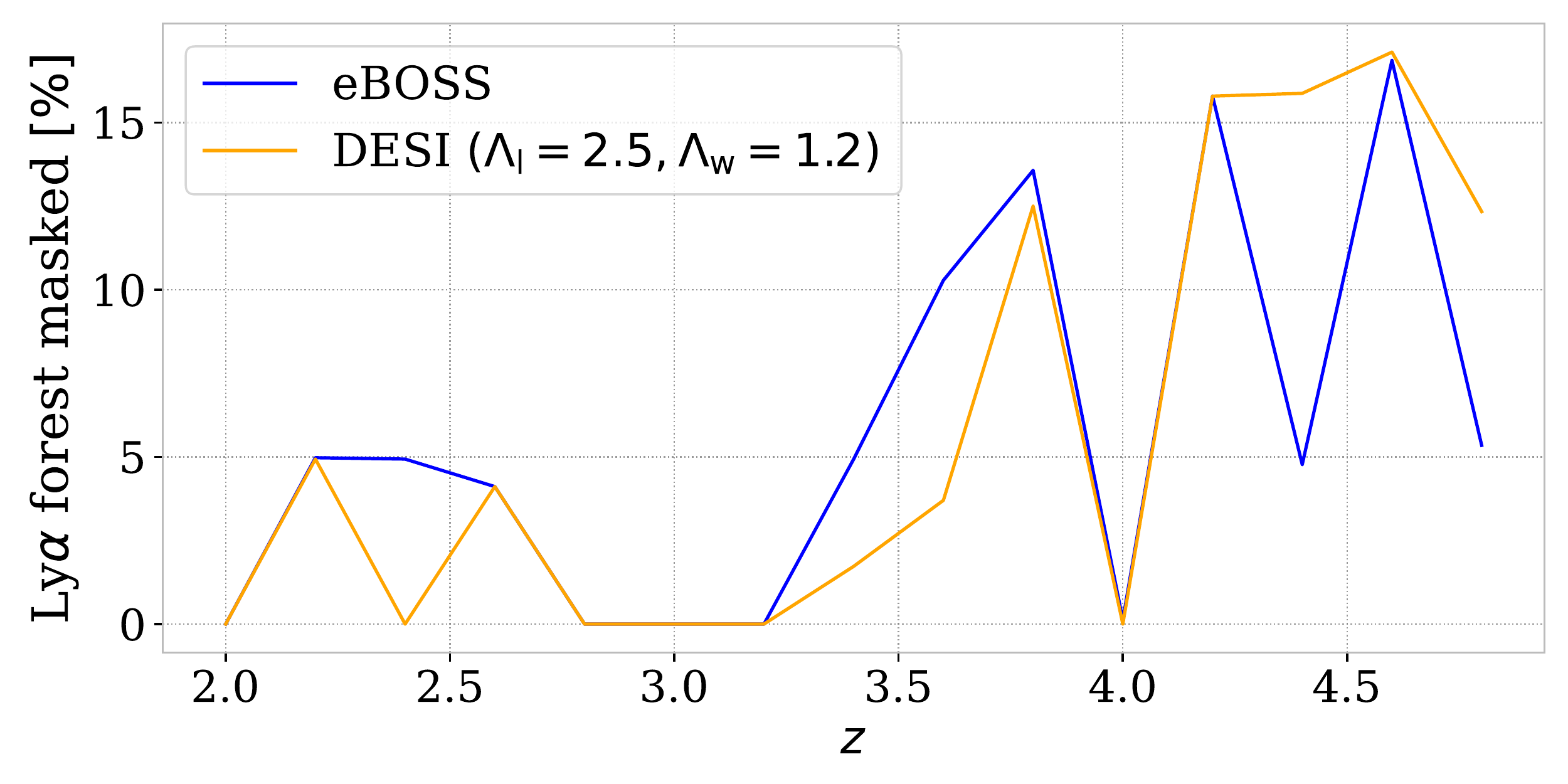}
    \caption{Percentage of spectral length masked by atmospheric lines as a function of redshift, for eBOSS$^{\ref{igmhub/picca/blob/master/etc/list_veto_line_Pk1D.txt}}$ and DESI $(\Lambda_{\mathrm{l}} = 2.5,\Lambda_{\mathrm{w}} = 1.2)^{\ref{corentinravoux/p1desi/blob/main/etc/skylines/list_mask_p1d_DESI_EDR.txt}}$ catalogs.}
    \label{fig:lines_masking_comparison}
\end{figure}

We develop an algorithm similar to \citet{lee_boss_2013} to compute an atmospheric line catalog adapted to the DESI instrument. A median smoothing $\mathcal{M}(d_{\lambda})$ of spectral width $d_{\lambda} = 160$ \AA~is applied on the average sky spectrum. Atmospheric lines are selected when the average sky spectrum is larger than the product of the smoothed sky flux, $\mathcal{M}(d_{\lambda}) \circledast \left \langle f_{\mathrm{sky}} \right \rangle$, by a threshold $\Lambda_{\mathrm{l}}$. In Fig. \ref{fig:sky_fiber}, the dashed red line represents this product for $\Lambda_{\mathrm{l}} = 2.5$ and $d_{\lambda} = 160~\text{\AA}$. 

A second threshold $\Lambda_{\mathrm{w}}=1.2$ defines the width of atmospheric lines. The upper and lower wavelength limits of an atmospheric line are defined as the first wavelengths on each side whose average sky spectrum is lower than $\Lambda_{\mathrm{w}} \times \mathcal{M}(d_{\lambda}) \circledast \left \langle f_{\mathrm{sky}} \right \rangle$. To remain conservative and prevent numerical effects potentially caused by masking at a spectrum pixel position, the line widths are increased by 1 \AA~on each side.

In this atmospheric line catalog, we also add the galactic absorption lines, which correspond to relatively broad absorptions made by dust in the Milky Way. We take the same lines as eBOSS: \caii~H and K lines at $3,968$ and $3,933$ \AA, and the NaD doublet at $5,893$ \AA. The DESI atmospheric emission line catalog built from this procedure is available online$^{\ref{corentinravoux/p1desi/blob/main/etc/skylines/list_mask_p1d_DESI_EDR.txt}}$.

We verify that the produced atmospheric line catalog correctly masks the DESI noise. For this purpose, we compute an average noise by stacking the pipeline noise $\sigma_{\mathrm{pip}}$ of $125,477$ objects categorized by \texttt{redrock} as quasars or luminous red galaxies on \svt~observations. This average noise with the DESI atmospheric line catalog such that $(\Lambda_{\mathrm{l}} = 2.5,\Lambda_{\mathrm{w}} = 1.2)$ is shown in the bottom panel of Fig. \ref{fig:vi_atmospheric_lines}. A zoom on an atmospheric line is shown in the top panels of Fig. \ref{fig:vi_atmospheric_lines} with eBOSS and DESI masks. The eBOSS mask is too wide for DESI stacked noise, which highlights the benefit of creating a new catalog. After a visual inspection of most atmospheric lines, we validate that all the spectrum pixels showing an increase in noise are masked by setting the width threshold to $\Lambda_{\mathrm{w}} = 1.2$.

Comparing the average noise (Fig. \ref{fig:vi_atmospheric_lines}) to the average sky flux (Fig. \ref{fig:sky_fiber}), there is a consistency between atmospheric emission lines and observed peaks in the pipeline noise. The feature at $4,360$ \AA~is an exception, as it appears wide and relatively high in the DESI noise and not in the average sky flux. Its wavelength is inside a known transmission dip around $4,400$ \AA~due to an issue with DESI's spectrograph collimator coating \citep{guy_spectroscopic_2022}. For this specific line, we take the same value as the eBOSS catalog and force the algorithm to consider it as an atmospheric line even if it does not pass the $\Lambda_{\mathrm{l}}$ requirement.

A comparison of the percentage of \lya~forest masked for eBOSS and DESI catalog, as a function of redshift, is given in Fig. \ref{fig:lines_masking_comparison}. To remain conservative, \CR{we chose the value of $\Lambda_{\mathrm{l}} = 2.5$ to obtain} a catalog of atmospheric emission lines with a spectral length masked similar to the eBOSS catalog. A complete study on synthetic data, out of this paper's scope, will be done to decrease the length masked, and consequently the impact on \pk.

\section{Synthetic data corrections}
\label{sec:synthetic_data}

Synthetic data (otherwise called mocks) are generated to characterize the impact on \pk~of continuum fitting, spectral resolution, noise modeling, and spectrum pixel masking. From this, we derive empirical corrections of these effects and apply them to the data measurement.

\subsection{Synthetic data sets}

We generated a set of \texttt{DESI-Lite} \citep{karacayli_optimal_2020} mocks, specifically designed for \pk. The full description of these mocks is given in \citet{karacayli_optimal_2023}. The \texttt{DESI-Lite} software produces uncorrelated \lya~forests that mimic the redshift and noise distribution of the \svtda~dataset. Ten independent realizations are generated with different initial conditions. For each realization, a random catalog of DLAs is created to follow the redshift and column density distribution of the latest eBOSS catalog \citep{chabanier_completed_2021}.

The \texttt{quickquasars} software \git{desihub/desisim/blob/main/py/desisim/scripts/quickquasars.py}{} \citep{quickquasars}, included in the \texttt{desisim} package$^{\ref{desihub/desisim}}$, transforms \lya~transmission into spectra with observational and astrophysical contaminants. For each realization, two sets of spectra are generated by imprinting DLAs according to the catalog aforementioned or not.

We run 5 different \pk~variations to study the impact of different \lya~contaminants: 

\begin{itemize}
    \item \textbf{\textit{TRUECONT}}:\enspace The true quasar continuum imposed by \texttt{quickquasars} is applied instead of the continuum fitting procedure described in Sec. \ref{subsec:delta_extraction}. In comparison to \textit{RAW} mocks, this realization is impacted by finite noise and resolution.
    \item \textbf{\textit{CONT}}:\enspace The \lya~contrasts are calculated using the pipeline detailed in Sec. \ref{subsec:delta_extraction}. This type of mocks includes the impact of continuum fitting.
    \item \textbf{\textit{DLAm}}:\enspace Realization for which the DLAs are not added to forests at the \texttt{quickquasars} stage, though we mask spectrum pixels as if they were present. The objective of this kind of mocks is to characterize the impact of DLA masking.
    \item \textbf{\textit{LINEm}}:\enspace Similarly to \textit{DLAm} but masking the atmospheric emission lines catalog built in Sec. \ref{subsec:atmospheric_lines_p1d}.
    \item \textbf{\textit{DLA}}:\enspace Realization for which the DLA are applied to the spectra without masking them. The objective of this mocks is to measure the impact of DLA to compute a DLA completeness systematic error in Sec. \ref{sec:uncertainties}.
\end{itemize}

For all the mocks, we take the same procedure as for the FFT calculation on observational data. In particular, the same \snr-weighting is applied, and the number of sub-forest for each realization is around $81,500$ with a small statistical variation between realizations. This is slightly larger than the number of sub-forest of the data sample given in Sec.~\ref{sec:results}.

In the next sections, all the results are shown for the combination of ten independent realizations. To decrease the error bars, we also performed a linear rebin that provides a wavenumber binning three times coarser than what we used for observational data. The error bars of the presented ratios are computed in quadrature.

\subsection{Continuum fitting correction}
\label{subsec:continuum_correction}

\begin{figure}
	\includegraphics[width=\columnwidth]{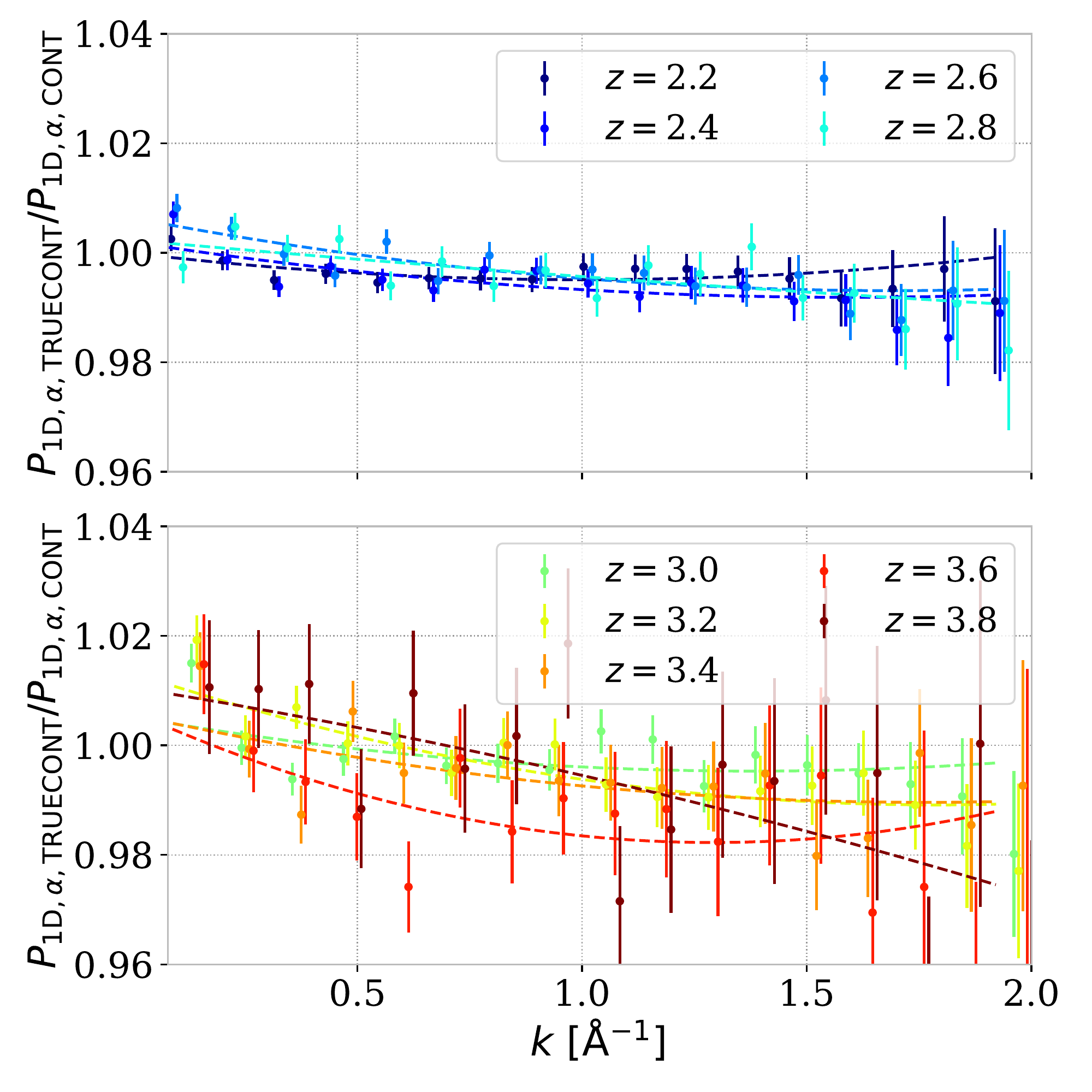}
    \caption{Ratios between the power spectrum obtained using true continuum (\textit{TRUECONT}) and the one derived with our pipeline (\textit{CONT}) on the combination of 10 mocks. Each power spectrum is re-binned by a factor 3 to reduce error bars. Fitting functions are represented by continuous lines, and used to correct the \pk~measurement. For clarity, we artificially offset the points corresponding to different redshifts.}
    \label{fig:continuum_correction}
\end{figure}

The continuum fitting procedure defined in Sec. \ref{subsec:delta_extraction} systematically distort the measured $C_{\mathrm{q}}(\lambda,z_{\mathrm{q}}) \overline{F}(\lambda)$ term by suppressing large-scale modes, and may bias the \pk~measurement. This is a well-known effect in BAO measurements \citep{bourboux_completed_2020}. To create a correction which contains this effect, we compare the mock computed using the true continuum (\textit{TRUECONT}) with the one which follows the standard continuum fitting procedure (\textit{CONT}):

\begin{equation}
\label{eq:cont}
    A_{\mathrm{cont}}(z,k) = \frac{P_{1\mathrm{D},\alpha,\mathrm{TRUECONT}}(k,z)}{P_{1\mathrm{D},\alpha, \mathrm{CONT}}(k,z)}\,.
\end{equation}

This correction is shown for the combination of ten \texttt{DESI-Lite} mocks in Fig. \ref{fig:continuum_correction}. We use a second-order polynomial function to fit this correction and apply it to the \pk~measurement. 

This correction differs in amplitude compared to the eBOSS measurement \citep{chabanier_one-dimensional_2019}. As for eBOSS, the one-dimensional power spectrum with continuum fitting is higher than that measured with the true continuum. However, in our case, the impact is much smaller than eBOSS, for which this ratio was near 4\% (without using a first-order polynomial function in the continuum fitting). Furthermore, we do not have a large-scale impact as significant as eBOSS.

\subsection{Spectrum pixel masking}
\label{subsec:masking_p1d}

\begin{figure}
	\includegraphics[width=\columnwidth]{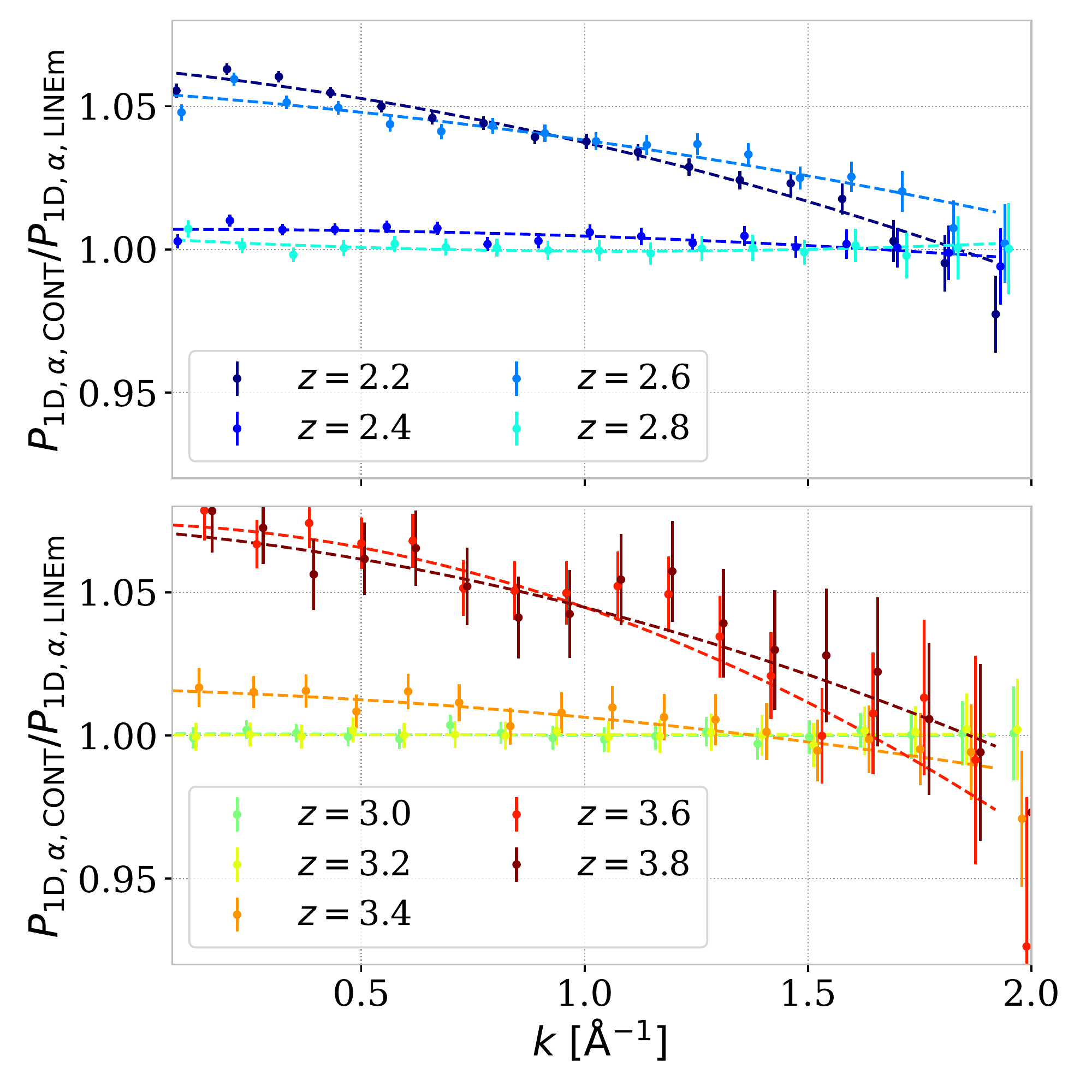}
    \caption{Ratio between the unmasked (\textit{CONT}) and masked (\textit{LINEm}) power spectra (equation (\ref{eq:skylines})) for atmospheric line masking on the combination of the 10 mocks. The DESI $(\Lambda_{\mathrm{l}} = 2.5,\Lambda_{\mathrm{w}} = 1.2)$ atmospheric line catalog is used. Each power spectrum is re-binned by a factor 3 to reduce error bars. Second-order polynomial functions are employed to fit the corrections in each redshift bin. For clarity, we artificially offset the points corresponding to different redshifts.}
    \label{fig:lines_masking}
\end{figure}

\begin{figure}
	\includegraphics[width=\columnwidth]{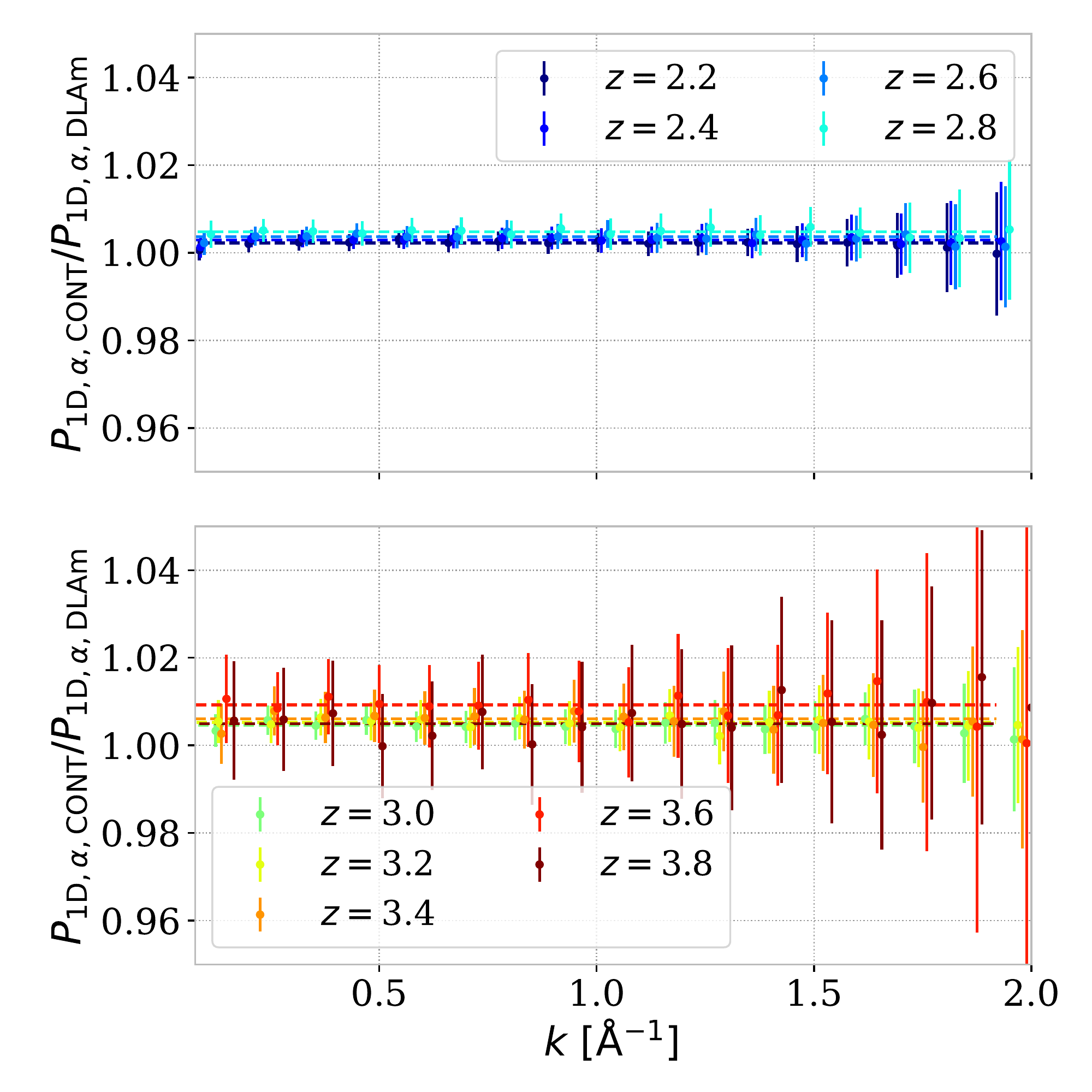}
    \caption{Ratio between the unmasked (\textit{CONT}) and masked (\textit{DLAm}) power spectra (equation (\ref{eq:dla})) for DLA masking on the combination of the 10 mocks. Each power spectrum is re-binned by a factor 3 to reduce error bars. Continuous lines shows constant fits are used for correction. For clarity, we artificially offset the points corresponding to different redshifts.}
    \label{fig:dla_masking}
\end{figure}

For both DLA and atmospheric line masking, we remove some data points from the measured spectra. This does not impact studies performed on real-space spectra, such as the continuum fitting or the quadratic maximum likelihood estimator for \pk~\citep{karacayli_optimal_2022}. On the other hand, the FFT calculation requires that spectrum pixels are equally-spaced. Consequently, when computing the Fourier transform, we impose a  value of $\delta_{F}=0$ (equivalent to mean transmitted flux fraction value for $F$) and infinite standard deviation to the masked spectrum pixels. This masking introduces a $k$-dependent bias, which we need to quantify.

In order to determine and correct this bias in our \pk~measurement, we compare mocks for which DLAs or atmospheric lines are masked (respectively \textit{DLAm} and \textit{LINEm}) with mocks where no masking is applied (\textit{CONT}). On those two mocks, the DLAs and \CR{atmospheric} emission lines are not imposed on spectra. We want to derive only the corrections of masking in order to apply them on data. The coefficients used for both masking corrections are defined as the ratio between the unmasked and the masked power spectra:

\begin{equation}
    \label{eq:skylines}
    A_{\mathrm{line}}(k,z) = \frac{P_{1\mathrm{D},\alpha,\mathrm{CONT}}(k,z)}{P_{1\mathrm{D},\alpha,\mathrm{LINEm}}(k,z)}\,.
\end{equation}

\begin{equation}
    \label{eq:dla}
    A_{\mathrm{dla}}(k,z) = \frac{P_{1\mathrm{D},\alpha,\mathrm{CONT}}(k,z)}{P_{1\mathrm{D},\alpha,\mathrm{DLAm}}(k,z)}\,.
\end{equation}

\subsubsection{Atmospheric emission lines}

The correction induced by the DESI atmospheric line mask, as defined in Sec. \ref{subsec:atmospheric_lines_p1d}, is shown in Fig. \ref{fig:lines_masking}. We verified that, for all redshifts where the masked \lya~forest length of DESI is close to eBOSS, the impact of masking is lower in the DESI case. It indicates that applying thinner masks to our measurement mitigates the impact of masking.

As expected, the correction roughly scales with the number of masked spectrum pixels. The effect of masking is a relatively smooth function of wavenumber, and its main impact is at low wavenumber. The most impacted redshift bins are $z=2.2$ (\caii~galactic absorption lines), $z=2.6$ (lines at $4,360$ \AA~in the transmission dip), and at high redshift for which many atmospheric lines need to be masked. As shown in Fig. \ref{fig:lines_masking_comparison}, redshifts $z=2.4,2.8,3.0,3.2$ have no masks applied, and only a few for $z=3.4$. It is also in agreement with the level of corrections.  The impact of atmospheric line masking is qualitatively in agreement with the eBOSS results in \citet{chabanier_one-dimensional_2019}. We choose to model $A_{\mathrm{line}}(k,z)$ by a second-order polynomial fit and use this correction in the final calculation of \pk.

\begin{figure*}
	\includegraphics[width=\textwidth]{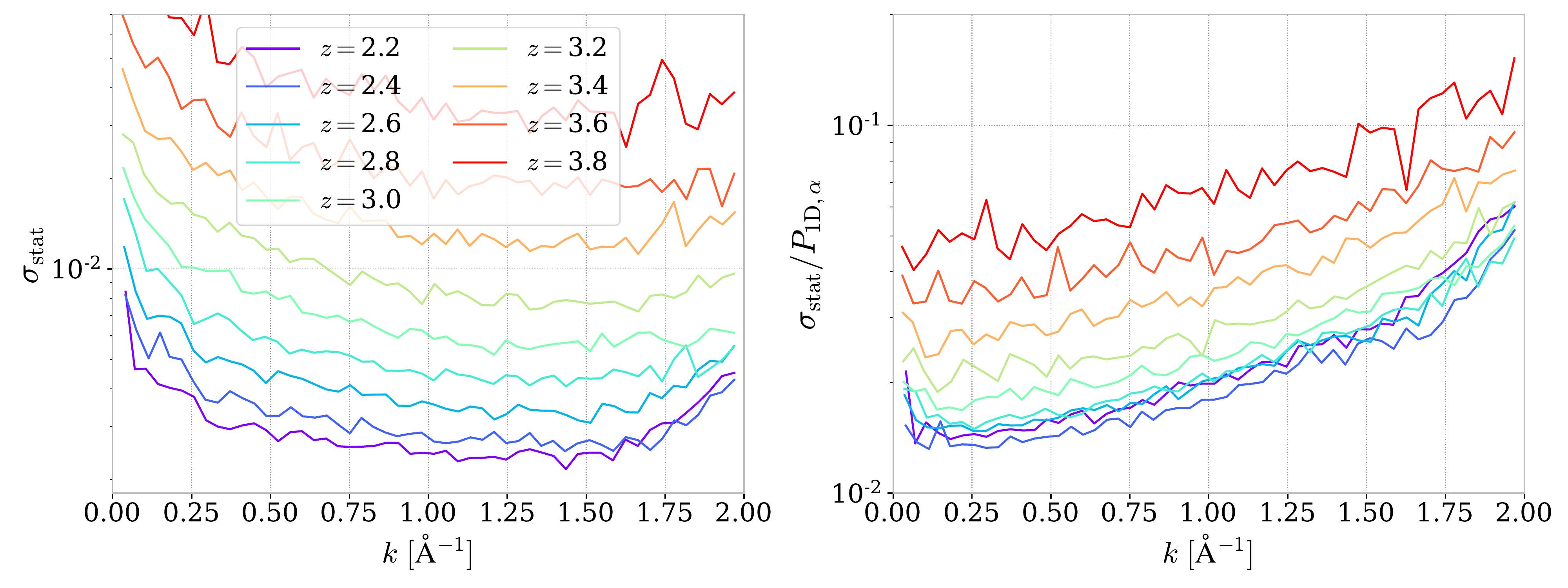}
    \caption{Statistical uncertainties (left) and its relative value with respect to \pk~(right) of the DESI \svtda~measurement in \AA, as a function of wavenumber.}
    \label{fig:uncertainties}
\end{figure*}

\subsubsection{DLA masking}

DLAs are added at random locations in the \lya~forest during the creation of the mocks. For this study, we do not attempt to characterize the completeness of the DLA finder applied to the data, and we use a "truth" DLA catalog for masking.

We mask the "truth" catalog with the same parameters as the DLA data catalog, i.e., for $N_{\mathrm{\hi}} > 10^{20.3}~\mathrm{cm}^{-2}$. The correction induced by the masking, $A_{\mathrm{dla}}(k,z)$, is represented in Fig. \ref{fig:dla_masking}. As it was already seen in the eBOSS measurement \citep{chabanier_one-dimensional_2019}, the DLA masking has a small impact compare to atmospheric emission lines. This is due to the random distribution of DLAs and the smaller masking in terms of \lya~forest length. As the impact is very similar for all wavenumbers, we apply a k-independent correction $A_{\mathrm{dla}}(k,z) = A_{\mathrm{dla}}(z)$, whose amplitude is $0.5\,\%$ on average.

\section{Uncertainty estimation}
\label{sec:uncertainties}

\begin{figure*}
	\includegraphics[width=0.92\textwidth]{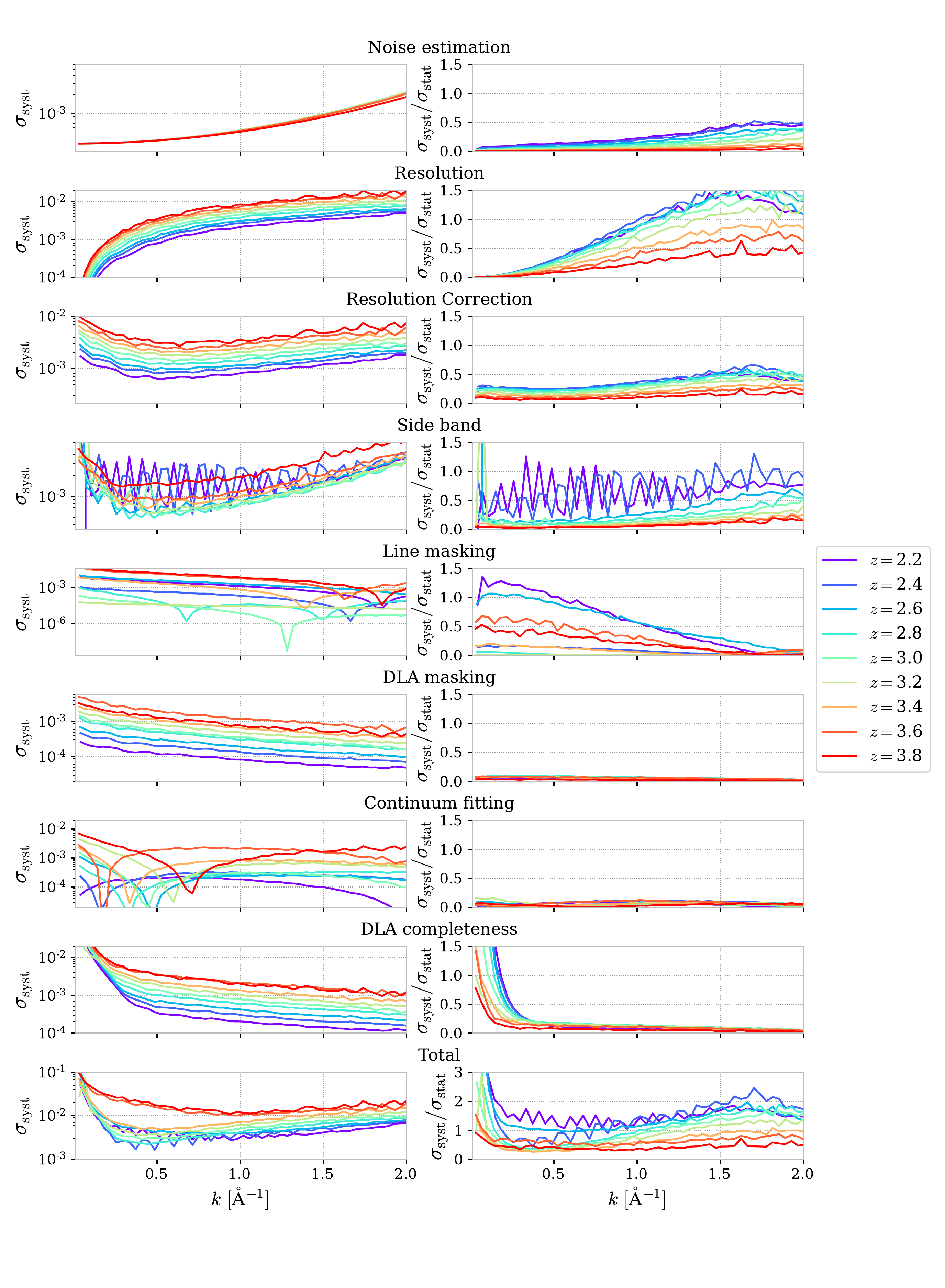}
    \caption{Value of the systematic uncertainties $\sigma_{\mathrm{syst}}$ in \AA, for different redshift bins, on the DESI \svtda~\pk~measurement. Each line is associated to a systematic considered in this article. The left panels show the absolute uncertainties, and the right panels, their relative values with respect to statistical uncertainties showed in Fig. \ref{fig:uncertainties}.}
    \label{fig:systematic_uncertainties}
\end{figure*}

\begin{figure*}
	\includegraphics[width=0.95\textwidth]{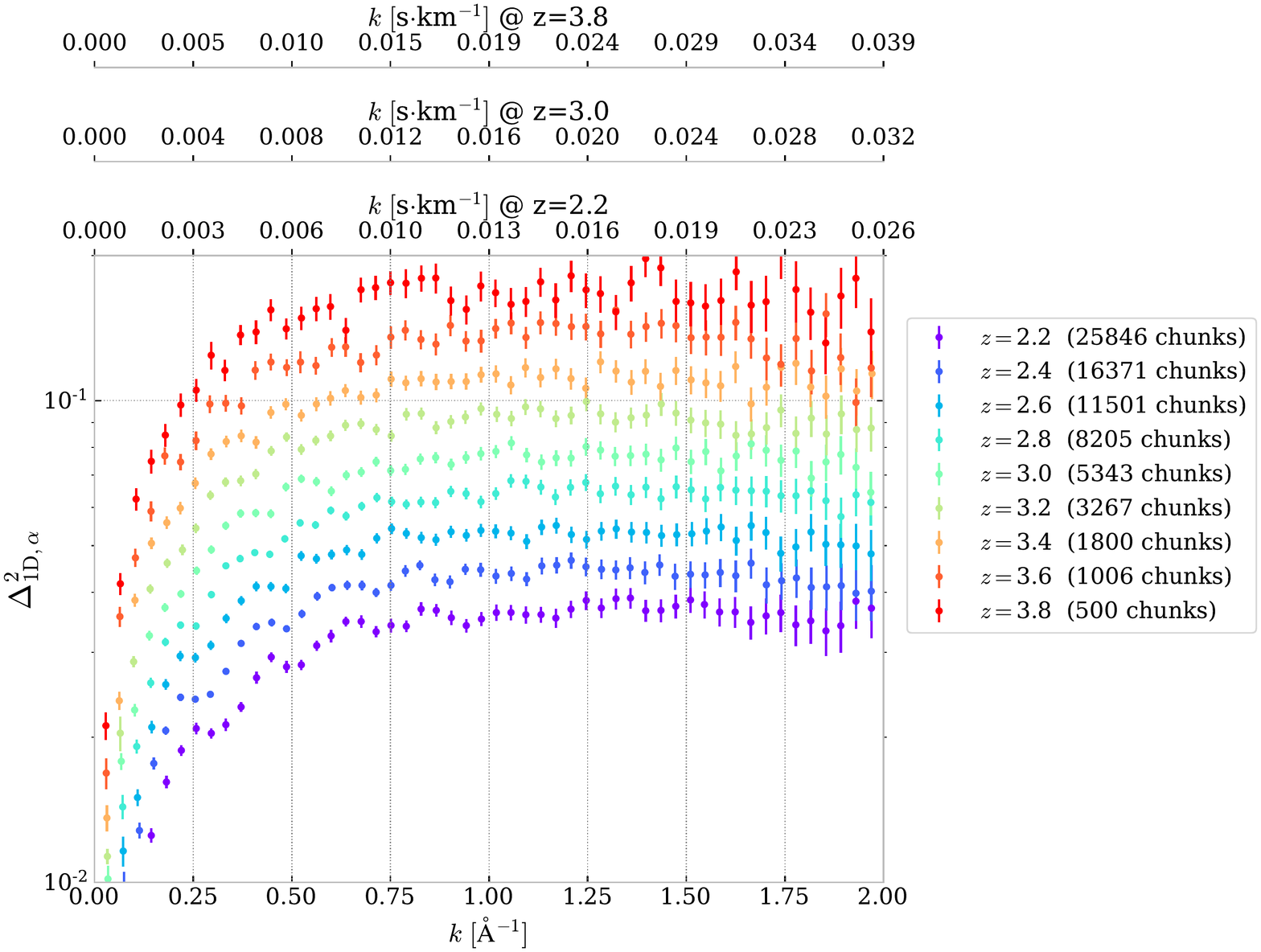}
    \caption{Normalized one-dimensional \lya~forest power spectrum ($\Delta_{1\mathrm{D},\alpha}(k)$) using the \svtda~data set, for redshift bins from $z=2.2$ to $z=3.8$. All the corrections given in equation (\ref{eq:final_p1d_estimator}) are applied to perform this measurement. As an illustration, wavenumbers in velocity space for different redshifts are represented at the top of the figure. Error bars are systematic and statistical uncertainties added in quadrature.}
    \label{fig:measurement}
\end{figure*}

The statistical uncertainty of our averaged \pk~measurement, noted $\sigma_{\mathrm{stat}}$, is obtained during the \snr~weighting scheme presented in appendix \ref{appendix:snr_weighting}. For each $(k,z)$ bin, a binned histogram of standard deviation as a function of \snr~is derived. Fitting this histogram provides a function $\sigma_{k,z}(\overline{\mathrm{SNR}})$ that is used to define the statistical uncertainty:

\begin{equation}
    \sigma_{\mathrm{stat}}(k,z) = \sqrt{\frac{1}{\sum_{i} \left(\sigma_{k,z}(\overline{\mathrm{SNR}}_{i}) \right)^{-2}}}\,,
\end{equation}

\noindent where $\sigma$ is the fitted function, and the $i$ index runs over the \snr~bins chosen.

The obtained statistical uncertainties are shown in Fig. \ref{fig:uncertainties}. Despite using a \snr-dependent weighting in the \pk~calculation compared to a redshift-dependent \snr~cut as eBOSS, we find similar trends as in \citet{chabanier_one-dimensional_2019}. The statistical uncertainty depends mainly on the number of sub-forests for each redshift bin, so that $\sigma_{\mathrm{stat}}$ is an increasing function of redshift. This error bar depends also on the power spectrum level, as pointed out by the right panel of Fig. \ref{fig:uncertainties}, for which low redshift bins are not separated. In the case of $z \sim 2.2 - 2.4$, at small scales ($k \gtrsim 1.5$ \invAA), the statistical uncertainties are crossing each other. It is caused by the large noise increase in the blue spectral band due to atmospheric absorptions. 

Looking at wavenumber dependence, $\sigma_{\mathrm{stat}}$ increases as a function of $k$ for small scales ($k \gtrsim 1.5$ \invAA). This is due to the resolution correction, which effectively increases the rms of individual \pk. At large scales ($k \lesssim 1$ \invAA), $\sigma_{\mathrm{stat}}$ is a decreasing function of $k$, mainly due to the decrease in Fourier modes available to compute \pk.

In our study, we characterized the impact of several instrumental and astrophysical contaminants. From this extensive study, we associate systematic errors, noted $\sigma_{\mathrm{syst}}$, to our \pk~measurement. Fig. \ref{fig:systematic_uncertainties} shows the systematic uncertainties for different redshift bins and their relative values with respect to statistical errors. Similarly to \citet{chabanier_one-dimensional_2019}, we made conservative choices to define those uncertainties:

\begin{itemize}
    \item\textbf{{Noise estimation}}:\enspace As presented in Sec. \ref{subsec:noise}, the pipeline noise is corrected using the $\alpha$ corrective term, which depends on the data set considered. We assign a systematic uncertainty equal to $30~\%$ of the average $\alpha$ for each redshift bin.
    \item\textbf{{Resolution}}:\enspace We fit the average resolution correction given in Fig. \ref{fig:mean_resolution} by a simplified model $\mathrm{exp}\left(-0.5(k \Delta \lambda)^{2}\right)\cdot \mathrm{sinc} (0.5k \Delta \lambda_{\mathrm{pix}})$ with $\Delta \lambda_{\mathrm{pix}}$ fixed to the DESI spectral pixel separation. This procedure allows to determine the effective spectral resolution $\Delta \lambda$ in \AA. Measurements of DESI PSF stability, shown in \citet{abareshi_overview_2022} (Fig. 30), indicate that its fractional change is less than $1~\%$ over all spectrographs. We therefore assign a conservative systematic uncertainty $\sigma_{\Delta\lambda}= 1~\% \Delta \lambda$. Using the abovementioned simplified resolution model, this translates into a \pk~uncertainty equal to $2k^{2} \Delta \lambda \sigma_{\Delta\lambda} \cdot P_{1\mathrm{D},\alpha}(k)$.
    \item\textbf{{Resolution correction}}:\enspace We apply a correction to the resolution modeling as presented in Sec. \ref{subsec:resolution}, by multiplying $A_{\mathrm{res}}$ to \pk. We add an associated systematic error defined as 30\% of this correction.
    \item\textbf{{Side-band}}:\enspace The fitted side-band power spectrum $P_{\mathrm{SB1,m}}$ measured in Sec. \ref{subsec:side_band} is subtracted to \pk~to account essentially for metal absorptions in the \lya~forest region. We associate to this correction a systematic uncertainty equal to the statistical errors of the measured SB1 power spectrum. This is a conservative choice, as the modeling performed in Sec. \ref{subsec:side_band} closely reproduces \ps.
    \item\textbf{{Spectrum pixel masking}}:\enspace The impact of masking DLAs and atmospheric emission lines on the \pk~measurement was determined with synthetic data in Sec. \ref{subsec:masking_p1d}. Spectrum pixel masking is corrected by multiplying $A_{\mathrm{line}}(z,k) \cdot A_{\mathrm{dla}}(z,k)$ to the \pk~estimator. We define the systematic error associated with each masking as $30~\%$ of this correction.
    \item\textbf{{Continuum fitting}}:\enspace Similarly, we assign a systematic error of $30~\%$ times the $A_{\mathrm{cont}}(z,k)$ correction computed in Sec. \ref{subsec:continuum_correction}.
    \item\textbf{{DLA completeness}}:\enspace Using the synthetic data described in Sec. \ref{sec:synthetic_data}, we derived the impact of DLA on the one-dimensional power-spectrum as the ratio between mock with DLA (\textit{DLA}) and without (\textit{CONT}). We fit this ratio with an adapted function provided by \citet{rogers_simulating_2017}. As detailed previously in Sec. \ref{subsec:catalogs}, our DLA catalog of data results from the combination of two finders. The trend of this ratio is reported on the \CR{penultimate} panel of Fig. \ref{fig:systematic_uncertainties}. In \citet{chabanier_completed_2021}, the authors perform a full study on eBOSS data and provide the completeness of the CNN finder. The completeness of this finder is higher than 85\% for $\log(N_{\mathrm{\hi}}) > 20.3$. To be conservative, we choose to associate an uncertainty of 15\% of the total impact of DLAs on \pk to the incompleteness of our catalog. We stress that this uncertainty is over-estimated, as the CNN finder has a higher completeness for DLAs with higher column density and since we are using an additional GP algorithm. 
\end{itemize}

For all the corrections we applied on our \pk~measurement, the choice of 30\% in the associated systematic uncertainties is motivated by the fact that we consider a shift randomly ranging between no correction and 100\% of the correction. It is described by a uniform distribution between 0 and 1. The standard deviation of the distribution, equal to 0.30, quantifies the spread among the possible values, leading to a systematic uncertainty equal to 30\% of the correction.

Opposite to eBOSS \citep{chabanier_one-dimensional_2019}, we chose to not account for the incompleteness of the BAL catalog in our analysis, as it is one of the weaker contaminants. The study of BAL catalog completeness will be performed on further studies. 

The general trends in Fig. \ref{fig:systematic_uncertainties} are similar to those of the eBOSS measurement \citep{chabanier_one-dimensional_2019}. Given our limited statistics, most of the systematic errors are smaller than the statistical uncertainties for all redshift bins and all scales. However, this is expected to change for future DESI measurements, which will offer unprecedented statistics\CR{, thus reducing the statistical errors.}

\CR{There is room for improvement for the major source of systematic uncertainties presented above. The noise modeling can be improved by understanding and correcting the source of unaccounted noise. Regarding the resolution modeling, the mathematical model and its verification with the relatively new CCD mocks presented in Sec.~\ref{subsec:resolution} can be improved with additional tests and larger datasets. Decreasing statistical error on the side-band power spectrum will directly reduce the associated systematic error. Concerning the spectrum pixel masking, especially for the atmospheric lines, thinner masks can be applied considering the improvement of atmospheric emission line subtraction in DESI (see \citet{guy_spectroscopic_2022}). Furthermore, a more complex but analytical mathematical correction could be derived for this regular masking. For the DLA completeness, a more advanced study as the one performed in \citet{chabanier_one-dimensional_2019} is needed to reduce the associated systematic uncertainties. Additionally, as shown in \citet{rogers_simulating_2017}, the impact and correction of objects with lower column density (sub-DLAs, Lyman limit systems...) should be accounted for in a future, more developed study. Finally, the systematics are defined from a simplified assumption of uncertainty propagation, along with conservative choices. We plan to improve the modeling of each systematic and obtain more reliable uncertainties by modeling them directly into large samples of synthetic data.}

\section{DESI measurement}
\label{sec:results}

\begin{table}
	\centering
    \caption{Number of sub-forest, average redshift, and signal-to-noise ratio for each redshift bins in the final data set sample used in this measurement.}
    \label{tab:data_properties}
    \begin{tabular}{cc@{\hskip 0.05in}c@{\hskip 0.05in}c@{\hskip 0.05in}c@{\hskip 0.05in}c@{\hskip 0.05in}c@{\hskip 0.05in}c@{\hskip 0.05in}c@{\hskip 0.05in}c@{\hskip 0.05in}}
    \hline
    \textbf{z bin} & 2.2 & 2.4 & 2.6 & 2.8 & 3.0 & 3.2 & 3.4 & 3.6 & 3.8 \\
    \textbf{\#} & 25,846 & 16,371 & 11,501 & 8,205 & 5,343 & 3,267 & 1,800 & 1,006 & 500 \\
    $\mathbf{\langle z \rangle}$ & 2.2 & 2.4 & 2.6 & 2.8 & 2.99 & 3.19 & 3.39 & 3.59 & 3.79 \\
    $\mathbf{\overline{\mathrm{SNR}}}$ & 2.79 & 3.0 & 3.08 & 3.13 & 3.28 & 3.27 & 3.14 & 3.07 & 3.21  \\
    \hline
    \end{tabular}
\end{table}

\begin{figure*}
    \centering
    \begin{subfigure}[t]{0.49\textwidth}
    \centering
	\includegraphics[width=\columnwidth]{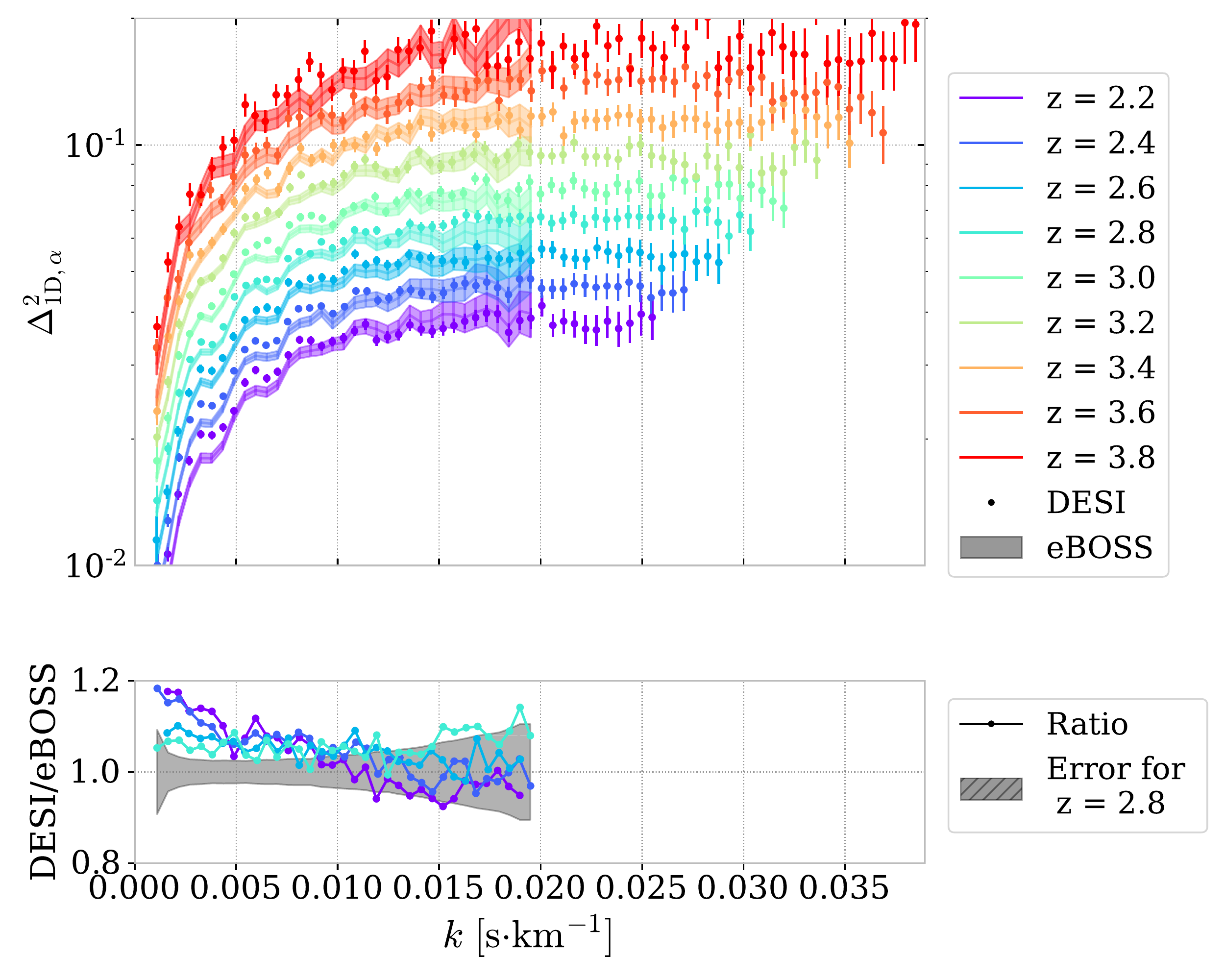}
    \end{subfigure}%
    ~ 
    \begin{subfigure}[t]{0.49\textwidth}
    \centering
	\includegraphics[width=\columnwidth]{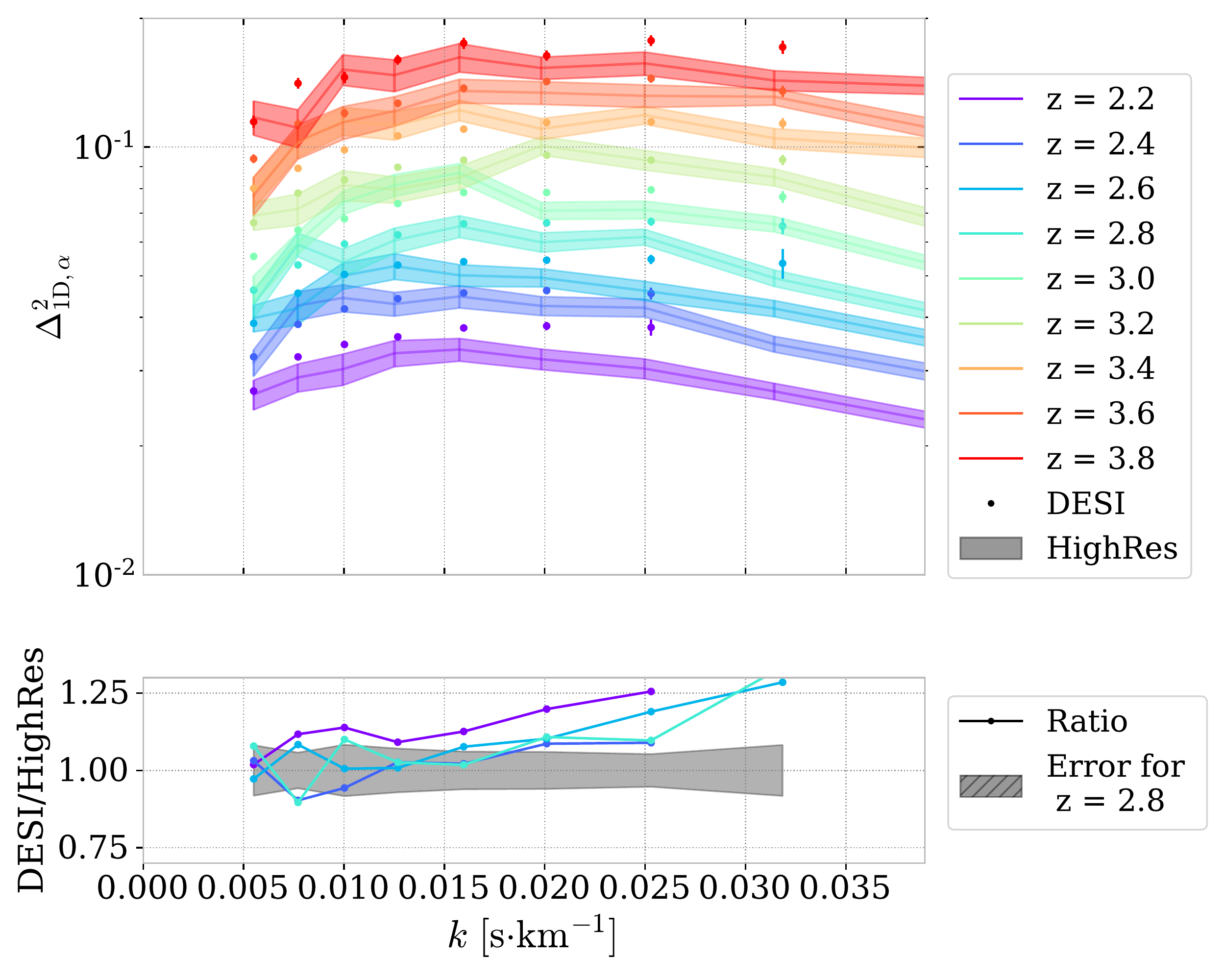}
    \end{subfigure}
    \caption{(left) Comparison between the measurement performed here and that of the eBOSS data \citep{chabanier_one-dimensional_2019}. For this comparison, all the analysis was performed in velocity units using the conversion given by equation (\ref{eq:conversion}). Both normalized \pk~are shown in the top panel (our measurement with points and the eBOSS measurement with shaded colored areas). The ratio between DESI and eBOSS measurements is shown in the bottom panel. The striped gray area in the bottom panel shows the centered error bar of the ratio averaged over all shown redshift bins. We remove high redshift bins for clarity. A description of the tests used to explain the difference between DESI and eBOSS can be found in appendix \ref{appendix:comparison_eboss}. (right) Same comparison with the high-resolution measurement obtained using the combination of KODIAQ, SQUAD, and XQ-100 surveys \citep{karacayli_optimal_2022}. In this case, our \pk~measurement and the error bars associated are rebinned to the wavenumber binning of the high-resolution measurement. For clarity, only the four first redshift bins are shown in the bottom panel of each figure, along with the error on the ratio for $z=2.8$ centered to unity.}
    \label{fig:comparison}
\end{figure*}

We apply the methodology and corrections described in previous sections to the combination of \svt, and \da~data sets, noted \svtda. We choose to remove the \svo~data set from this measurement due to considerations on the noise power spectrum shown in Sec. \ref{subsec:noise} and measurements performed in the appendix \ref{appendix:data_set_comparison}.

The \pk~measurement is done using the pipeline and parameters presented in Sec. \ref{sec:analysis}. Considering all the corrections defined in the previous section, the final \pk~estimator is defined by

\begin{equation}
\label{eq:final_p1d_estimator}
\begin{aligned}
&P_{1\mathrm{D},\alpha}(k) = A_{\mathrm{line}}(z,k) \cdot A_{\mathrm{dla}}(z,k) \cdot A_{\mathrm{cont}}(z,k) \cdot A_{\mathrm{res}}(k) \cdot \\
&~\left(\left\langle\left[P_{\mathrm{raw}}(k)-P_{\mathrm {pipeline}}(k) - \alpha \right] \cdot \mathbf{R}^{-2}(k) \right\rangle - P_{\mathrm{SB1,m}}(k) \right)\,.
\end{aligned}
\end{equation}

Fig. \ref{fig:measurement} presents the normalized \pk~measurement such that $\Delta^{2}_{1\mathrm{D},\alpha} = k P_{1\mathrm{D},\alpha} / \pi$. This observable is shown for 9 redshift bins ranging from $2.2$ to $3.8$, for wavenumbers $0.145 \leq k \leq 2$ \invAA, and using a total of $73,839$ sub-forests extracted from $26,330$ quasar spectra. The represented error bars are the statistical and systematic uncertainties added in quadrature. The details of sub-forest properties for each redshift bin are given in Tab. \ref{tab:data_properties}.

\subsection{Comparison with other measurements}

We perform a comparison with past measurements. The comparison of the DESI \pk~with the last moderate-resolution measurement (eBOSS survey \citet{chabanier_one-dimensional_2019}) is shown in Fig. \ref{fig:comparison} (left). We also compare our measurement with the last high-resolution measurement obtained using the combination of KODIAQ, SQUAD, and XQ-100 surveys \citep{karacayli_optimal_2022} in Fig. \ref{fig:comparison} (right).

Both moderate- and high-resolution measurements were expressed in velocity units ($\kms$). The conversion between this unit system and the one used for DESI (\AA$^{-1}$) is defined by:

\begin{equation}
    \label{eq:conversion}
    k \left[\kms\right] = k\left[\text{\AA}^{-1}\right] \times \lambda_{\alpha} (1+z) / c \,.
\end{equation}

This conversion is performed at the stage of the FFT estimation by converting the terms in equation (\ref{eq:p1d_estimator}) in velocity units before performing the ensemble average to compute \pk. We found that doing this conversion on the averaged \pk~significantly shift \pk~because of the redshift term in the equation (\ref{eq:conversion}). This is caused by the non-uniform redshift distribution in each z-bins, pointed out by Fig. \ref{fig:quasar_redshift_hist}. Similarly, we converted all corrections terms appearing in equation (\ref{eq:final_p1d_estimator}) in velocity units. For the comparison with eBOSS measurement, we compute \pk~in velocity units with the same binning. For the high-resolution measurement, we rebin our DESI measurement to the same wavenumber binning, and account this rebinning in the calculation of error bars, to obtain a fair comparison.

The comparison with the eBOSS measurement in Fig. \ref{fig:comparison} (left) yields a $\sim 15\%$ difference at small wavenumber ($k < 0.01~\invkms$). To investigate this discrepancy, we performed a detailed investigation by varying most of the parameters of our analysis, which are susceptible to impact small scales. The discussion concerning those tests is detailed in appendix \ref{appendix:comparison_eboss}. On the other scales ($0.01~\invkms< k < 0.02~\invkms$), our measurement agrees with eBOSS considering the error bars. 

The major improvement lays at large wavenumber ($k > 0.015~\invkms$) where the improved DESI resolution and noise modeling allows us reaching much smaller scales than eBOSS, especially for high redshifts. We are able to conservatively reach the wavenumber $k_{\mathrm{max}} = 2.0$ \invAA~for all redshift bins. In comparison, the eBOSS measurement \citep{chabanier_one-dimensional_2019} achieved maximal wavenumber of $k_{\mathrm{max}} = 1.54$ \invAA~for $z= 2.2$ and $k_{\mathrm{max}} = 1.03$ \invAA~for $z= 3.8$. At large wavenumbers, the eBOSS measurement is highly contaminated by noise and resolution. We consider that our measurement is more suitable to probe those scales, considering the improvement in resolution and noise estimation.

We compare our measurement with the high-resolution measurements in Fig. \ref{fig:comparison} (right). This measurement \citep{karacayli_optimal_2022} is performed with a statistically smaller sample of quasars but with a very high spectral resolution ($5,000\lesssim R \lesssim 60,000$) and \snr. The high-resolution \pk~measurements thus allow reaching very small scales with large error bars. Our agreement shows a 20\% disagreement at small scales ($k > 0.02~\invkms$), mostly for the lower redshifts measured. It indicates that there is still room for improvement in our measurement's noise and resolution modeling. We note that considering error bars, both measurements agree on intermediate scales.

\section{Conclusion and prospects}
\label{sec:conclusion}

We performed the first measurement of the one-dimensional power spectrum (\pk) with DESI data. The main objective of this paper is to carefully characterize the different contaminants of \pk~with regard to DESI instrument. In particular, we modeled the noise and spectral resolution of DESI. In comparison to the previous eBOSS measurement \citep{chabanier_one-dimensional_2019}, we improved the analysis of side-band power spectrum and atmospheric emission lines. 

We used adapted synthetic data to correct the impact of spectrum pixel masking, continuum fitting and spectral resolution modeling. We performed a complete review of the systematic uncertainties linked to the \pk~pipeline and compared the DESI measurement with previous moderate- and high-resolution measurements. We find a relatively good agreement, except for a slight difference at large scales with the eBOSS measurement, partially due to the different residual correction we apply. Our estimation of \pk~is also compared with the QMLE method in a companion paper~\cite{karacayli_optimal_2023}. Measurements with FFT and QMLE methods agree at 1\% level precision up to half the Nyquist frequency.

The DESI spectral resolution is approximately two times better than SDSS. Consequently, our \pk~measurement is of high scientific interest to probe the small scales of the intergalactic medium. However, the data sets we exploited remains inferior to eBOSS in terms of statistics. If we apply the same \snr~cut as in \citet{chabanier_one-dimensional_2019}, our sample contains $17,333$ sub-forests compared to $94,558$ for eBOSS. However, we expect future DESI data sets to provide high increase of statistics (up to $1$ million \lya~forest, thus almost $3$ milion sub-forests). This unprecedented dataset will allow obtaining a sub-percent precision measurement. The resulting \pk~measurement will provide stringent constraints on the sum of neutrinos masses, warm dark matter models, and on the parameters of the intergalactic medium~\citep{desi_collaboration_desi_2016,valluri_snowmass_2022}.

We plan to improve our analysis to keep the level of systematic error close to the statistical one. First, applying stricter constraints on a larger \lya~forest sample will be beneficial to reduce systematical uncertainties. Furthermore, we also plan to improve the treatment of contaminants presented in Sec. \ref{sec:uncertainties}. In particular, we plan to test extensively the resolution and noise estimations on pixel-level simulations of the CCD camera. 

High-resolution hydrodynamical simulations, associated with Gaussian processes emulator such as in \citet{pedersen_emulator_2020,walther_simulating_2021} will be employed with next DESI \pk~measurement to obtain constraints on cosmological and intergalactic medium parameters.

\section*{Acknowledgments}

The authors acknowledge support from grant ANR-16-CE31-0021. The project leading to this publication has received funding from Excellence Initiative of Aix-Marseille University - A*MIDEX, a French “Investissements d’Avenir” programme (AMX-20-CE-02 - DARKUNI).

This material is based upon work supported by the U.S. Department of Energy (DOE), Office of Science, Office of High-Energy Physics, under Contract No. DE–AC02–05CH11231, and by the National Energy Research Scientific Computing Center, a DOE Office of Science User Facility under the same contract. Additional support for DESI was provided by the U.S. National Science Foundation (NSF), Division of Astronomical Sciences under Contract No. AST-0950945 to the NSF’s National Optical-Infrared Astronomy Research Laboratory; the Science and \CR{Technology} Facilities Council of the United Kingdom; the Gordon and Betty Moore Foundation; the Heising-Simons Foundation; the French Alternative Energies and Atomic Energy Commission (CEA); the National Council of Science and Technology of Mexico (CONACYT); the Ministry of Science and Innovation of Spain (MICINN), and by the DESI Member Institutions: \url{https://www.desi.lbl.gov/collaborating-institutions}. Any opinions, findings, and conclusions or recommendations expressed in this material are those of the author(s) and do not necessarily reflect the views of the U. S. National Science Foundation, the U. S. Department of Energy, or any of the listed funding agencies.

The authors are honored to be permitted to conduct scientific research on Iolkam Du’ag (Kitt Peak), a mountain with particular significance to the Tohono O’odham Nation.

\section*{Data Availability}

The DESI spectra from \svo~and \svt~will be publicly available in the Early Data Release (EDR) in 2023. We added the first two months of the main survey to improve our statistics. The main survey spectra will be made publicly available as part of Year 1 data release in the future. All the data points of the figures in this article are made available according to the data management policy of DESI\footnote{\label{zenodo_data_release}\url{https://doi.org/10.5281/zenodo.7993570}}. \CR{All the outputs and plots of this article are generated with \texttt{picca}$^{\ref{igmhub/picca}}$ (8.1.2) and \texttt{p1desi}\git{corentinravoux/p1desi}{} (1.0.0).}

\bibliographystyle{mnras}
\bibliography{bibli}

\begin{thebibliography}{}
\makeatletter
\relax
\def\mn@urlcharsother{\let\do\@makeother \do\$\do\&\do\#\do\^\do\_\do\%\do\~}
\def\mn@doi{\begingroup\mn@urlcharsother \@ifnextchar [ {\mn@doi@}
  {\mn@doi@[]}}
\def\mn@doi@[#1]#2{\def\@tempa{#1}\ifx\@tempa\@empty \href
  {http://dx.doi.org/#2} {doi:#2}\else \href {http://dx.doi.org/#2} {#1}\fi
  \endgroup}
\def\mn@eprint#1#2{\mn@eprint@#1:#2::\@nil}
\def\mn@eprint@arXiv#1{\href {http://arxiv.org/abs/#1} {{\tt arXiv:#1}}}
\def\mn@eprint@dblp#1{\href {http://dblp.uni-trier.de/rec/bibtex/#1.xml}
  {dblp:#1}}
\def\mn@eprint@#1:#2:#3:#4\@nil{\def\@tempa {#1}\def\@tempb {#2}\def\@tempc
  {#3}\ifx \@tempc \@empty \let \@tempc \@tempb \let \@tempb \@tempa \fi \ifx
  \@tempb \@empty \def\@tempb {arXiv}\fi \@ifundefined
  {mn@eprint@\@tempb}{\@tempb:\@tempc}{\expandafter \expandafter \csname
  mn@eprint@\@tempb\endcsname \expandafter{\@tempc}}}

\bibitem[\protect\citeauthoryear{Abareshi et~al.,}{Abareshi
  et~al.}{2022}]{abareshi_overview_2022}
Abareshi B.,  et~al., 2022, Overview of the {Instrumentation} for the {Dark}
  {Energy} {Spectroscopic} {Instrument}, \url {http://arxiv.org/abs/2205.10939}

\bibitem[\protect\citeauthoryear{Ahumada et~al.,}{Ahumada
  et~al.}{2020}]{ahumada_sixteenth_2020}
Ahumada R.,  et~al., 2020, \mn@doi [arXiv:1912.02905 [astro-ph]]
  {10.3847/1538-4365/ab929e}

\bibitem[\protect\citeauthoryear{Alexander et~al.,}{Alexander
  et~al.}{2022}]{alexander_desi_2022}
Alexander D.~M.,  et~al., 2022, The {DESI} {Survey} {Validation}: {Results}
  from {Visual} {Inspection} of the {Quasar} {Survey} {Spectra}, \url
  {http://arxiv.org/abs/2208.08517}

\bibitem[\protect\citeauthoryear{Armengaud, Palanque-Delabrouille, Y{\`e}che,
  Marsh  \& Baur}{Armengaud et~al.}{2017}]{armengaud_constraining_2017}
Armengaud E.,  Palanque-Delabrouille N.,  Y{\`e}che C.,  Marsh D. J.~E.,   Baur
  J.,  2017, \mn@doi [Monthly Notices of the Royal Astronomical Society]
  {10.1093/mnras/stx1870}, 471, 4606

\bibitem[\protect\citeauthoryear{Bailey et~al.}{Bailey
  et~al.}{2023}]{bailey_2023}
Bailey S.,  et~al., 2023, in preparation

\bibitem[\protect\citeauthoryear{Bajtlik, Duncan  \& Ostriker}{Bajtlik
  et~al.}{1988}]{bajtlik_quasar_1988}
Bajtlik S.,  Duncan R.~C.,   Ostriker J.~P.,  1988, \mn@doi [The Astrophysical
  Journal] {10.1086/166217}, 327, 570

\bibitem[\protect\citeauthoryear{Baur, Palanque-Delabrouille, Y{\`e}che,
  Magneville  \& Viel}{Baur et~al.}{2016}]{baur_lyman-alpha_2016}
Baur J.,  Palanque-Delabrouille N.,  Y{\`e}che C.,  Magneville C.,   Viel M.,
  2016, \mn@doi [Journal of Cosmology and Astroparticle Physics]
  {10.1088/1475-7516/2016/08/012}, 2016, 012

\bibitem[\protect\citeauthoryear{Baur, Palanque-Delabrouille, Y{\`e}che,
  Boyarsky, Ruchayskiy, Armengaud  \& Lesgourgues}{Baur
  et~al.}{2017}]{baur_constraints_2017}
Baur J.,  Palanque-Delabrouille N.,  Y{\`e}che C.,  Boyarsky A.,  Ruchayskiy
  O.,  Armengaud {\'E}.,   Lesgourgues J.,  2017, \mn@doi [Journal of Cosmology
  and Astroparticle Physics] {10.1088/1475-7516/2017/12/013}, 2017, 013

\bibitem[\protect\citeauthoryear{Bautista et~al.,}{Bautista
  et~al.}{2017}]{bautista_measurement_2017}
Bautista J.~E.,  et~al., 2017, \mn@doi [Astronomy \& Astrophysics]
  {10.1051/0004-6361/201730533}, 603, A12

\bibitem[\protect\citeauthoryear{Blanton et~al.,}{Blanton
  et~al.}{2017}]{blanton_sloan_2017}
Blanton M.~R.,  et~al., 2017, \mn@doi [arXiv:1703.00052 [astro-ph]]
  {10.3847/1538-3881/aa7567}

\bibitem[\protect\citeauthoryear{Boera, Becker, Bolton  \& Nasir}{Boera
  et~al.}{2019}]{boera_revealing_2019}
Boera E.,  Becker G.~D.,  Bolton J.~S.,   Nasir F.,  2019, \mn@doi [The
  Astrophysical Journal] {10.3847/1538-4357/aafee4}, 872, 101

\bibitem[\protect\citeauthoryear{Bolton \& Schlegel}{Bolton \&
  Schlegel}{2010}]{bolton_spectro-perfectionism_2010}
Bolton A.~S.,  Schlegel D.~J.,  2010, \mn@doi [Publications of the Astronomical
  Society of the Pacific] {10.1086/651008}, pp 100119133735095--000

\bibitem[\protect\citeauthoryear{Bolton, Puchwein, Sijacki, Haehnelt, Kim,
  Meiksin, Regan  \& Viel}{Bolton et~al.}{2017}]{bolton_sherwood_2017}
Bolton J.~S.,  Puchwein E.,  Sijacki D.,  Haehnelt M.~G.,  Kim T.-S.,  Meiksin
  A.,  Regan J.~A.,   Viel M.,  2017, \mn@doi [Monthly Notices of the Royal
  Astronomical Society] {10.1093/mnras/stw2397}, 464, 897

\bibitem[\protect\citeauthoryear{Borde, Palanque-Delabrouille, Rossi, Viel,
  Bolton, Y{\`e}che, LeGoff  \& Rich}{Borde et~al.}{2014}]{borde_new_2014}
Borde A.,  Palanque-Delabrouille N.,  Rossi G.,  Viel M.,  Bolton J.,
  Y{\`e}che C.,  LeGoff J.-M.,   Rich J.,  2014, \mn@doi [Journal of Cosmology
  and Astroparticle Physics] {10.1088/1475-7516/2014/07/005}, 2014, 005

\bibitem[\protect\citeauthoryear{Brodzeller et~al.,}{Brodzeller
  et~al.}{2023}]{brodzeller_performance_2023}
Brodzeller A.,  et~al., 2023, Performance of the {Quasar} {Spectral}
  {Templates} for the {Dark} {Energy} {Spectroscopic} {Instrument}, \url
  {http://arxiv.org/abs/2305.10426}

\bibitem[\protect\citeauthoryear{Busca \& Balland}{Busca \&
  Balland}{2018}]{busca_quasarnet_2018}
Busca N.,  Balland C.,  2018, {QuasarNET}: {Human}-level spectral
  classification and redshifting with {Deep} {Neural} {Networks}, \url
  {http://arxiv.org/abs/1808.09955}

\bibitem[\protect\citeauthoryear{Chabanier et~al.,}{Chabanier
  et~al.}{2019}]{chabanier_one-dimensional_2019}
Chabanier S.,  et~al., 2019, \mn@doi [Journal of Cosmology and Astroparticle
  Physics] {10.1088/1475-7516/2019/07/017}, 2019, 017

\bibitem[\protect\citeauthoryear{Chabanier et~al.,}{Chabanier
  et~al.}{2021}]{chabanier_completed_2021}
Chabanier S.,  et~al., 2021, \mn@doi [arXiv:2107.09612 [astro-ph,
  physics:physics]] {10.3847/1538-4365/ac366e}

\bibitem[\protect\citeauthoryear{Chabanier et~al.,}{Chabanier
  et~al.}{2022}]{chabanier_modeling_2022}
Chabanier S.,  et~al., 2022, Modeling the {Lyman}-\ensuremath{\alpha} forest
  with {Eulerian} and {SPH} hydrodynamical methods, \url
  {http://arxiv.org/abs/2207.05023}

\bibitem[\protect\citeauthoryear{Chaussidon et~al.,}{Chaussidon
  et~al.}{2022}]{chaussidon_target_2022}
Chaussidon E.,  et~al., 2022, Target {Selection} and {Validation} of {DESI}
  {Quasars}, \url {http://arxiv.org/abs/2208.08511}

\bibitem[\protect\citeauthoryear{Croft, Weinberg, Katz  \& Hernquist}{Croft
  et~al.}{1998}]{croft_recovery_1998}
Croft R. A.~C.,  Weinberg D.~H.,  Katz N.,   Hernquist L.,  1998, \mn@doi [The
  Astrophysical Journal] {10.1086/305289}, 495, 44

\bibitem[\protect\citeauthoryear{Croft, Weinberg, Bolte, Burles, Hernquist,
  Katz, Kirkman  \& Tytler}{Croft et~al.}{2002}]{croft_towards_2002}
Croft R. A.~C.,  Weinberg D.~H.,  Bolte M.,  Burles S.,  Hernquist L.,  Katz
  N.,  Kirkman D.,   Tytler D.,  2002, \mn@doi [The Astrophysical Journal]
  {10.1086/344099}, 581, 20

\bibitem[\protect\citeauthoryear{DESI}{DESI}{2016a}]{desi_collaboration_desi_2016}
DESI 2016a, The {DESI} {Experiment} {Part} {I}: {Science},{Targeting}, and
  {Survey} {Design}, \url {http://arxiv.org/abs/1611.00036}

\bibitem[\protect\citeauthoryear{DESI}{DESI}{2016b}]{desi_collaboration_desi_2016-1}
DESI 2016b, The {DESI} {Experiment} {Part} {II}: {Instrument} {Design}, \url
  {http://arxiv.org/abs/1611.00037}

\bibitem[\protect\citeauthoryear{DESI}{DESI}{2023b}]{desi_collaboration_early_2023}
DESI 2023b, The {Early} {Data} {Release} of the {Dark} {Energy} {Spectroscopic}
  {Instrument}, \url {http://arxiv.org/abs/2306.06308}

\bibitem[\protect\citeauthoryear{DESI}{DESI}{2023a}]{desi_collaboration_validation_2023}
DESI 2023a, Validation of the {Scientific} {Program} for the {Dark} {Energy}
  {Spectroscopic} {Instrument}, \url {http://arxiv.org/abs/2306.06307}

\bibitem[\protect\citeauthoryear{Dawson et~al.,}{Dawson
  et~al.}{2016}]{dawson_sdss-iv_2016}
Dawson K.~S.,  et~al., 2016, \mn@doi [The Astronomical Journal]
  {10.3847/0004-6256/151/2/44}, 151, 44

\bibitem[\protect\citeauthoryear{Day, Tytler  \& Kambalur}{Day
  et~al.}{2019}]{day_power_2019}
Day A.,  Tytler D.,   Kambalur B.,  2019, \mn@doi [Monthly Notices of the Royal
  Astronomical Society] {10.1093/mnras/stz2214}, 489, 2536

\bibitem[\protect\citeauthoryear{Dey et~al.,}{Dey
  et~al.}{2019}]{dey_overview_2019}
Dey A.,  et~al., 2019, \mn@doi [The Astronomical Journal]
  {10.3847/1538-3881/ab089d}, 157, 168

\bibitem[\protect\citeauthoryear{Farr, Font-Ribera  \& Pontzen}{Farr
  et~al.}{2020}]{farr_optimal_2020}
Farr J.,  Font-Ribera A.,   Pontzen A.,  2020, \mn@doi [arXiv:2007.10348
  [astro-ph]] {10.1088/1475-7516/2020/11/015}

\bibitem[\protect\citeauthoryear{Gaikwad, Srianand, Haehnelt  \&
  Choudhury}{Gaikwad et~al.}{2021}]{gaikwad_consistent_2021}
Gaikwad P.,  Srianand R.,  Haehnelt M.~G.,   Choudhury T.~R.,  2021, \mn@doi
  [Monthly Notices of the Royal Astronomical Society] {10.1093/mnras/stab2017},
  506, 4389

\bibitem[\protect\citeauthoryear{Gunn \& Peterson}{Gunn \&
  Peterson}{1965}]{gunn_density_1965}
Gunn J.~E.,  Peterson B.~A.,  1965, American Astronomical Society

\bibitem[\protect\citeauthoryear{Gunn et~al.,}{Gunn
  et~al.}{2006}]{gunn_25_2006}
Gunn J.~E.,  et~al., 2006, \mn@doi [The Astronomical Journal] {10.1086/500975},
  131, 2332

\bibitem[\protect\citeauthoryear{Guy et~al.,}{Guy
  et~al.}{2022}]{guy_spectroscopic_2022}
Guy J.,  et~al., 2022, The {Spectroscopic} {Data} {Processing} {Pipeline} for
  the {Dark} {Energy} {Spectroscopic} {Instrument}, \url
  {http://arxiv.org/abs/2209.14482}

\bibitem[\protect\citeauthoryear{Herrera-Alcantar et~al.}{Herrera-Alcantar
  et~al.}{2023}]{quickquasars}
Herrera-Alcantar H.,  et~al., 2023, in preparation

\bibitem[\protect\citeauthoryear{Ho, Bird  \& Garnett}{Ho
  et~al.}{2021}]{ho_damped_2021}
Ho M.-F.,  Bird S.,   Garnett R.,  2021, \mn@doi [Monthly Notices of the Royal
  Astronomical Society] {10.1093/mnras/stab2169}, 507, 704

\bibitem[\protect\citeauthoryear{Ir{\v s}i{\v c} et~al.,}{Ir{\v s}i{\v c}
  et~al.}{2016}]{irsic_lyman-alpha_2016}
Ir{\v s}i{\v c} V.,  et~al., 2016, \mn@doi [Monthly Notices of the Royal
  Astronomical Society] {10.1093/mnras/stw3372}, p. stw3372

\bibitem[\protect\citeauthoryear{Ir{\v s}i{\v c}, Viel, Haehnelt, Bolton  \&
  Becker}{Ir{\v s}i{\v c} et~al.}{2017}]{irsic_first_2017}
Ir{\v s}i{\v c} V.,  Viel M.,  Haehnelt M.~G.,  Bolton J.~S.,   Becker G.~D.,
  2017, \mn@doi [Physical Review Letters] {10.1103/PhysRevLett.119.031302},
  119, 031302

\bibitem[\protect\citeauthoryear{Kara{\c c}ayl{\i}, Font-Ribera  \&
  Padmanabhan}{Kara{\c c}ayl{\i} et~al.}{2020}]{karacayli_optimal_2020}
Kara{\c c}ayl{\i} N.~G.,  Font-Ribera A.,   Padmanabhan N.,  2020, \mn@doi
  [Monthly Notices of the Royal Astronomical Society] {10.1093/mnras/staa2331},
  497, 4742

\bibitem[\protect\citeauthoryear{Kara{\c c}ayl{\i} et~al.,}{Kara{\c c}ayl{\i}
  et~al.}{2022}]{karacayli_optimal_2022}
Kara{\c c}ayl{\i} N.~G.,  et~al., 2022, \mn@doi [Monthly Notices of the Royal
  Astronomical Society] {10.1093/mnras/stab3201}, 509, 2842

\bibitem[\protect\citeauthoryear{{Kara{\c{c}}ayl{\i}}
  et~al.,}{{Kara{\c{c}}ayl{\i}} et~al.}{2023a}]{karacayli_optimal_2023}
{Kara{\c{c}}ayl{\i}} N.~G.,  et~al., 2023a, Optimal {1D}
  {Ly}\${\textbackslash}alpha\$ {Forest} {Power} {Spectrum} {Estimation} --
  {III}. {DESI} early data, \url {http://arxiv.org/abs/2306.06316}

\bibitem[\protect\citeauthoryear{{Kara{\c{c}}ayl{\i}}
  et~al.,}{{Kara{\c{c}}ayl{\i}} et~al.}{2023b}]{karacayli_framework_2023}
{Kara{\c{c}}ayl{\i}} N.~G.,  et~al., 2023b, \mn@doi [\mnras]
  {10.1093/mnras/stad1363}, \href
  {https://ui.adsabs.harvard.edu/abs/2023MNRAS.522.5980K} {522, 5980}

\bibitem[\protect\citeauthoryear{Khaire et~al.,}{Khaire
  et~al.}{2019}]{khaire_power_2019}
Khaire V.,  et~al., 2019, \mn@doi [Monthly Notices of the Royal Astronomical
  Society] {10.1093/mnras/stz344}, 486, 769

\bibitem[\protect\citeauthoryear{Kim, Viel, Haehnelt, Carswell  \&
  Cristiani}{Kim et~al.}{2004}]{kim_power_2004}
Kim T.-S.,  Viel M.,  Haehnelt M.~G.,  Carswell R.~F.,   Cristiani S.,  2004,
  \mn@doi [Monthly Notices of the Royal Astronomical Society]
  {10.1111/j.1365-2966.2004.07221.x}, 347, 355

\bibitem[\protect\citeauthoryear{Kramida, {Yu.~Ralchenko}, Reader  \& {and NIST
  ASD Team}}{Kramida et~al.}{2021}]{kramida_nist_2021}
Kramida A.,  {Yu.~Ralchenko} Reader J.,   {and NIST ASD Team} 2021, {NIST
  Atomic Spectra Database (ver. 5.9), [Online]. Available:
  {\tt{https://physics.nist.gov/asd}} [2022, May 17]. National Institute of
  Standards and Technology, Gaithersburg, MD.}

\bibitem[\protect\citeauthoryear{Lee et~al.,}{Lee et~al.}{2013}]{lee_boss_2013}
Lee K.-G.,  et~al., 2013, \mn@doi [The Astronomical Journal]
  {10.1088/0004-6256/145/3/69}, 145, 69

\bibitem[\protect\citeauthoryear{Lesgourgues \& Pastor}{Lesgourgues \&
  Pastor}{2006}]{lesgourgues_massive_2006}
Lesgourgues J.,  Pastor S.,  2006, \mn@doi [Physics Reports]
  {10.1016/j.physrep.2006.04.001}, 429, 307

\bibitem[\protect\citeauthoryear{Lesgourgues \& Pastor}{Lesgourgues \&
  Pastor}{2012}]{lesgourgues_neutrino_2012}
Lesgourgues J.,  Pastor S.,  2012, \mn@doi [Advances in High Energy Physics]
  {10.1155/2012/608515}, 2012, 1

\bibitem[\protect\citeauthoryear{{Levi} et~al.,}{{Levi}
  et~al.}{2013}]{levi_desi_2013}
{Levi} M.,  et~al., 2013, arXiv e-prints, \href
  {https://ui.adsabs.harvard.edu/abs/2013arXiv1308.0847L} {p. arXiv:1308.0847}

\bibitem[\protect\citeauthoryear{Lopez et~al.,}{Lopez
  et~al.}{2016}]{lopez_xq-100_2016}
Lopez S.,  et~al., 2016, \mn@doi [Astronomy \& Astrophysics]
  {10.1051/0004-6361/201628161}, 594, A91

\bibitem[\protect\citeauthoryear{Luki{\'c}, Stark, Nugent, White, Meiksin  \&
  Almgren}{Luki{\'c} et~al.}{2015}]{lukic_lyman-alpha_2015}
Luki{\'c} Z.,  Stark C.,  Nugent P.,  White M.,  Meiksin A.,   Almgren A.,
  2015, \mn@doi [arXiv:1406.6361 [astro-ph]] {10.1093/mnras/stu2377}

\bibitem[\protect\citeauthoryear{Lynds}{Lynds}{1971}]{lynds_absorption-line_1971}
Lynds R.,  1971, \mn@doi [The Astrophysical Journal] {10.1086/180695}, 164, L73

\bibitem[\protect\citeauthoryear{McDonald, Miralda-Escude, Rauch, Sargent,
  Barlow, Cen  \& Ostriker}{McDonald et~al.}{2000}]{mcdonald_observed_2000}
McDonald P.,  Miralda-Escude J.,  Rauch M.,  Sargent W. L.~W.,  Barlow T.~A.,
  Cen R.,   Ostriker J.~P.,  2000, \mn@doi [The Astrophysical Journal]
  {10.1086/317079}, 543, 1

\bibitem[\protect\citeauthoryear{McDonald, Seljak, Cen, Bode  \&
  Ostriker}{McDonald et~al.}{2005}]{mcdonald_physical_2005}
McDonald P.,  Seljak U.,  Cen R.,  Bode P.,   Ostriker J.~P.,  2005, \mn@doi
  [Monthly Notices of the Royal Astronomical Society]
  {10.1111/j.1365-2966.2005.09141.x}, 360, 1471

\bibitem[\protect\citeauthoryear{McDonald et~al.,}{McDonald
  et~al.}{2006}]{mcdonald_lyman-alpha_2006}
McDonald P.,  et~al., 2006, \mn@doi [The Astrophysical Journal Supplement
  Series] {10.1086/444361}, 163, 80

\bibitem[\protect\citeauthoryear{McQuinn}{McQuinn}{2016}]{mcquinn_evolution_2016}
McQuinn M.,  2016, \mn@doi [Annual Review of Astronomy and Astrophysics]
  {10.1146/annurev-astro-082214-122355}, 54, 313

\bibitem[\protect\citeauthoryear{Meiksin}{Meiksin}{2009}]{meiksin_physics_2009}
Meiksin A.~A.,  2009, \mn@doi [Reviews of Modern Physics]
  {10.1103/RevModPhys.81.1405}, 81, 1405

\bibitem[\protect\citeauthoryear{Miller et~al.,}{Miller
  et~al.}{2023}]{miller_optical_2023}
Miller T.~N.,  et~al., 2023, The {Optical} {Corrector} for the {Dark} {Energy}
  {Spectroscopic} {Instrument}, \url {http://arxiv.org/abs/2306.06310}

\bibitem[\protect\citeauthoryear{Murphy, Kacprzak, Savorgnan  \&
  Carswell}{Murphy et~al.}{2018}]{murphy_uves_2018}
Murphy M.~T.,  Kacprzak G.~G.,  Savorgnan G. A.~D.,   Carswell R.~F.,  2018,
  \mn@doi [arXiv:1810.06136 [astro-ph]] {10.1093/mnras/sty2834}

\bibitem[\protect\citeauthoryear{Myers et~al.,}{Myers
  et~al.}{2022}]{myers_target_2022}
Myers A.~D.,  et~al., 2022, The {Target} {Selection} {Pipeline} for the {Dark}
  {Energy} {Spectroscopic} {Instrument}, \url {http://arxiv.org/abs/2208.08518}

\bibitem[\protect\citeauthoryear{O{\textquoteright}Meara}{O{\textquoteright}Meara}{2017}]{omeara_second_2017}
O{\textquoteright}Meara J.~M.,  2017, The Astronomical Journal, p.~5

\bibitem[\protect\citeauthoryear{O{\textquoteright}Meara
  et~al.,}{O{\textquoteright}Meara et~al.}{2015}]{omeara_first_2015}
O{\textquoteright}Meara J.~M.,  et~al., 2015, \mn@doi [The Astronomical
  Journal] {10.1088/0004-6256/150/4/111}, 150, 111

\bibitem[\protect\citeauthoryear{Palanque-Delabrouille
  et~al.,}{Palanque-Delabrouille
  et~al.}{2013}]{palanque-delabrouille_one-dimensional_2013}
Palanque-Delabrouille N.,  et~al., 2013, \mn@doi [Astronomy \& Astrophysics]
  {10.1051/0004-6361/201322130}, 559, A85

\bibitem[\protect\citeauthoryear{Palanque-Delabrouille
  et~al.,}{Palanque-Delabrouille
  et~al.}{2015}]{palanque-delabrouille_neutrino_2015}
Palanque-Delabrouille N.,  et~al., 2015, \mn@doi [Journal of Cosmology and
  Astroparticle Physics] {10.1088/1475-7516/2015/11/011}, 2015, 011

\bibitem[\protect\citeauthoryear{Palanque-Delabrouille, Y{\`e}che,
  Sch{\"o}neberg, Lesgourgues, Walther, Chabanier  \&
  Armengaud}{Palanque-Delabrouille
  et~al.}{2020}]{palanque-delabrouille_hints_2020}
Palanque-Delabrouille N.,  Y{\`e}che C.,  Sch{\"o}neberg N.,  Lesgourgues J.,
  Walther M.,  Chabanier S.,   Armengaud E.,  2020, \mn@doi [Journal of
  Cosmology and Astroparticle Physics] {10.1088/1475-7516/2020/04/038}, 2020,
  038

\bibitem[\protect\citeauthoryear{Parks, Prochaska, Dong  \& Cai}{Parks
  et~al.}{2017}]{parks_deep_2017}
Parks D.,  Prochaska J.~X.,  Dong S.,   Cai Z.,  2017, \mn@doi
  [arXiv:1709.04962 [astro-ph]] {10.1093/mnras/sty196}

\bibitem[\protect\citeauthoryear{Pedersen, Font-Ribera, Rogers, McDonald,
  Peiris, Pontzen  \& Slosar}{Pedersen et~al.}{2020b}]{pedersen_emulator_2020}
Pedersen C.,  Font-Ribera A.,  Rogers K.~K.,  McDonald P.,  Peiris H.~V.,
  Pontzen A.,   Slosar A.,  2020b, arXiv:2011.15127 [astro-ph]

\bibitem[\protect\citeauthoryear{Pedersen, Font-Ribera, Kitching, McDonald,
  Bird, Slosar, Rogers  \& Pontzen}{Pedersen
  et~al.}{2020a}]{pedersen_massive_2020}
Pedersen C.,  Font-Ribera A.,  Kitching T.~D.,  McDonald P.,  Bird S.,  Slosar
  A.,  Rogers K.~K.,   Pontzen A.,  2020a, \mn@doi [arXiv:1911.09596
  [astro-ph]] {10.1088/1475-7516/2020/04/025}

\bibitem[\protect\citeauthoryear{Pedersen, Font-Ribera  \& Gnedin}{Pedersen
  et~al.}{2023}]{pedersen_compressing_2023}
Pedersen C.,  Font-Ribera A.,   Gnedin N.~Y.,  2023, \mn@doi [The Astrophysical
  Journal] {10.3847/1538-4357/acb433}, 944, 223

\bibitem[\protect\citeauthoryear{Pieri et~al.,}{Pieri
  et~al.}{2014}]{pieri_probing_2014}
Pieri M.~M.,  et~al., 2014, \mn@doi [Monthly Notices of the Royal Astronomical
  Society] {10.1093/mnras/stu577}, 441, 1718

\bibitem[\protect\citeauthoryear{Pontzen et~al.,}{Pontzen
  et~al.}{2008}]{pontzen_damped_2008}
Pontzen A.,  et~al., 2008, \mn@doi [Monthly Notices of the Royal Astronomical
  Society] {10.1111/j.1365-2966.2008.13782.x}

\bibitem[\protect\citeauthoryear{Puchwein et~al.,}{Puchwein
  et~al.}{2023}]{puchwein_sherwood-relics_2023}
Puchwein E.,  et~al., 2023, \mn@doi [Monthly Notices of the Royal Astronomical
  Society] {10.1093/mnras/stac3761}, 519, 6162

\bibitem[\protect\citeauthoryear{Raichoor et~al.}{Raichoor
  et~al.}{2023}]{raichoor_2023}
Raichoor A.,  et~al., 2023, in preparation

\bibitem[\protect\citeauthoryear{Ramírez-Pérez et~al.,}{Ramírez-Pérez
  et~al.}{2023}]{ramirez-perez_lyman-alpha_2023}
Ramírez-Pérez C.,  et~al., 2023, The {Lyman}-alpha forest catalog from the
  {Dark} {Energy} {Spectroscopic} {Instrument} {Early} {Data} {Release}, \url
  {http://arxiv.org/abs/2306.06312}

\bibitem[\protect\citeauthoryear{Rogers, Bird, Peiris, Pontzen, Font-Ribera  \&
  Leistedt}{Rogers et~al.}{2017}]{rogers_simulating_2017}
Rogers K.~K.,  Bird S.,  Peiris H.~V.,  Pontzen A.,  Font-Ribera A.,   Leistedt
  B.,  2017, \mn@doi [Monthly Notices of the Royal Astronomical Society]
  {10.1093/mnras/stx2942}, 474, 3032

\bibitem[\protect\citeauthoryear{Schlafly et~al.,}{Schlafly
  et~al.}{2023}]{schlafly_survey_2023}
Schlafly E.~F.,  et~al., 2023, Survey {Operations} for the {Dark} {Energy}
  {Spectroscopic} {Instrument}, \url {http://arxiv.org/abs/2306.06309}

\bibitem[\protect\citeauthoryear{Schlegel et~al.}{Schlegel
  et~al.}{2023}]{schlegel_2023}
Schlegel D.,  et~al., 2023, in preparation

\bibitem[\protect\citeauthoryear{Seljak, Slosar  \& McDonald}{Seljak
  et~al.}{2006}]{seljak_cosmological_2006}
Seljak U.,  Slosar A.,   McDonald P.,  2006, \mn@doi [Journal of Cosmology and
  Astroparticle Physics] {10.1088/1475-7516/2006/10/014}, 2006, 014

\bibitem[\protect\citeauthoryear{Silber et~al.,}{Silber
  et~al.}{2022}]{silber_robotic_2022}
Silber J.~H.,  et~al., 2022, Technical Report arXiv:2205.09014, The {Robotic}
  {Multi}-{Object} {Focal} {Plane} {System} of the {Dark} {Energy}
  {Spectroscopic} {Instrument} ({DESI}), \url
  {http://arxiv.org/abs/2205.09014}.
arXiv, \mn@doi{10.48550/arXiv.2205.09014}, \url
  {http://arxiv.org/abs/2205.09014}

\bibitem[\protect\citeauthoryear{Smee et~al.,}{Smee
  et~al.}{2013}]{smee_multi-object_2013}
Smee S.~A.,  et~al., 2013, \mn@doi [The Astronomical Journal]
  {10.1088/0004-6256/146/2/32}, 146, 32

\bibitem[\protect\citeauthoryear{{Valluri} et~al.,}{{Valluri}
  et~al.}{2022}]{valluri_snowmass_2022}
{Valluri} M.,  et~al., 2022, arXiv e-prints, \href
  {https://ui.adsabs.harvard.edu/abs/2022arXiv220307491V} {p. arXiv:2203.07491}

\bibitem[\protect\citeauthoryear{Viel, Lesgourgues, Haehnelt, Matarrese  \&
  Riotto}{Viel et~al.}{2005}]{viel_constraining_2005}
Viel M.,  Lesgourgues J.,  Haehnelt M.~G.,  Matarrese S.,   Riotto A.,  2005,
  \mn@doi [Physical Review D] {10.1103/PhysRevD.71.063534}, 71, 063534

\bibitem[\protect\citeauthoryear{Viel, Becker, Bolton, Haehnelt, Rauch  \&
  Sargent}{Viel et~al.}{2008}]{viel_how_2008}
Viel M.,  Becker G.~D.,  Bolton J.~S.,  Haehnelt M.~G.,  Rauch M.,   Sargent W.
  L.~W.,  2008, \mn@doi [Physical Review Letters]
  {10.1103/PhysRevLett.100.041304}, 100, 041304

\bibitem[\protect\citeauthoryear{Viel, Becker, Bolton  \& Haehnelt}{Viel
  et~al.}{2013}]{viel_warm_2013}
Viel M.,  Becker G.~D.,  Bolton J.~S.,   Haehnelt M.~G.,  2013, \mn@doi
  [Physical Review D] {10.1103/PhysRevD.88.043502}, 88, 043502

\bibitem[\protect\citeauthoryear{Walther, Hennawi, Hiss, O{\~n}orbe, Lee, Rorai
   \& O'Meara}{Walther et~al.}{2018}]{walther_new_2018}
Walther M.,  Hennawi J.~F.,  Hiss H.,  O{\~n}orbe J.,  Lee K.-G.,  Rorai A.,
  O'Meara J.,  2018, \mn@doi [The Astrophysical Journal]
  {10.3847/1538-4357/aa9c81}, 852, 22

\bibitem[\protect\citeauthoryear{Walther, Armengaud, Ravoux,
  Palanque-Delabrouille, Y{\`e}che  \& Luki{\'c}}{Walther
  et~al.}{2021}]{walther_simulating_2021}
Walther M.,  Armengaud E.,  Ravoux C.,  Palanque-Delabrouille N.,  Y{\`e}che
  C.,   Luki{\'c} Z.,  2021, \mn@doi [Journal of Cosmology and Astroparticle
  Physics] {10.1088/1475-7516/2021/04/059}, 2021, 059

\bibitem[\protect\citeauthoryear{Yang, Zheng, Bourboux, Dawson, Pieri, Rossi,
  Schneider  \& de~la Macorra}{Yang et~al.}{2022}]{yang_metal_2022}
Yang L.,  Zheng Z.,  Bourboux H. d. M.~d.,  Dawson K.,  Pieri M.~M.,  Rossi G.,
   Schneider D.~P.,   de~la Macorra A.,  2022, \mn@doi [The Astrophysical
  Journal] {10.3847/1538-4357/ac7b2e}, 935, 121

\bibitem[\protect\citeauthoryear{{Y{\`e}che}, Palanque-Delabrouille, Baur  \&
  BourBoux}{{Y{\`e}che} et~al.}{2017}]{yeche_constraints_2017}
{Y{\`e}che} C.,  Palanque-Delabrouille N.,  Baur J.,   BourBoux H. d. M.~d.,
  2017, \mn@doi [Journal of Cosmology and Astroparticle Physics]
  {10.1088/1475-7516/2017/06/047}, 2017, 047

\bibitem[\protect\citeauthoryear{{Y{\`e}che} et~al.,}{{Y{\`e}che}
  et~al.}{2020}]{yeche_preliminary_2020}
{Y{\`e}che} C.,  et~al., 2020, \mn@doi [Research Notes of the American
  Astronomical Society] {10.3847/2515-5172/abc01a}, \href
  {https://ui.adsabs.harvard.edu/abs/2020RNAAS...4..179Y} {4, 179}

\bibitem[\protect\citeauthoryear{{Zou} et~al.,}{{Zou}
  et~al.}{2017}]{zou_project_2017}
{Zou} H.,  et~al., 2017, \mn@doi [\pasp] {10.1088/1538-3873/aa65ba}, \href
  {https://ui.adsabs.harvard.edu/abs/2017PASP..129f4101Z} {129, 064101}

\bibitem[\protect\citeauthoryear{Zou et~al.}{Zou
  et~al.}{2023}]{zou_statistical_2023}
Zou J.,  et~al., 2023, in preparation

\bibitem[\protect\citeauthoryear{du~Mas~des Bourboux et~al.,}{du~Mas~des
  Bourboux et~al.}{2020}]{bourboux_completed_2020}
du~Mas~des Bourboux H.,  et~al., 2020, \mn@doi [The Astrophysical Journal]
  {10.3847/1538-4357/abb085}, 901, 153

\bibitem[\protect\citeauthoryear{du~Mas~des Bourboux et~al.,}{du~Mas~des
  Bourboux et~al.}{2021}]{du_mas_des_bourboux_picca_2021}
du~Mas~des Bourboux H.,  et~al., 2021, Astrophysics Source Code Library, p.
  ascl:2106.018

\makeatother
\end{thebibliography}

\appendix

\section{DESI data set comparison}
\label{appendix:data_set_comparison}

\begin{figure}
	\includegraphics[width=\columnwidth]{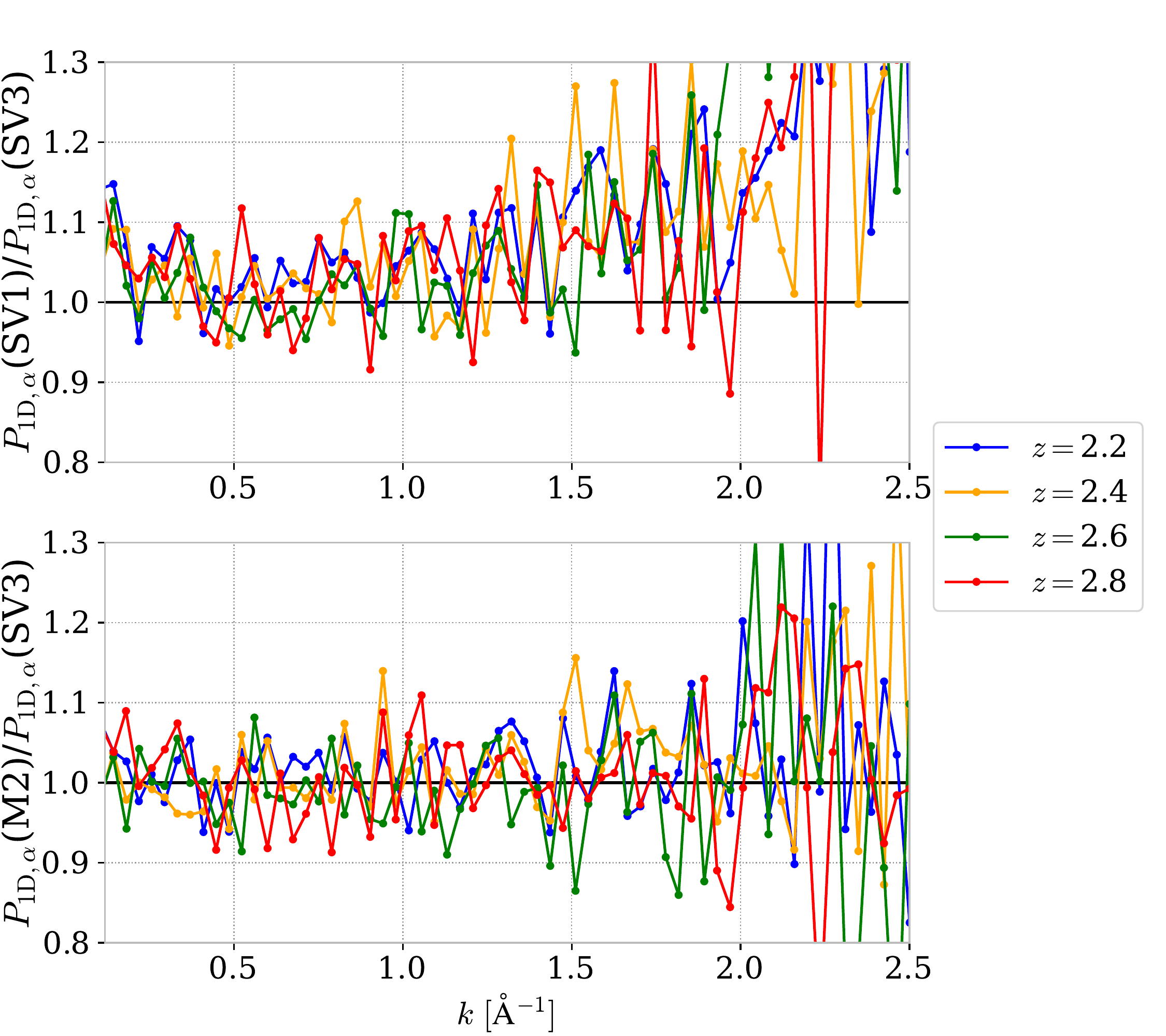}
    \caption{Ratio of the \pk, measured with the same parameters, between \svo~and \svt~(top), and \da~and \svt~(bottom), for four redshift bins.}
    \label{fig:comparison_dataset}
\end{figure}

\svo, \svt, and \da~are very different data sets: they have different target selections and exposure times and were collected for different states of the DESI instrument. From a \pk~point of view, the noise properties of \svo~are a potential issue. While our initial goal was to analyze the full data sample available, we choose first to compare the measured \pk~on the separate \svo, \svt, and \da~data sets. Fig. \ref{fig:comparison_dataset} shows their respective ratios, on the four redshift bins with largest statistics. It appears that the measurement of \pk~on \svo~is biased compared to the other two data sets. In particular, we believe that the difference at $k \gtrsim 1.0$ \invAA~is due to an imperfection in the noise correction presented in Sec. \ref{subsec:noise}. Consequently, we decide to remove the \svo~data set in this study to remain conservative.

\section{SNR weighting}
\label{appendix:snr_weighting}

For this \pk~measurement, we keep all \lya~sub-forests available in the measurement sample, independently of their \snr, unlike what was done in eBOSS \citep{chabanier_one-dimensional_2019}, reminding that only one \snr~cut is applied at earlier stages of the analysis, during the continuum fitting procedure as described in Sec. \ref{subsec:delta_extraction}.

Individual \lya~power spectra, falling into the same wavenumber bin, do not have the same dispersion which varies as function of the \snr. In our analysis, we account for this effect at the last step of our FFT estimator pipeline, by weighting each of the \lya~sub-forests by a \snr~dependent factor, while averaging over the full measurement sample.

First, for each \lya~sub-forest (i), the variance of individual \pk $_{,i}$ is fitted according to the following:

\begin{equation}
\label{eq:P1D_dispersion}
    \sigma^2 (P_{1\mathrm{D},\alpha,\mathrm{i}}) = \frac{\mathrm{a}}{(\overline{\mathrm{SNR}}_{\mathrm{i}} - 1)^2} + \mathrm{b}\,,
\end{equation}

\noindent where \snr$_{i}$ is the mean signal to noise ratio of the \lya~sub-forest (i), defined in equation (\ref{eq:mean_snr}).

Hence, the weighting factor is:
\begin{equation}
    \mathrm{W}_{i} = \frac{1}{\sigma^2 (P_{1\mathrm{D},\alpha,\mathrm{i}})}\,.
\end{equation} 

The employed fitting model works well empirically with the measured $\sigma^2 (P_{1\mathrm{D},\alpha,\mathrm{i}})$. Also according to equation (\ref{eq:P1D_dispersion}), W$_{i}$ tends to 0 as \snr$_{i}$ tends to 1, consistently with the applied \snr~cut = 1. 

\begin{figure}
	\includegraphics[width=\columnwidth]{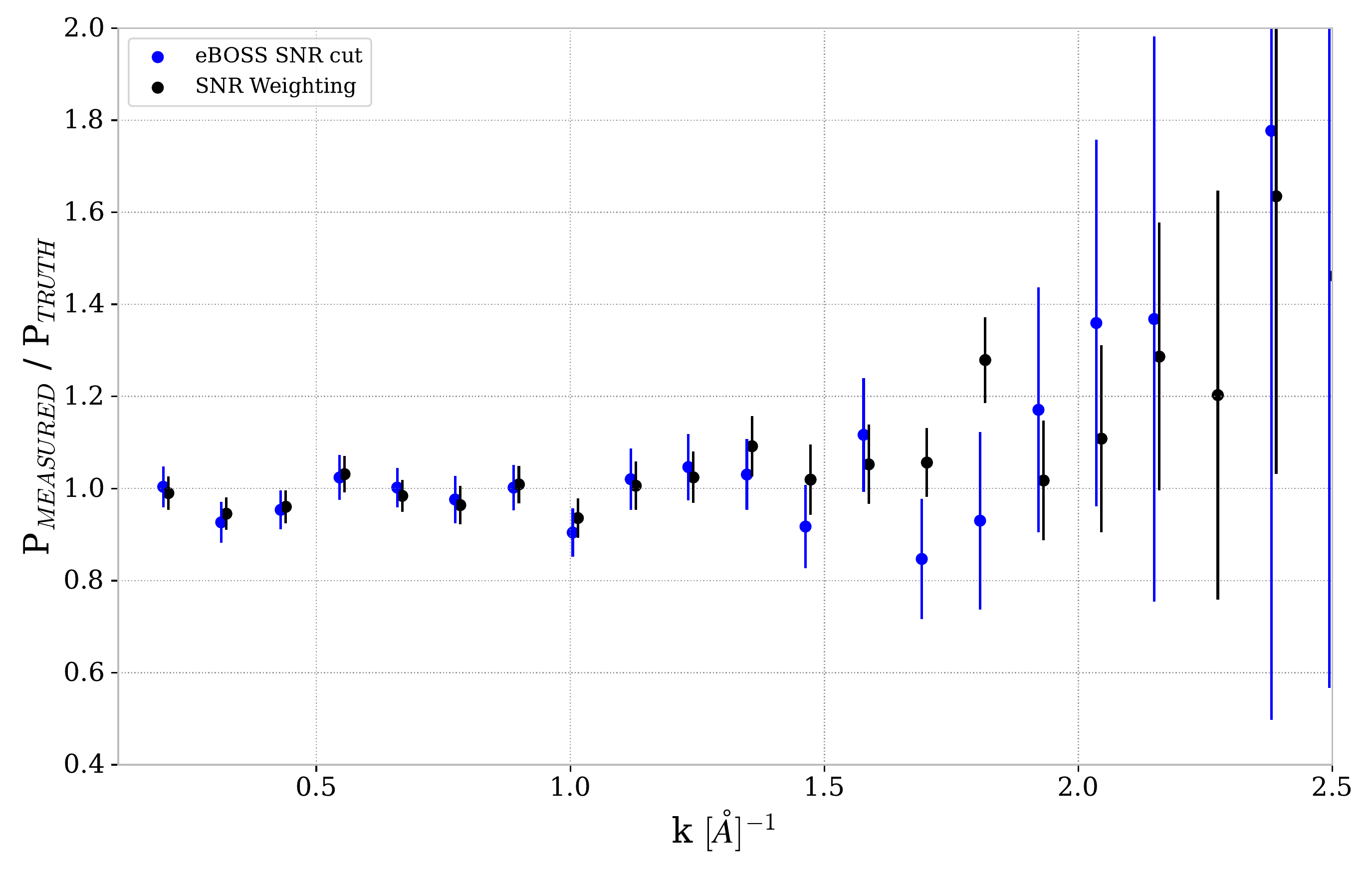}
    \caption{Comparison between measured \pk~and \texttt{DESI-Lite} mocks truth power spectrum, for the redshift bin z = 3.8, for both \pk~measured with eBOSS \snr~cut method, and the \snr~weighting method.}
    \label{fig:snr_weighting_comp}
\end{figure}

Tests for this \snr~weighting method were done on \texttt{DESI-Lite} mocks that mimic the \svo+\svt~data set described in Sec. \ref{sec:instrument_data}, and are specifically designed for \pk~measurement. A comparison between measured \pk~and mocks truth power spectrum is represented in Fig. \ref{fig:snr_weighting_comp}, for both \pk~measured with eBOSS \snr~cut method, and our \snr~weighting method.

Fig. \ref{fig:snr_weighting_comp} shows that we have an improvement compared to the eBOSS method, especially at large wavenumber and redshift values, where we are mostly limited by the statistics, as well as at low wavenumber and redshift values, while for the eBOSS measurement, there was no possible optimization at both small and large wavenumber ranges at the same time.

\section{Details on noise estimation}
\label{appendix:noise_estimators}

\subsection{Pipeline noise}

The noise associated to each spectrum is computed with the spectroscopic pipeline presented in Sec. \ref{subsec:desi_pipeline}. It is modeled as the addition of effects at the CCD scale. Each contribution is calculated by measuring the spatial variance on the CCD image from the associated noise source. It is assumed that this noise comes from the following four sources:

\begin{itemize}
    \item\textbf{Poisson noise}:\enspace Measuring photons with a CCD is a statistical process. It creates a noise source which is directly linked to the input flux, and particularly dominant for low fluxes. For DESI, this noise is estimated by modeling the CCD. 
    \item\textbf{Over-scan}:\enspace The over-scan measures the bulk offset, i.e., the average level of all CCD pixels. It is used to remove small variations in the bias. Over-scan suppression introduces noise. 
    \item\textbf{Bias}:\enspace Noise due to the response of the CCD to a minimal exposition time. It emerges from parasite electron or CCD pixel defects. The master bias estimates this noise.
    \item\textbf{Dark current}:\enspace Readout noise due to the thermal motion of the atoms composing CCD material which induces charge deposit. Dark current is estimated using the master dark. In DESI, the modeling of noise is improved to account for Poisson noise in the dark frames.
\end{itemize}

All those noise sources have all been corrected for their dependence on the CCD position. By adding these four terms, we obtain a CCD noise estimator which is propagated to the spectra by the "spectroperfectionism" formalism. A pipeline noise estimator $\sigma_{\mathrm{pip}}$ is then obtained.

The noise power spectrum estimated from pipeline (\pp) is computed from the standard deviation $\sigma_{\delta_{F}}$ linked to $\sigma_{\mathrm{pip}}$ by equation (\ref{eq:error_delta}). For each unmasked spectrum pixel, a contrast $\delta_{\mathrm{pipeline}}$ is generated following a normal probability distribution such that:

\begin{equation}
   \delta_{\mathrm{pipeline}}(\lambda) \hookrightarrow \mathcal{N}\left(0,\sigma_{\delta_{F}}(\lambda)\right)\,.
\end{equation}

This procedure is repeated $N_{\mathrm{G}}$ times ($N_{\mathrm{G}} = 2500$) to obtain a converged noise power spectrum. For each quasar, the associated noise power spectrum is the average of the $N_{\mathrm{G}}$ noise contrasts after Fourier transformation:

\begin{equation}
   P_{\mathrm{pipeline}}(k) = \left\langle \left|\delta_{\mathrm{pipeline}}\right|^2 \right \rangle_{N_{\mathrm{G}}}\,.
\end{equation}

\subsection{Exposure difference noise}

Another noise estimation can be done using the difference between exposures of the same quasar, when several exposures are available for the same object. The difference between exposures removes the physical signal, leaving only the fluctuations due to noise. We implemented a noise power spectrum estimator using this principle. We define the difference coadd of a quasar of index $j$ by separating half of its exposures in the even category ($N_{\mathrm{even}}$ exposures) and the other half in the odd category ($N_{\mathrm{odd}}$ exposures) such that:

\begin{equation}
    \Delta f_{j} = \frac{1}{2} \left( \frac{\sum_{k=1}^{N_{\mathrm{even}}} \left(\sigma_{\mathrm{pip},k}\right)^{-2} f_{k}}{\sum_{k=1}^{N_{\mathrm{even}}} \left(\sigma_{\mathrm{pip},k}\right)^{-2}} -  \frac{\sum_{k=1}^{N_{\mathrm{odd}}} \left(\sigma_{\mathrm{pip},k}\right)^{-2} f_{k}}{\sum_{k=1}^{N_{\mathrm{odd}}} \left(\sigma_{\mathrm{pip},k}\right)^{-2}}\right)\,,
\end{equation}

\noindent where $\sigma_{\mathrm{pip,k}}^2$ is the pipeline noise of the exposure $k$ for quasar $j$. In the case where the total number of exposures is even, $N_{\mathrm{even}} = N_{\mathrm{odd}}$. The standard deviation of $\Delta f_{j}$ can be calculated from the variances of individual exposures:

\begin{equation}
    \sigma_{\Delta f_{j}} = \frac{1}{2} \sqrt{\frac{1}{\sum_{k=1}^{N_{\mathrm{even}}} \left(\sigma_{\mathrm{pip},k}\right)^{-2}} + \frac{1}{\sum_{k=1}^{N_{\mathrm{odd}}} \left(\sigma_{\mathrm{pip},k}\right)^{-2}}}\,.
\end{equation}

This difference coadd is unbiased, i.e., of zero average, whatever the values of the sum of the inverse variance for both exposure populations. Finally, this estimator does not necessarily need an even total number of exposures. 

To derive an estimator of \pn, the variance of $\Delta f_{j}$ must be equal to that of the coadded flux defined by:

\begin{equation}
    \label{eq:coadd}
    f _{j}= \frac{\sum_{k} \left(\sigma_{\mathrm{pip},k}\right)^{-2} f_{k}}{\sum_{k} \left(\sigma_{\mathrm{pip},k}\right)^{-2}}\,,
\end{equation}

To obtain the same variance, we multiply $\Delta f_{j}$ by the ratio $\sigma_{f_{j}} / \sigma_{\Delta f_{j}}$:

\begin{equation}
    \label{eq:diff_exp_corr}
    \Delta f_{j}^{\mathrm{corr}} = 2 \frac{\frac{1}{\sqrt{\sum_{k=1}^{N_{\mathrm{tot}}} \left(\sigma_{\mathrm{pip},k}\right)^{-2}}}}{ \sqrt{\frac{1}{\sum_{k=1}^{N_{\mathrm{even}}} \left(\sigma_{\mathrm{pip},k}\right)^{-2}} + \frac{1}{\sum_{k=1}^{N_{\mathrm{odd}}} \left(\sigma_{\mathrm{pip},k}\right)^{-2}}}} \Delta f_{j}\,,
\end{equation}

\noindent where $N_{\mathrm{tot}}$ is the total number of exposure for the quasar $j$ ($N_{\mathrm{tot}} = N_{\mathrm{even}} + N_{\mathrm{odd}}$).

In SDSS analysis \citep{mcdonald_lyman-alpha_2006,palanque-delabrouille_one-dimensional_2013,chabanier_completed_2021}, the variance of all exposures for a given object was considered equal. In this case, $\sigma_{\Delta f_{j}}$ can be simplified, and a correction was applied only in the case of an odd number of exposures. In the case of a constant exposure variance, the difference coadd in equation (\ref{eq:diff_exp_corr}) is equal to the one derived in \citet{palanque-delabrouille_one-dimensional_2013,chabanier_one-dimensional_2019}. Our new estimator corrects the variance for any exposure time. It is essential in the case of DESI first data, for which the exposure times can be very variable compared to SDSS.

The exposure difference coadd is computed in \texttt{picca}$^{\ref{igmhub/picca}}$ \citep{du_mas_des_bourboux_picca_2021}. We obtain an estimator of \pn~called difference power spectrum and noted \pd, such that:

\begin{equation}
P_{\mathrm{diff}}(k) = \left\rvert \mathcal{F} \left[ \frac{\Delta f_{j}^{\mathrm{corr}} (\lambda)}{\overline{F}(\lambda) C_{\mathrm{q}}(\lambda)} - 1 \right] \right\vert^2 \,.
\end{equation}

\subsection{Additional considerations on noise estimation}
\label{subsec:noise_add}

\begin{figure}
	\includegraphics[width=\columnwidth]{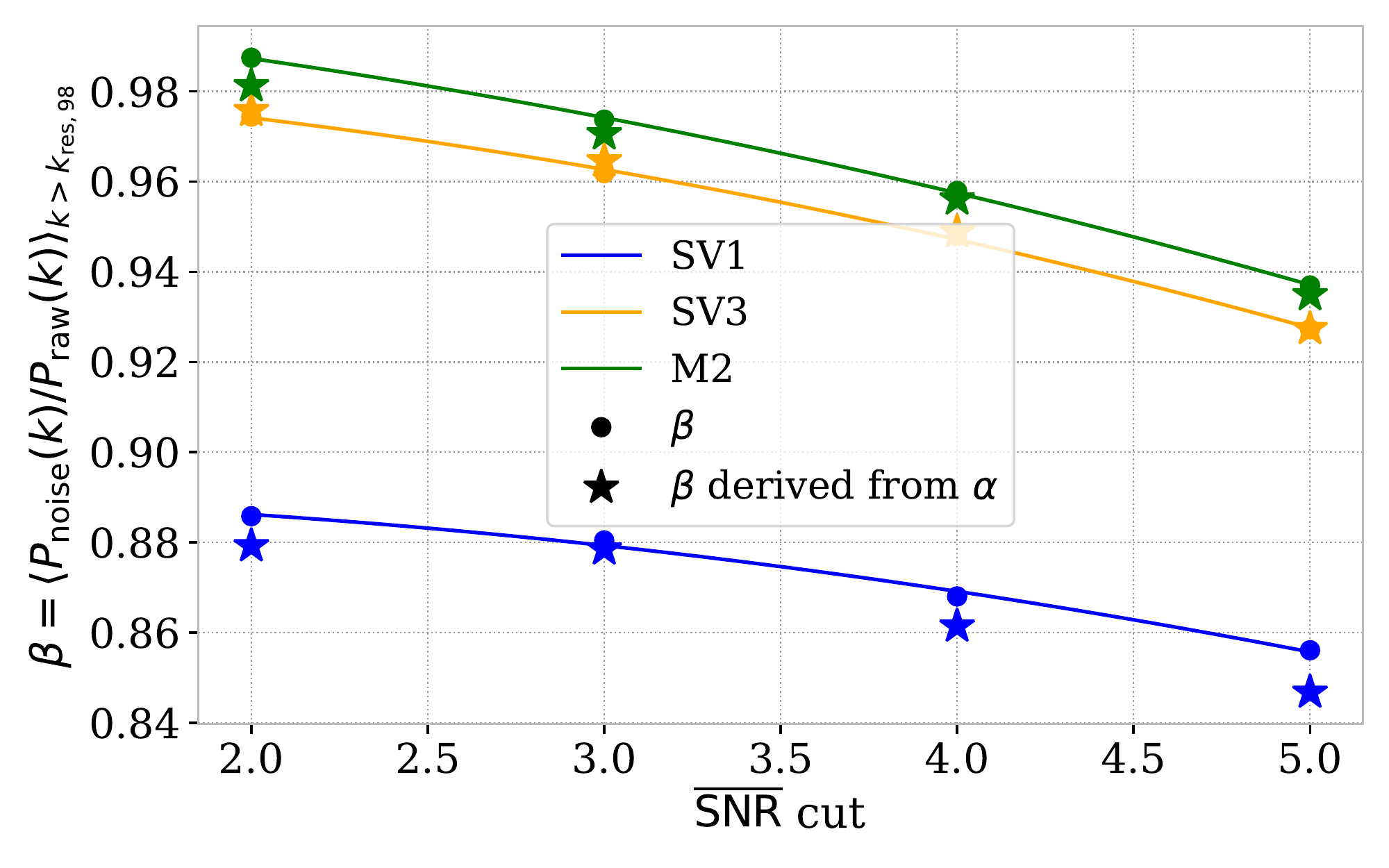}
    \caption{Asymptotic ratios $\beta$ between the noise and raw power spectra for \svo~(blue), \svt~(yellow), and \da~(green) data sets, as a function of the minimal \snr~cut, for the pipeline noise. Points give the direct estimation of $\beta$. Stars represent $\beta$ as derived from the asymptotic differences in Fig. \ref{fig:diff_pipeline_comparison}. Second-order polynomial fits are shown only for representation.}
    \label{fig:noise_study_beta}
\end{figure}

\begin{figure}
	\includegraphics[width=\columnwidth]{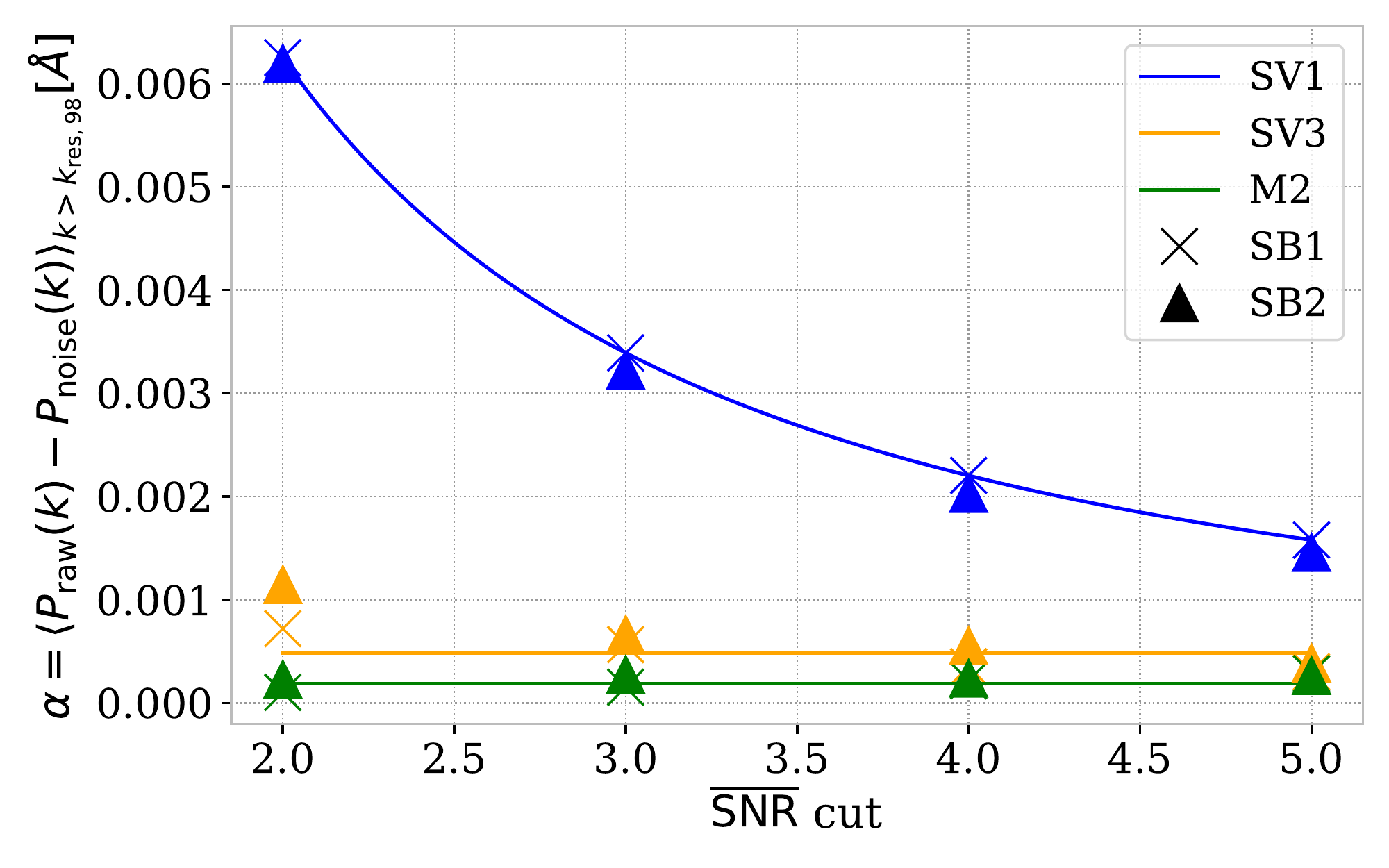}
    \caption{Asymptotic differences $\alpha$ for \svo~(blue), \svt~(yellow), and \da~(green) data sets in the side-band regions SB1 (crosses) and SB2 (triangles), for the pipeline noise. The continuous lines are fits of the $\alpha$ values, whose parameters are given in Tab. \ref{tab:noise_correction}}
    \label{fig:noise_study_sb}
\end{figure}

Fig. \ref{fig:noise_study_beta} shows the asymptotic ratios $\beta$ for the three data sets. We remark that for \svt~and \da, the \snr~dependence of $\beta$ is much more pronounced than that of $\alpha$. The absolute noise level is the main parameter which varies when changing the minimal \snr~cut. It indicates that the residual noise source is additive rather than multiplicative. To support this hypothesis, we computed the $\beta$ values derived from the $\alpha$ of Fig. \ref{fig:diff_pipeline_comparison}, using the mean value of $P_{\mathrm{noise}}$ for all redshift. They are shown as stars in Fig. \ref{fig:noise_study_beta}, and exhibit similar trends to the direct $\beta$ computation, which corroborates that the missing noise is additive. Consequently, we decide to correct \pp~using an additive term $\alpha$ ($P_{\mathrm{noise}} = P_{\mathrm{pipeline}} + \alpha$).

The same noise study is performed on side-band regions SB1 and SB2 for which the astrophysical signal, i.e. absorption from intergalactic elements, is much lower than the \lya~band. The $\alpha$ values are shown in Fig. \ref{fig:noise_study_sb}. For side-bands, the overall missing noise level exhibits similar trends as a function of the minimal \snr~cut, but is lower than for the \lya~band. This is likely due to the use of different quasar populations employed for side-band and \lya~measurements. Indeed, the DESI observation strategy is different for low redshift quasars (used for side-band study) and \lya~quasars. On average, the number of exposures is larger for \lya~than low-redshift quasars. Consequently, and in accordance with our previous interpretations, the misestimation of noise is larger for \lya~measurement than side-bands. The difference might also be due to the much lower astrophysical signal in the side-bands, allowing an improved estimation of asymptotic noise level. As the results for SB1 and SB2 are very similar, we decide to apply the same correction for these two bands.

\section{Comparison with eBOSS measurement on large scales}
\label{appendix:comparison_eboss}

We performed a series of tests to investigate the discrepancy at large scales ($k < 0.01~\invkms$) between our measurement and the eBOSS measurement in~\citet{chabanier_one-dimensional_2019} as seen in Fig.~\ref{fig:comparison} (left).

We first focused on reproducing the eBOSS measurement with the \texttt{picca} software used in our analysis and the eBOSS parameters. Starting from SDSS spectra, we applied the pipeline used in~\citet{chabanier_one-dimensional_2019}, and successively replaced each step (continuum fitting in Sec.~\ref{subsec:delta_extraction}, Fourier transform and averaging in Sec.~\ref{subsec:fft_estimator}) by the new \texttt{picca} software. We assessed that the version used in our analysis could reproduce the eBOSS measurement without noticeable bias at all scales.

Compared to the eBOSS measurement, some parameters are changed for the continuum fitting presented in Sec.~\ref{subsec:delta_extraction}. For eBOSS, this pipeline step was performed separately on sub-forests instead of the total \lya~forest. We checked that performing our continuum fitting on sub-forest does not modify the large-scales level of \pk. Additionally, we have performed the following changes in the continuum analysis. We removed the smoothing of the common continuum in equation \ref{eq:smoothing_continuum}, which was not used in~\citet{chabanier_one-dimensional_2019}. We changed the polynomial order in equation \ref{eq:continuum_quasar} to zero as in eBOSS. Additionally, we tested to modify parameters of the continuum fitting procedure, which were used as eBOSS but could potentially change the level of \pk~at large scales. We also applied non-constant weights in equation \ref{eq:noise_delta_extraction}, removed the forcing to zero of the \lya~contrast stack, or changed the observed wavelength to a smaller range. The conclusion of those continuum fitting tests is that none of those effects could be responsible for the $\sim 15\%$ discrepancy visible on the eBOSS comparison.
Furthermore, the corrections we derived from the mocks in Sec.~\ref{sec:synthetic_data} are different than the eBOSS corrections and these differences could explain the disagreement between the two measurements. To check the impact of corrections, we computed an eBOSS measurement without any corrections, and we realized the comparison in Fig.~\ref{fig:comparison} (left), adding the corrections successively. This test yields that part of the disagreement (between $4$ and $6$ \%) is due to the continuum fitting correction, which is different from eBOSS (see Fig. 7 in~\citet{chabanier_one-dimensional_2019}). 

Another effect that could impact the largest scale is the possible impurity and incompleteness of the DLA catalog. In order to eliminate effects due to differences in the DLA catalog, we created a common set of quasars and DLAs by merging eBOSS and DESI catalogs. We found that the results from this common set of quasars are not significantly different, which indicates that missing DLAs cannot fully explain the disagreement. Finally, we varied the DLA and BAL catalog used by varying confidence levels, or column density $N_{\mathrm{\hi}}$ to include sub-DLAs. Masking and correcting the damping wings for $\log(N_{\mathrm{\hi}}) < 20.3$ systems decreases the discrepancy only for the very first wavenumber bins ($k < 0.004~\invkms$).

To conclude, the large-scale disagreement between our measurement and eBOSS~\citep{chabanier_one-dimensional_2019} cannot be fully explained for the moment. The continuum correction is responsible for a portion of this discrepancy. A complete study on DLA completeness or a comparison at the spectrum level between eBOSS and DESI will be performed in future studies to investigate in detail this discrepancy

\bsp
\label{lastpage}
\end{document}